\documentclass[a4paper,12pt]{article}
\usepackage{jheppub} % for details on the use of the package, please see the JINST-author-manual
\usepackage{lineno}
\usepackage{amsmath}
\usepackage{tikz}
\usepackage{spectralsequences}
\usepackage{hyperref}
\usepackage{jheppub}
\usepackage{xcolor}
\usepackage{tikz-cd}
\usepackage{pgfplots}
\usepackage{tikz-3dplot}
\usepackage{braket}
\usepackage{tikz,lipsum,lmodern}
\usepackage[most]{tcolorbox}
\tcbuselibrary{skins}
\AtBeginDocument{%
  \setlength{\abovedisplayskip}{12pt plus 3pt minus 3pt}%
  \setlength{\belowdisplayskip}{12pt plus 3pt minus 3pt}%
  \setlength{\abovedisplayshortskip}{6pt plus 2pt}%
  \setlength{\belowdisplayshortskip}{8pt plus 2pt minus 2pt}%
}

\usetikzlibrary{matrix}
\newcommand*\Z{\mathbb{Z}}

\usepackage{tikz}

\definecolor{MfiveBlue}{HTML}{69A9DC}
\definecolor{MfiveEdge}{HTML}{22577A}
\definecolor{MtwoOrange}{HTML}{F4A261}
\definecolor{MtwoEdge}{HTML}{B45309}

\tikzset{
    m5brane/.style={
        fill=MfiveBlue,
        fill opacity=0.32,
        draw=MfiveEdge,
        line width=1pt
    },
    m2brane/.style={
        fill=MtwoOrange,
        fill opacity=0.72,
        draw=MtwoEdge,
        line width=1pt
    },
    m2boundary/.style={
        draw=MtwoEdge,
        line width=2.2pt,
        line cap=round
    }
}
\usetikzlibrary{calc,arrows.meta}

\definecolor{DpBlue}{RGB}{66,135,200}
\definecolor{VBOrange}{RGB}{240,153,62}
\definecolor{VBPurple}{RGB}{150,105,190}
\definecolor{SigmaRed}{RGB}{190,45,55}

\tikzset{
    dpface/.style={
        draw=DpBlue!80!black,
        fill=DpBlue!25,
        fill opacity=0.32,
        draw opacity=1,
        line width=0.8pt,
        line join=round
    },
    dpedge/.style={
        draw=DpBlue!80!black,
        line width=0.8pt
    },
    interfaceA/.style={
        draw=VBOrange!85!black,
        fill=VBOrange!75,
        fill opacity=0.72,
        draw opacity=1,
        line width=0.9pt
    },
    interfaceB/.style={
        draw=VBPurple!85!black,
        fill=VBPurple!65,
        fill opacity=0.72,
        draw opacity=1,
        line width=0.9pt
    },
    twistline/.style={
        draw=SigmaRed,
        line width=2.2pt,
        line cap=round
    },
    morphismlabel/.style={
        font=\small,
        align=center
    }
}
\usepackage{booktabs}
\usepackage{tabularx}
\usepackage{array}
\usepackage[table]{xcolor}

\definecolor{BeforeBlue}{HTML}{DCEAF7}
\definecolor{GaugingPurple}{HTML}{E9DDF2}
\newcolumntype{Y}{>{\raggedright\arraybackslash}X}
\definecolor{HeaderBlue}{HTML}{356A9A}
\definecolor{BareBlue}{HTML}{E8F2FA}
\definecolor{BraneOrange}{HTML}{FCEBDD}

\usepackage{tikz}
\usepackage{float}
\usetikzlibrary{arrows.meta,calc}

\definecolor{braneB}{RGB}{45,105,185}
\definecolor{braneC}{RGB}{190,55,55}
\definecolor{created}{RGB}{220,135,35}

\tikzset{
    brane/.style={
        line width=2.2pt
    },
    motion/.style={
        ->,
        thick,
        gray!75
    },
    createdbrane/.style={
        line width=2.8pt,
        created
    },
    blueplane/.style={
        draw=braneB!80!black,
        fill=braneB,
        fill opacity=0.18,
        draw opacity=0.65,
        line width=0.6pt
    },
    redplane/.style={
        draw=braneC!80!black,
        fill=braneC,
        fill opacity=0.16,
        draw opacity=0.65,
        line width=0.6pt
    }
}

\usetikzlibrary{arrows.meta}

\definecolor{BoundaryBlue}{HTML}{DCEAF7}
\definecolor{BoundaryEdge}{HTML}{477A9F}
\definecolor{BulkLineColor}{HTML}{30343B}
\definecolor{ProbeRed}{HTML}{C94C4C}

\tikzset{
    boundary plane/.style={
        draw=BoundaryEdge,
        fill=BoundaryBlue,
        line width=0.8pt
    },
    bulk Wilson line/.style={
        draw=BulkLineColor,
        line width=1.6pt
    },
    probe Wilson line/.style={
        draw=ProbeRed,
        line width=1.7pt
    },
    vertex operator/.style={
        circle,
        fill=black,
        inner sep=2.2pt
    },
    braid arrow/.style={
        draw=black,
        line width=0.9pt,
        -{Stealth[length=5pt,width=5pt]}
    }
}

\usepackage{booktabs}
\usepackage{tabularx}
\usepackage{array}
\usepackage[table]{xcolor}
\definecolor{BeforeColor}{HTML}{DCEAF7}
\definecolor{AfterColor}{HTML}{F8E2CF}
\usepackage{graphicx}
\usepackage{mathtools}
\usepackage{arydshln}

\usepackage{multirow}
\newcolumntype{Y}{>{\raggedright\arraybackslash}X}
\title{\boldmath Non-Invertible Charges for M-Theory}

\author{Kenneth Intriligator, Aiden Sheckler, Eric Y. Yang}
\affiliation{Department of Physics, University of California, San Diego,\\
9500 Gilman Drive, La Jolla CA 92093-0319, USA}

\abstract{The gauge charges of M-theory M2 and M5 branes {\it almost} resemble the charges of a 6-group. The charges of such a 6-group are expected to be modeled by operators with non-invertible fusion. We here consider non-invertible charge defects for continuous 2-groups and 6-groups, aiming to capture the gauge charges of M2 and M5 branes. We find a model for charges of a {\it pseudo 6-group} $\frak G$ which possesses remarkable similarities with known features of M2 and M5 branes, including: Hanany-Witten moves, the $\frak a_{N-1}$ charges, higher-form symmetry, and singleton decomposition of M5-brane worldvolume theories, and couplings of M5/anti-M5 states. }
\begin{document}

\maketitle

\section{Introduction}

Extended operators and excitations are important components of quantum field theory and string theory. A breakthrough insight of generalized symmetries is the realization that such objects may be charged under “higher-form” symmetries \cite{Gaiotto:2014kfa}.\footnote{See \cite{Brennan:2023mmt, McGreevy:2022oyu, Costa:2024wks, Bhardwaj:2023kri, Shao:2023gho} for a review.} As a result, the powerful tools of global symmetry in quantum theories have been extended and applied to illuminate the nature of extended objects in QFT. It is also useful to consider gauging such generalized symmetries, in which case the extended operators may carry gauge charge.\footnote{This is especially crucial for theories of quantum gravity (including M-theory) which are expected to possess no global symmetries \cite{Banks:1988yz, Kallosh:1995hi, Banks:2010zn, Harlow:2018tng}.} Although a gauged symmetry is simply a useful redundancy, and may not be considered fundamental, gauge charges still provide convenient labels on operators which can be leveraged to organize interactions. \\

We here consider aspects of the gauge charges of M2 and M5 branes of M-theory. This is a rich subject with an enormous literature, spanning decades of development in physical mathematics; a partial list of foundational references that are relevant for our present considerations include~\cite{Witten:1996md, Witten:1996hc, Freed:1998tg, Diaconescu:2000wz, Hopkins:2002rd, Diaconescu:2003bm, Moore:2004jv, Freed:2004yc, Freed:2006yc, Fiorenza:2019usl, Banerjee:2026wop, Golub:2026mfm} and references therein. Our purpose is to bring to bear some useful developments in the study of generalized symmetries on the investigation of these fundamental objects of M-theory. We here propose a model which captures the nature of the gauge charges, using recent developments in the study of higher group and non-invertible symmetries. We show that this model captures key aspects of M2 and M5 brane charges, and may be useful for further development of our understanding of both M-theory and generalized symmetries. \\

\subsection{Gauge Charges of M-Theory}

To set the stage, let us review what is meant by the gauge charges of M2 and M5 branes. We give a more thorough and precise review in \hyperref[sec22]{Section 2.2}. The worldvolume theory of an M2-brane is known to host a 3-form gauge field $C_3$, with coupling given by the holonomy over the worldvolume $\gamma_3$:
\begin{equation}\label{c}
    S_{M2}\supset \int_{\gamma_3}C_3
\end{equation}
On the other hand, the M5-brane sources magnetic flux for this gauge field, so that the field strength $G_4$ integrates to unit flux around any 4-sphere $S^4$ linking the brane worldvolume $\Sigma_6$. However there is an additional subtlety which is visible from the low-energy effective theory of 11d supergravity. The supergravity action is known to host a Chern-Simons term which is cubic in the gauge field:
\begin{equation}
    S_{\text{sugra}}\supset \int_{M_{11}}\frac 16C_3G_4G_4-C_3I_8
\end{equation}
The curious $\frac 16$ quantization is especially subtle, and at first seems ill-defined. This was beautifully resolved by Witten \cite{Witten:1996md}, and plays an important role in our analysis. We review the full definition in \hyperref[appa]{Appendix A}. A consequence of this Chern-Simons term is that the worldvolume theory of an M5-brane is not itself gauge invariant under gauge transformations $C_3\to C_3+d\lambda_2$, but rather has an anomalous variation which is canceled by the bulk action \cite{Witten:1996hc}:
\begin{equation}\label{gt}
    \delta_{\lambda_2} \mathcal S_{M5}\sim -\frac{1}{4\pi}\int_{\Sigma_6}\lambda_2\wedge G_4
\end{equation}
The properties \hyperref[c]{(1.1)} and \hyperref[gt]{(1.3)} are schematically what we refer to as characterizing the gauge charges of the branes. The key observation we consider is that these charges are \textit{almost} the same as the charges of a 6-group $\mathbb G^{(6)}$, defined by extending a $U(1)^{(2)}$ 2-form symmetry by a $U(1)^{(5)}$ 5-form symmetry with Postnikov class $\hat \kappa=1$. Indeed, the gauge fields of this 6-group may be decomposed into a 3-form $C_3$ and a 6-form $B_6$. The higher-group structure requires modified gauge transformations:
\begin{equation}
    \begin{gathered}
        C_3\to C_3+d\lambda_2 \\ 
        B_6\to B_6-\frac {\hat\kappa}{2\pi}\lambda_2G_4+d\Lambda_5
    \end{gathered}
\end{equation}
One sees immediately the similarity to the gauge charges \hyperref[c]{(1.1)} and \hyperref[gt]{(1.3)}, however there are several subtle differences. An obvious difference is the factor of $\frac 12$ in the gauge variation which will play an important role, but another difference lies in a nuance of the M-theory $C_3$ field. Due to fermionic contributions on the M2-brane worldvolume, this field is not an ordinary abelian gauge field, but must in fact have a shifted flux quantization condition, so that $\frac{1}{2\pi}\int G_4=\int \frac 12 w_4(TM_{11})\mod \mathbb Z$. There are also gravitational anomalies which play a crucial role. We give a more thorough overview and comparison in \hyperref[sec2]{Section 2}. \\

The fundamental inspiration for this work is to leverage the great advancements which have been made in recent years on understanding higher groups and their charges in quantum field theory to this problem of the M2 and M5 gauge charges. In particular, the charges of a 6-group $\mathbb G^{(6)}$ gauge symmetry are expected to be modeled by 3-dim and 6-dim defect operators, where the 6d operators are \textit{non-invertible}, and thus described by a higher-categorical structure. Our ambition here is to derive and study these charges, and use them as a model for the M2 and M5 brane gauge charges. To this end, we hope to capture key features of the quantum nature of these objects, which may provide a pathway towards deeper understanding of the structure of M-theory. \\

\subsection{Higher Groups and Charges}

Before presenting our results, let us first specify more precisely what we mean by higher groups and their charges in the context of quantum field theory. 
Just as different ordinary (0-form) symmetries may combine into rich structures, for example via group extensions or mixed anomalies, so too can higher-form symmetries combine to participate in more intricate symmetry structures. In fact the generalization to higher-form symmetries allows for an even richer landscape, because now higher symmetries of \textit{different degrees} can mix nontrivially. If we require that the overall symmetry remains invertible, then a generic consistent structure that higher-form symmetries can organize into is a \textit{higher group}. For example, an “extension” of a 0-form symmetry $G^{(0)}$ by a 1-form symmetry $H^{(1)}$ can form a 2-group:
\begin{equation}
    H^{(1)}\to \mathbb G^{(2)}\to G^{(0)}
\end{equation}
More generally, higher-form symmetries of degree up to $p-1$ may organize to form a $p$-group $\mathbb G^{(p)}$. Higher groups appear in diverse corners of quantum field theory and string theory. Higher group global symmetries in field theory have recently been studied \cite{Cordova:2018cvg, Benini:2018reh, Cordova:2020tij, Kapustin:2013uxa, Brennan:2020ehu, Hidaka:2020iaz, Harder:2026csj,Delcamp:2019fdp} (see also \cite{Brito:2026eqi} for a recent lattice realization of 2-group symmetry). The Green-Schwarz mechanism of supergravity gauge fields in string theory also resembles a higher-group structure. We give a brief review of higher groups and their gauge fields in \hyperref[sec21]{Section 2.1}. \\

In the case of 0-form symmetries, an extension between two symmetries is a valuable tool, since it only augments the algebraic structure available to us to constrain or organize the theory. In particular, the extension data has dramatic implications for the \textit{charges} of the symmetry. Quantum mechanically, any charged operator must transform under a representation of the symmetry. We typically identify the \textit{charge} $R$ of an operator $\mathcal O$ under a symmetry $G$ as the specific representation $R\in \text{Rep}(G)$ that it transforms under. In this way, we may discuss $\text{Rep}(G)$ as the algebra of charges. This can also be generalized to higher-form and higher-group symmetries \cite{Bartsch:2022mpm, Bartsch:2022ytj,Bhardwaj:2023wzd, Bhardwaj:2022lsg, Bhardwaj:2022yxj, Bhardwaj:2022kot, Bhardwaj:2022maz, Decoppet:2023bay, Bartsch:2023pzl, Delcamp:2023kew}. Generalized symmetries provides the appropriate framework to work concretely with the operators representing both the symmetries of $G$ and the charges $\text{Rep}(G)$. As we review in \hyperref[sec3]{Section 3}, both a global symmetry $G$ and charges $\text{Rep}(G)$ may be represented by (possibly extended) operators. The unitary operators generating global symmetry $G$ transformations are topological operators, while the operators carrying charge $R\in \text{Rep}(G)$ are typically non-topological. However it is often possible to find a \textit{different} theory where the operators associated to $R\in \text{Rep}(G)$ are topological. For example, for finite symmetries this can be achieved by gauging $G$. Another useful framework is to work with the SymTFT \cite{Gaiotto:2020iye, Freed:2022qnc, Bhardwaj:2023ayw, Kaidi:2022cpf, Thorngren:2019iar}, which is a topological theory in one-higher dimension which simultaneously hosts topological operators representing $G$ and $\text{Rep}(G)$. These operators may be used to study the full symmetry algebra and charge algebra respectively, which generally may be formalized as (higher) fusion categories. \\

When the symmetry $G$ is an extension, this extension data can be used to organize the rich structure of $\text{Rep}(G)$. In particular when the extension is non-split, this algebra can be categorical, in which case $\text{Rep}(G)$ is identified as a non-invertible higher fusion category\footnote{This is neither a necessary nor sufficient condition for non-invertibility. For example, in 2D sequentially gauging a $\mathbb{Z}_N^{(0)}$ and a $\mathbb{Z}_2$ which acts on $\mathbb{Z}_N^{(0)}$ as charge conjugation produces a $\text{Rep}(D_{N})$ symmetry, which is non-invertible. }. This is particularly well-appreciated for finite 0-form symmetries \cite{Tachikawa:2017gyf}. In this case, the topological operators representing $\text{Rep}(G)$ can have intricate fusion relations. This includes non-invertible fusion $\mathcal D\otimes \mathcal D^\dag \neq 1$, as well as more complicated fusion channels involving junctions of higher-codimension, which are categorically formalized as higher morphisms. \\

The same is true of higher groups. For a finite higher group $\mathbb G^{(p)}$ with nontrivial “extension class”, the resulting higher fusion category $\text{Rep}(\mathbb G^{(p)})$ may be a non-invertible fusion category. Such examples have received comparatively less attention than representation categories of finite 0-form symmetries. As a simple example, the minimal abelian TQFTs $\mathcal A^{N,p}$ which describe a 2+1d abelian topological order with anomalous 1-form symmetry \cite{Hsin:2018vcg}, may be used to construct the non-invertible defects for the representation category $\text{Rep}(\mathbb G^{(3)}_N)$ of a finite 3-group $\mathbb G^{(3)}_N$. We review this example in \hyperref[sec32]{Section 3.2}. Such representation categories in general may be immensely intricate structures. Nevertheless, given the extension data of $\mathbb G^{(p)}$, there is sometimes an ad-hoc procedure for studying at least some of the non-invertible defects in $\text{Rep}(\mathbb G^{(p)})$. \\

\subsection{Modeling M2 and M5 Brane Charges}

The observation of the higher structure of M2 and M5 gauge charges has been noted many times throughout the literature \cite{DAuria:1982uck, Henneaux:1984mh, Townsend:1995gp, Candiello:1993di, deAlwis:1997gq, Bandos:1997gd, Bandos:2004xw, Fiorenza:2013nha, Giotopoulos:2024xcg, Fiorenza:2012mr, Sati:2024klw, Giotopoulos:2024sit, Fiorenza:2019ckz, Kalkkinen:2003gq, Fiorenza:2018ekd, Giotopoulos:2024ovz, Armas:2024caa, Rosabal:2025zry}, dating back to \cite{DAuria:1982uck}. In particular, relation to a possible 6-group structure was noted in \cite{Armas:2024caa}. There has been vigorous work over the years studying the precise characterization of this higher-symmetry structure, and proposed models for the M2 and M5 brane charges. Our approach here will remain more or less agnostic to the details of these formalizations, however we survey various possibilities and proposals in \hyperref[sec55]{Section 5.5}. For now, we simply refer to this mysterious “pseudo-6-group” structure which reproduces the gauge charges \hyperref[c]{(1.1)} and \hyperref[gt]{(1.3)} as $\frak G$. \\

Our approach will instead pursue the following logic: in 11d the charges of a true 6-group $\mathbb G^{(6)}$ will include 3-dim and 6-dim defects. In particular the 6-dim defects are expected to be represented by non-invertible topological operators due to the nontrivial Postnikov class. Our goal will be to construct these defects and study their fusion. Along the way, we will find that it is in fact most convenient to \textit{first} construct defects which have precisely the properties of charges for $\frak G$. It will turn out that this approach allows for construction of the charge defects for \textit{both} $\frak G$ and $\mathbb G^{(6)}$. After constructing the 3d and 6d charge defects of $\frak G$ and studying their fusion, we would like to treat them as a \textbf{model} for the gauge charges of M2 and M5 branes. We will then empirically compare the results of our construction to known features of M2 and M5 branes.  \\

Our results show remarkable correspondence with known phenomena of M2 and M5 branes. We find that the topological defects representing charges of $\frak G$ act as models of M2 and M5-branes which reproduce:
\begin{itemize}
    \item Junctions of an M2-brane ending on an M5-brane. 
    \item The organization of string charges on a stack of $N$ M5-branes along the root lattice of the $\frak a_{N-1}$ Lie algebra, and the defect group representing the $\mathbb Z_N^{(2)}$ 2-form symmetry.
    \item A clear decomposition of the charges on a stack of $N$ M5-branes into an $\frak a_{N-1}$ charge sector and a singleton sector capturing the center-of-mass mode. 
    \item The gauge coupling of M5/anti-M5 brane states.
    \item Hanany-Witten moves, where an M5-brane crossing another M5-brane yields a stretched M2-brane between them.
    \item Procedures for analyzing higher-dimensional junctions of branes. For example, 4d junctions of M5-branes representing $\frac 12$-BPS codimension-2 defects of the worldvolume theory. 
\end{itemize}

In the course of our analysis of the (pseudo) 6-group charges of the 11d theory, we will find that there is an almost exact analog to 2-group charges in 5d, which might be considered as modeling the charges of 5d Maxwell-Chern-Simons. This setting is almost an exact parallel of the 11d case, except all the dimensions are shifted to a lower degree. The pseudo-6-group gauge fields $(C_3,B_6)$ modeling M2 and M5 branes are replaced by 2-group gauge fields $(A_1,B_2)$ modeling Wilson lines and 't Hooft surfaces. We will therefore find it comfortable to first construct the charges for a 2-group $\mathbb H^{(2)}$ and pseudo-2-group $\frak H$ in 5d in \hyperref[sec41]{Section 4.1} before directly generalizing to the 6-group $\mathbb G^{(6)}$ and pseudo-6-group $\frak G$ of 11d in \hyperref[sec42]{Section 4.2}.  \\

The rest of the paper is organized as follows. In \hyperref[sec2]{Section 2} we first review the relevant features of higher groups and discuss our examples of interest, followed by a review of the relevant features of M2 and M5 branes and their coupling to the gauge field. We then compare these features and give motivation for our proposed model. We also review some other hints of the pseudo-6-group structure in string and M-theory which have been previously studied. In \hyperref[sec3]{Section 3} we review the generalized symmetry formalism used for studying symmetries $G$  and their charges $\text{Rep}(G)$ as higher fusion categories, including their relationship under gauging, and the utility of the SymTFT. We then review in detail an explicit and surprisingly relevant example of a finite 3-group and the construction of its non-invertible charges. In \hyperref[sec4]{Section 4} we take up the task of constructing and studying the non-invertible charges for the continuous 6-group $\mathbb G^{(6)}$ and the related pseudo-6-group $\frak G$. We begin with a simpler but nearly equivalent example of a 2-group in 5d, before generalizing to a 6-group in 11d. In \hyperref[sec5]{Section 5} we study the resulting fusion rules of our non-invertible defects and compare with known features of M2 and M5-branes. We also discuss several other points of comparison, such as other connections between branes and non-invertible defects, similarities between the M5-brane action and the defect action, other appearances and hints of the pseudo-6-group structure, and a comparison with previous work on the charges and symmetries of 11d supergravity. In \hyperref[sec6]{Section 6} we conclude with a discussion of our results and the many interesting future directions.

\section{Higher Groups and M-Theory Charges}\label{sec2}

We would like study the relationship between higher groups and the M2 and M5 brane charges of M-theory. We first review some necessary background on higher groups and their gauge fields. We then review the gauge charges of M2 and M5 branes. This then allows us to observe a concrete relationship between them which inspires our approach in the following sections. 

\subsection{Review: Higher groups and their gauge fields}\label{sec21}

We begin with a brief review of higher groups. To remain concrete, let us start with the simplest case of a 2-group. There are multiple equivalent approaches one may take to define a 2-group, but to gain intuition let us consider the context of a 2-group global symmetry in a quantum field theory \cite{Cordova:2018cvg, Benini:2018reh, Kapustin:2013uxa}. Intuitively, a 2-group global symmetry $\mathbb G^{(2)}$ can be thought of as an extension of a 0-form symmetry $G^{(0)}$ by a 1-form symmetry $H^{(1)}$:
\begin{equation}
    1 \to H^{(1)} \to \mathbb G^{(2)}\to G^{(0)}\to 1
\end{equation}
From this perspective, the elementary data defining a 2-group is the tuple $(G,H,\rho,\alpha)$ where $G$ is a group, $H$ is an abelian group, $\rho\in H^3(BG,H) $ is a cohomology class known as the \textit{Postnikov class} which can be thought of as the extension class, and $\alpha : G\to \text{Aut}(H)$ is a group action which can be thought of as the twist. We will primarily be interested in 2-group gauge fields, which are defined as connections on a 2-group bundle over spacetime. It will therefore be useful to repeatedly resort to the above decomposition, since it allows us to decompose a 2-group gauge field into an ordinary 1-form $G$ gauge field $A_1$ and a 2-form $H$ gauge field $B_2$ with modified gauge transformations realizing the Green-Schwarz mechanism. \\

More formally, a 2-group may be defined as a weak 2-category with a single object, weakly-invertible morphisms, and invertible 2-morphisms. This data is compactly summarized (up to suitable equivalence) in a tuple known as a crossed module. For an overview see \cite{baez2004higherdimensionalalgebrav2groups, Baez:2010ya}. Since we will be primarily interested in dealing with 2-group gauge fields we will skip directly to 2-connections on 2-group bundles \cite{Baez:2004in, bartels2006highergaugetheoryi}. \\

Let us begin with flat connections. Recall that a flat and connection on an ordinary abelian $G$-bundle over spacetime $M$ corresponds to a homomorphism $\text{Hom}(\pi_1(M),G)$. In the case of connected $M$ and $G=U(1)$ the topologically trivial flat connections define a torus $\mathcal W^1(M)=H^1(M)\otimes \mathbb{R/Z}$. More generally, flat connections can also be topologically nontrivial, in which case the full space of flat connections is the extension:
\begin{equation}
    1\to \mathcal W^1(M)\to H^1(M,\mathbb {R/Z})\to \text{Tors}(H^2(M,\mathbb Z))\to 1
\end{equation}
For a 2-group $\mathbb G$, then a flat 2-connection can be defined as a 2-homomorphism from the fundamental 2-groupoid $\Pi_2(M) $ which encodes both $\pi_1(M)$ and $\pi_2(M)$. \\

We can represent such a 2-bundle and flat connection concretely in the familiar simplicial definition of bundles on $M$. We first specify $G$-valued transition functions $A_{ij}$ on overlaps $U_{ij}$ of open charts $U_i, U_j$, a collection of which cover $M$. These $A_{ij}$ define a 1-cocycle which label edges of the Poincare-dual complex. Now on triple intersections $U_{ijk}$ we specify a 2-cochain $B\in C^2(M,H)$ which satisfies the twisted cocycle condition on the faces of the dual simplex \cite{Benini:2018reh}:
\begin{equation}
    \alpha(A_{ij})B_{jkl}-B_{ikl}+B_{ijl}-B_{ijk}=\rho(A_{ij},A_{jk},A_{kl})
\end{equation}
The components $(A_1,B_2)$ together form a flat 2-group connection. The above condition relates these components by telling us that the $\alpha$-twisted differential $d_\alpha B$ must equal the pullback of the Postnikov class:
\begin{equation}\label{flat}
    d_{\alpha}B=A^*\rho
\end{equation}
These components also have modified gauge transformations \cite{Benini:2018reh}:
\begin{equation}
    \begin{gathered}
        A_{ij}\to \tilde A_{ij}\equiv f_iA_{ij}f^{-1}_j \\ 
        B\to \alpha(f)B+\zeta(A,f)+d_\alpha\gamma 
    \end{gathered}
\end{equation}
where $\zeta(A,f)$ is a 2-cochain trivialization $d_\alpha\zeta(A,f)=\tilde A^*\rho - \alpha(f)A^*\rho$. In particular, the choice of representative $A^*\rho $ is not gauge invariant. A transformation $\rho \to \rho +d_\alpha\lambda$ must be compensated for by a simultaneous shift $B\to B+A^*\lambda$. \\ 

In this way 2-group bundles are classified by a Cech cohomology group with coefficients in $\mathbb G$ \cite{bartels2006highergaugetheoryi}. As with ordinary bundles, there also exists a convenient classification using a 2-group classifying space $B\mathbb G$ \cite{baez2009classifyingspacetopological2group}, which encodes the homotopy type of $\mathbb G$. It can be explicitly constructed from the bottom up by a simplicial category, which corresponds to constructing the geometric realization of the nerve of the 2-category $\mathbb G$. Moreover, this classifying space fits into a fibration of the classifying spaces for $G$ and $H$. This is achieved by defining a 2-group functor $\pi : B\mathbb G\to BG$ which defines a fibration with fiber $B^2H$:
\begin{equation}
    B^2H \to B\mathbb G\xrightarrow{\pi}BG
\end{equation}
This fibration is topologically classified by the Postnikov class $\rho \in H^3_{SM}(BG,H)$ which is valued in Segal-Mitchison group cohomology \cite{Schommer_Pries_2011}. 2-group bundles can then be equivalently classified by homotopy classes of maps from spacetime into $B\mathbb G$, which we denote as $[M,B\mathbb G]$ \cite{baez2009classifyingspacetopological2group}. \\

More generally, the classifying space of a $k$-group $\mathbb G^{(k)}$ may be constructed by a tower:
\[\begin{tikzcd}[column sep=tiny, row sep=small]
	{B^kH^{(k-1)}} & {B\mathbb  G^{(k)}} \\
	& {B\mathbb G^{(k-1)}} \\
	& \vdots \\
	{B^3H^{(2)}} & {B\mathbb G^{(3)}} \\
	{B^2H^{(1)}} & {B\mathbb G^{(2)}} \\
	& {BG^{(0)}}
	\arrow[from=1-1, to=1-2]
	\arrow[from=1-2, to=2-2]
	\arrow[from=2-2, to=3-2]
	\arrow[from=3-2, to=4-2]
	\arrow[from=4-1, to=4-2]
	\arrow[from=4-2, to=5-2]
	\arrow[from=5-1, to=5-2]
	\arrow[from=5-2, to=6-2]
\end{tikzcd}\]
where each stage is topologically classified by a Postnikov invariant $\rho_{n+2} \in H^{n+2}(B \mathbb G^{(n)},H^{(n)})$. Once again, $k$-group bundles are topologically classified by homotopy classes of maps $[M,B\mathbb G^{(k)}]$.  In this work, we will only be interested in $k$-groups with 2 stages, so that the whole fibration tower takes the form:
\[\begin{tikzcd}[sep=scriptsize]
	{B^{k} H} & {B\mathbb G^{(k)}} \\
	& {BG^{(p)}}
	\arrow[from=1-1, to=1-2]
	\arrow[from=1-2, to=2-2]
\end{tikzcd}\]
with Postnikov class $\rho \in H^{k+1}(B^pG, H)$. \\

Quantum field theories may possess a $k$-group global symmetry $\mathbb G^{(k)}$.  In these cases, we can couple the theory to background fields $A_1, A_2, \dots ,A_k$ representing the components of a generically non-flat $\mathbb G^{(k)}$ connection.  \\

\newcounter{example}[section]
\renewcommand{\theexample}{\thesection.\arabic{example}}
\newcommand{\examplehead}[1]{%
  \par\medskip\noindent
  \refstepcounter{example}%
  \tcbox[on line, colback=pink!20, colframe=black!65, boxrule=0.5pt,
         arc=2pt, left=4pt, right=4pt, top=2pt, bottom=2pt]
        {\textbf{Example~\theexample:} #1}%
  \par\smallskip}

% usage
\examplehead{Abelian 2-Group}\label{2group}

\bigskip

As an example, consider the case of an untwisted abelian 2-group $\mathbb G^{(2)}$ with $G =U(1)$ and $H=U(1)$ and Postnikov class $\kappa \in H^3(BU(1),U(1))\cong \mathbb Z$. Using the isomorphism $H^3_{SM}(BU(1),U(1)) \cong H^4(BU(1),\mathbb Z)$ for Segal-Mitchison cohomology then we can identify the Postnikov class with the Chern class $\kappa \cdot  c_1^2$. The background fields $A_1, B_2$ couple to 1-form and 2-form Noether currents $j_{d-1}, J_{d-2}$ respectively, and have modified gauge transformations:
\begin{equation}
    \begin{gathered}
        A_1\to A_1+d\lambda_0 \\ 
        B_2\to B_2+d\Lambda_1-\frac{\kappa}{2\pi}\lambda_0dA_1
    \end{gathered}
\end{equation}
which realize the Green-Schwarz mechanism. Since this connection is not necessarily flat, it can have nontrivial 2-group curvature represented by the gauge-invariant field strength:
\begin{equation}
    H_3=dB_2+\frac{\kappa}{2\pi}A_1dA_1
\end{equation}
For a recent discussion of the full Cech cocycle definition for a non-flat 2-group connection see \cite{Harder:2026csj}. Notice in the flat case when $H_3=0$ we recover the flat cocycle condition \hyperref[flat]{(2.4)}. \\

The Wilson line operator measuring $A_1$ holonomy is:

$$W_n[\gamma] = \exp\Big(in\int_\gamma A_1\Big)$$
However the naive Wilson surface $V_n[\Sigma]=\exp\Big( i n \int_{\Sigma} B_2\Big)$ for the 2-form $B_2$ is not gauge invariant:
\begin{equation}
    V_n[\Sigma_2]=\exp\Big(in\int_\Sigma B_2\Big) \xrightarrow[\delta_{\lambda}]{}\exp\Big({i n\int_\Sigma B_2-\frac\kappa{2\pi} \lambda_0 dA_1} \Big)
\end{equation}
A non-genuine but gauge-invariant holonomy operator can be constructed on a disk $M_3$ with boundary $\partial M_3=\Sigma_2$:
\begin{equation}
    D_n[M_3]\equiv \exp\Big( in\int_\Sigma B_2+in\frac{\kappa }{2\pi}\int_{M_3} A_1dA_1 \Big) \cong \exp\Big (in \int_{M_3}H_3\Big)
\end{equation} \\ 

\examplehead{Abelian 6-Group}\label{6groupex}

\bigskip

Another example which will shortly become relevant is the 6-group analog of the above 2-group. This 6-group is defined to have only two stages:
\begin{equation} \label{6group}
     B^6U(1) \to B\mathbb G^{(6)} \to B^3 U(1)
\end{equation}
with Postnikov class $\kappa \in H^7(B^3U(1),U(1))\cong \mathbb Z$. Once again we can consider the background gauge fields $C_3$ and $B_6$ which have the modified gauge transformations:
\begin{equation}
    \begin{gathered}
        C_3\to C_3+d\lambda_2 \\
        B_6\to B_6+d\Lambda_5-\frac{\kappa}{2\pi}\lambda_2dC_3
    \end{gathered}
\end{equation}
We define the gauge-invariant 7-form curvature:
\begin{equation}
    H_7 = dB_6+\frac{\kappa}{2\pi}C_3dC_3
\end{equation}
and the Wilson holonomy operators:
\begin{equation}\label{G6charge}
    \begin{gathered}
        W_n[\gamma_3]\equiv \exp\Big(in\int_{\gamma_3}C_3\Big)\\
        V_n[\Sigma_6] \equiv \exp\Big(i n \int_{\Sigma_6}B_6 \Big) \\
        D_n[M_7] \equiv \exp\Big(in \int_{M_7}H_7\Big)
    \end{gathered}
\end{equation}
The minimal bare operator $V_1$ has anomalous variation:
\begin{equation}\label{vanom}
    V_1[\Sigma_6]=\exp\Big(i\int_{\Sigma_6}B_6\Big) \xrightarrow{\delta_{\lambda_2}}\exp\Big(i\int_{\Sigma_6}B_6-\frac\kappa{2\pi} \lambda_2dC_3\Big)
\end{equation}

while the gauge invariant operators $W_n[\gamma]$ and $D_n[M]$ may be considered as charges of $\mathbb G^{(6)}$. We will elaborate more on this perspective later in \hyperref[sec3]{Section 3}. \\

\subsection{Charges of M2 and M5 Branes} \label{sec22}

Now that we have an understanding of higher groups and their gauge fields, let us see their relevance to M-theory by reviewing the gauge charges of M2 and M5 branes. \\

M-theory is known to host a 3-form abelian gauge field $C_3$ which couples to the worldvolume of the M2-brane. In the low energy limit this is identified with the 3-form gauge field of 11d supergravity. The gauge sector of the supergravity action involving this field takes the form:
\begin{equation}
    S \sim \int \frac{1}{2l^3}G_4\wedge *G_4+\frac 16 C_3G_4G_4-C_3I_8
\end{equation}
where $G_4=dC_3$ is the field strength and $I_8= \frac{1}{48}\Big( p_2(TM)-\lambda^2\Big)$ with $\lambda \equiv \frac 12p_1(TM)$. The reason we have used $\sim$ is to emphasize that this presentation is not globally well-defined when the gauge field $C_3$ is topologically nontrivial. Using the notation of \cite{Diaconescu:2003bm} we write the full Chern Simons topological term as $\Phi(C)$ and heuristically write:
\begin{equation}
    \Phi(C)\sim \exp\Big(2\pi i\int \frac 16 C_3G_4G_4-C_3I_8\Big)
\end{equation}
The curious $\frac 16$ quantization seems problematic at first, however this Chern-Simons form was shown by Witten to be well-defined on Spin manifolds using a model for $C_3$ as the Chern-Simons form for an $E_8$ gauge field \cite{Witten:1996md}. We review the full definition in \hyperref[appa]{Appendix A}.  \\

We are interested in the coupling of $C_3$ to the fundamental objects of M-theory: namely M2 and M5 branes. The coupling to an M2 brane on a 3d worldvolume $\Sigma_3$ is straightforward: it is given simply by the holonomy
\begin{equation}
    \exp\Big(i\int_{\Sigma_3}C_3\Big) \subset Z_{M2}[\Sigma_3]
\end{equation}
However there is a global topological subtlety to this coupling. It turns out that the above holonomy alone is not well-defined if one takes into account the worldvolume fermions of the M2-brane, as shown in \cite{Witten:1996md}. The relevant topological contribution to the membrane is actually:
\begin{equation}
    \sqrt{\det D_{S(N)}}\exp\Big(i\int_{\Sigma_3}C_3\Big)
\end{equation}
The first factor is the Pfaffian contribution from the path integral over the worldvolume fermions, where $S(N)$ is a chiral spinor bundle associated to the rank-8 normal bundle $N$ of the embedding $\Sigma_3\subset M_{11}$. This factor has a sign ambiguity which the holonomy factor must cancel for the total coupling to be well-defined. This is achieved by requiring that $C_3$ has a shifted flux quantization, so that the field strength $G_4$ satisfies:
\begin{equation}
    \frac{1}{2\pi}\int_{Z_4}G_4 \mod \mathbb Z=\frac 12 \int_{Z_4}w_4(TM)
\end{equation}
On a spin manifold one can take a canonical integral lift of $w_4$ to the class $\lambda = \frac 12 p_1\in H^4(TM,\mathbb Z)$. Therefore the M-theory gauge field $C_3$ is not an ordinary 3-form gauge field, but is required to have its flux shifted by this gravitational term dependent on the spacetime spin manifold.  \\

The coupling of the $C_3$ field to the M5-brane is considerably more subtle. This is because M5-branes are solitonic objects, which source magnetic charge for the field strength $G_4$, so that:
\begin{equation}
    \frac{1}{2\pi}\int_{S^4} G_4 = 1
\end{equation}
for a sphere $S^4$ linking a single fivebrane. There is no obvious local action for their worldvolume partition function\footnote{See \cite{Bandos:1997ui, Pasti:1997gx, Aganagic:1997zq, Bandos:1997gm, Howe:1996yn} for several proposed constructions.}. Indeed, the worldvolume theory for a stack of $k$ M5 branes sourcing $k$ units of flux is the 6d $\mathcal N=(2,0)$ superconformal theory of type $\frak a_{k-1}$ (along with an overall singleton tensor multiplet), which has no known Lagrangian description. \\

Instead, we can study the worldvolume theory using anomaly inflow from the bulk, as described in \cite{Witten:1996hc}. The effective action for an M5-brane on a worldvolume $\Sigma_6$ in a non-extremal background spacetime $M_{11}$ is defined by excising a small tubular neighborhood $T_{11}$ and imposing the magnetic flux condition around the small 10d boundary $N_{10} = \partial (M_{11}\setminus T_{11})$. For a nonsingular M5-brane this $N_{10}$ defines a sphere bundle over the worldvolume:
\begin{equation}
    S^4\hookrightarrow N_{10}\to \Sigma_6
\end{equation}
Along any such sphere the field strength $G_4$ integrates to 1. However the presence of the Chern-Simons term $\Phi(C)$ in the supergravity action implies that the bulk has an anomalous variation on this boundary. Therefore the boundary must be stacked with the M5-brane partition function which has the opposite anomalous variation in order to cancel the gauge anomaly by inflow. Heuristically, under a small gauge transformation $C_3\to C_3+d\lambda_2$ this variation is:
\begin{equation}
    \delta_{\lambda_2}\mathcal Z_{M5}[\Sigma_6] \sim e^{-i\frac 1{4\pi}\int_{\Sigma_6}\lambda_2G_4 }\mathcal Z_{M5}[\Sigma_6]
\end{equation}
where we have momentarily dropped the contribution from the gravitational anomaly. More precisely, the partition function is defined as the unique holomorphic section of a line bundle over the moduli space of $C_3$ configurations on $\Sigma_6$ which is defined by a 7d Wu-Chern-Simons theory on a bounding 7d manifold. We will elaborate on this definition in \hyperref[sec5]{Section 5}. This provides a natural explanation for the chiral 2-forms with self-dual field strength in the 6d tensor multiplet living on $\Sigma_6$. \\

Let us pause to consolidate these results. The coupling of the $C_3$ field to the M2-brane is given by the holonomy:
\begin{equation}
    Z_{M2}[\gamma_3]\sim \exp\Big(i\int_{\gamma_3}C_3\Big)
\end{equation}
while the coupling to the M5-brane is characterized by the anomalous variation: 
\begin{equation}
    \mathcal Z_{M5}[\Sigma_6]\xrightarrow{\delta_{\lambda_2}} e^{-i\frac{1}{4\pi}\int \lambda_2dC_3}
    \mathcal Z_{M5}[\Sigma_6]
\end{equation}
Already we can not help but compare to \hyperref[G6charge]{(2.14)} and \hyperref[vanom]{(2.15)}. However, before making a direct comparison, let us first be more precise with the above definitions. \\

To account for all topological subtleties, it is most useful to use differential cohomology. For concise reviews see \cite{Freed:2006yc, Moore:2025tmt, Debray_2025, Freed:2006ya, Szabo:2012hc}. The full gauge-invariant content of an ordinary 3-form gauge field $\check C_3$ can be defined as an element of $\check H^4(M_{11})$. For our purposes we will primarily be interested in the tuple of data $\check C_3\sim(g_4, G_4, C_3)$ where $g_4\in H^4(M,\mathbb Z)$ is the characteristic class defining the topological class of the bundle, $G_4\in \Omega^4(M)$ is the field strength, and $C_3\in C^3(M,\mathbb R)$ is a choice of cochain trivialization satisfying:
\begin{equation}
    \exp\Big(i\int_{\gamma_3}C_3\Big)=\exp\Big(i\int_{\Gamma_4}G_4\Big)
\end{equation}
whenever $\gamma_3=\partial \Gamma_4$. \\

However recall that the $C_3$ field of M-theory has a shifted flux quantization condition, so that the field strength is shifted by $\frac 12 \lambda = \frac 14 p_1(TM)$. We can formalize this by considering $\check C_3$ as an element of shifted differential cohomology $\check H^4_{\frac 12\lambda}(M) $ as discussed in \cite{Hopkins:2002rd, Monnier:2017klz, Monnier:2016jlo}. The relation between this model and the $E_8$ model of the $C_3$ field was further developed in \cite{Diaconescu:2003bm}. \\

We can also properly define the Chern-Simons term in differential cohomology. As reviewed in \hyperref[appa]{Appendix A} the differential cohomology group $\check H^{p+1}(M)$ has a natural pairing when the dimension of $M$ is $d=2p+1$, defined by:
\begin{equation}
    (\check A_p, \check B_p) \equiv \int_M \check A_p\star \check B_p \in U(1)
\end{equation}
This pairing gives a global definition for the Chern-Simons action. Indeed, for topologically trivial $A$ we have
\begin{equation}
    (\check A, \check A)= \frac{1}{2\pi}\int_M A_pdA_p
\end{equation}
On manifolds supporting a Spin structure (or more generally a Wu structure) this pairing has a quadratic refinement $Q(\check A)$ satisfying:
\begin{equation}
    Q(\check A+\check B)-Q(\check A)-Q(\check B)+Q(0)=(\check A, \check B)
\end{equation}
This quadratic refinement defines spin Chern-Simons theory in 3d, and more generally defines Wu Chern-Simons theory in higher dimensions. We review the definitions of these quadratic refinements in \hyperref[appa]{Appendix A}. \\ 

In this language, we can more precisely express the anomalous variation of the M5-brane partition function as:
\begin{equation}
    \mathcal Z_{M5} \xrightarrow{\delta_{\lambda_2}} e^{-i\delta_{\lambda_2}Q(\check C)}\mathcal Z_{M5}
\end{equation}
where $\delta_{\lambda_2}Q(\check C)$ denotes the anomalous gauge variation of $Q(\check C)$ on a bounding 7-manifold. Importantly, this expression includes  contributions from the gravitational term $I_8$ that we previously dropped. An easy proof of this variational formula is given in \hyperref[appa]{Appendix A}, and uses the fact that the supergravity Chern-Simons term $\Phi(\check C)$ in fact defines a \textit{cubic} refinement of a \textit{trilinear} form on differential cohomology \cite{Diaconescu:2003bm, Freed:2019sco, Han:2020lxd}. 

\subsection{The Comparison}\label{sec23}

It is now easy to see the resemblance between the gauge couplings of M2 and M5 branes and the Wislon line charges of the 6-group $\mathbb G^{(6)}$ defined in \hyperref[6group]{(2.11)}. The minimal 6-group Wilson surfaces were:  
\begin{equation}\label{6charges}
    \begin{gathered}
        W_1[\gamma_3] = \exp\Big(i\int_{\gamma_3}C_3\Big)  \\
        V_1[\Sigma_6] = \exp\Big(i\int_{\Sigma_6}B_6\Big) \xrightarrow{\delta_{\lambda_2}}\exp\Big(i\int_{\Sigma_6}B_6-\frac 1{2\pi}\lambda_2G_4\Big)
    \end{gathered}
\end{equation}

While the gauge charges of the M-theory branes were:
\begin{equation}\label{Mcharges}
    \begin{split}
        Z_{M2}[\gamma_3] &\sim \exp(i\int_{\gamma_3} C_3) \\
        Z_{M5}[\Sigma_6] &\xrightarrow{\delta_{\lambda_2}} e^{i\delta_{\lambda_2}Q(\check{C})}Z_{M5}[\Sigma_6]\simeq e^{-\frac{i}{4\pi}\int\lambda_2 dC_3}Z_{M5}[\Sigma_6]
    \end{split}
\end{equation}
%======

% M-theory Brane Charge

%======

This is the kernel of our motivation for this work. The charges of a nontrivial higher group $\mathbb H^{(k)}$ in a quantum field theory generically may be described by a non-invertible higher fusion category $\text{Rep}(\mathbb H^{(k)})$. Thus, by studying the charges of the 6-group $\mathbb G^{(6)}$ we hope to model key features of M2 and M5 branes, and their mutual relationships. \\

\begin{table}[t]
\centering
\renewcommand{\arraystretch}{1.3}
\setlength{\extrarowheight}{1pt}
\setlength{\tabcolsep}{1pt}

\begin{tabular}{
    |>{\centering\arraybackslash}m{0.47\textwidth}
    |>{\centering\arraybackslash}m{0.47\textwidth}|
}
\hline

\cellcolor{BeforeBlue}
\textbf{Minimal \(6\)-group operators}
&
\cellcolor{pink}
\textbf{M-theory brane gauge charges}
\\
\hline

\makebox[\linewidth][l]{%
    \makebox[5.5em][r]{%
        \(W_1(\gamma_3)\)
    }
    \(\;=\;
        e^{\,i\int_{\gamma_3}C_3}
    \)
}
&
\makebox[\linewidth][l]{%
    \makebox[5.5em][r]{%
        \(Z_{M2}[\gamma_3]\)
    }
    \(\;\sim\;
        e^{\,i\int_{\gamma_3}C_3}
    \)
}
\\[-3pt]
\hline

\cellcolor{black!4}
\makebox[\linewidth][l]{%
    \rule[-2ex]{0pt}{5.5ex}%
    \makebox[5.5em][r]{%
        \(V_1(\Sigma_6)\)
    }
    \(\;=\;
        e^{\,i\int_{\Sigma_6}B_6}
    \)
}
&
\cellcolor{black!4}
\makebox[\linewidth][l]{%
    \rule[-2ex]{0pt}{5.5ex}%
    \makebox[5.5em][r]{%
        \(Z_{M5}[\Sigma_6]\)
    }
    \(\xrightarrow{\delta_{\lambda_2}}
        e^{\,i\delta_{\lambda_2}Q(\check{C})}
        Z_{M5}[\Sigma_6]
    \)
}
\\

\cellcolor{black!4}
\rule[-1.2ex]{0pt}{5ex}%
\raisebox{5.5pt}[0pt][0pt]{%
    \makebox[\linewidth][l]{%
        \makebox[5.5em][r]{%
            \(V_1(\Sigma_6)\)
        }
        \(\xrightarrow{\delta_{\lambda_2}}
            e^{\,-\frac{i}{2\pi}
            \int_{\Sigma_6}\lambda_2\,dC_3}
            V_1(\Sigma_6)
        \)
    }
}
&
\cellcolor{black!4}
\rule[-1.2ex]{0pt}{5ex}%
\raisebox{5.5pt}[0pt][0pt]{%
    \makebox[\linewidth][l]{%
        \makebox[5.5em][r]{}
        \(\;\simeq\;
            e^{\,-\frac{i}{4\pi}
            \int_{\Sigma_6}\lambda_2\,dC_3}
            Z_{M5}[\Sigma_6]
        \)
    }
}
\\[-2pt]
\hline

\end{tabular}

\caption{
Comparison between the minimal bare \(6\)-group operators and the
corresponding gauge charges carried by the M2- and M5-branes.
}
\label{tab:bare-and-brane-charges}
\end{table}

We must highlight two primary differences between \hyperref[6charges]{(2.31)} and \hyperref[Mcharges]{(2.32)}:
\begin{enumerate}
    \item The M-theory $C_3$ field is not an ordinary 3-form gauge field, but has a shifted flux quantization condition $G_4+\frac 14 p_1(TM)$
    \item The variation of the 6-group charge $e^{i\int B_6}$ is the same as an ordinary level-2 bosonic Chern-Simons theory $\exp(\frac{i}{2\pi}\int C_3dC_3) $, while the variation of the M5 partition function is that of a level-1 Wu Chern-Simons theory $\exp(iQ(C)) \sim \exp(\frac{i}{4\pi}\int C_3(dC_3+\nu))$. 
\end{enumerate}
Both of these differences reflect the fact that the gauge charges of M-theory inherently depend on the tangential structure of the spacetime manifold. \\

In particular, recall that the Postnikov class of $\mathbb G^{(6)}$ was defined to be the Chern class $  G_4^2 \in H^8(B^3U(1),\mathbb Z)$. On the other hand, there is no obvious  Postnikov class in Eilenberg-MacLane cohomology related to the quadratic refinement $Q(C)$, which is defined using the index of a Dirac operator coupled an $E_8$ gauge bundle. A close relative which lives purely in spacetime is the mysterious class $\Theta \in H^8(N_{10},\mathbb Z)$ considered in \cite{Diaconescu:2003bm} which was proposed to capture the topological class of the electric charge of a pure $C$-field configuration on a spatial slice $N_{10}$. The de Rham refinement of this class is:
\begin{equation}
    [\Theta]_{dR}=\frac 12G_4^2-I_8
\end{equation}
We expand more upon the relationship in \hyperref[sec55]{Section 5.5}. There is therefore no obvious candidate 6-group which models \hyperref[Mcharges]{(2.32)}. The correct topological characterization of the M-theory charges has been vigorously studied over the years. We consider various possible interpretations, and review some proposed perspectives later in \hyperref[sec55]{Section 5.5}. For now we remain agnostic about the overall correct formalization of the structure, and instead take \hyperref[Mcharges]{(2.32)} as the defining characteristics and treat them directly. In the absence of a genuine 6-group structure, we use $\frak G$ to denote this pseduo-6-group structure defined by \hyperref[Mcharges]{(2.32)}  \\ 

Nevertheless, we will find that our study of the charges for the genuine 6-group $\mathbb G^{(6)}$ bears fruit in this direction. Indeed, in the process of studying these charges in \hyperref[sec4]{Section 4}, we will find that it is in fact most natural to construct charge operators which precisely capture the features of the $\frak G$ charges in \hyperref[Mcharges]{(2.32)}. As a \textit{consequence} they will also give rise to a construction of the $\mathbb G^{(6)}$ charges. In this sense, our construction models charges for both $\mathbb G^{(6)}$ and $\frak G$. Moreover, the categorical features of these defect operators, such as their fusions and higher morphisms, precisely reflect known features of M2 and M5 branes. \\

\subsection{Other Hints of Higher Groups}\label{sec24}

The higher-form structure of the 11d supergravity gauge field $C_3$ has been studied extensively over the years \cite{DAuria:1982uck, Henneaux:1984mh, Townsend:1995gp, Candiello:1993di, deAlwis:1997gq, Bandos:1997gd, Bandos:2004xw, Fiorenza:2013nha, Giotopoulos:2024xcg, Fiorenza:2012mr, Sati:2024klw, Giotopoulos:2024sit, Fiorenza:2019ckz, Kalkkinen:2003gq, Fiorenza:2018ekd, Giotopoulos:2024ovz, Armas:2024caa, Rosabal:2025zry}. The above core features have been approached from many different perspectives, and using a variety of different formalisms. Here we highlight an especially elementary hint of the pseudo-6-group structure which has appeared previously, while saving a more thorough discussion for \hyperref[sec55]{Section 5.5}. \\ 

Even in the absence of branes, the 11d supergravity Chern-Simons interaction modifies the equations of motion for the electric and magnetic field strengths to take the form:
\begin{equation}
    \begin{gathered}
        dG_7=\frac 12G_4G_4-I_8 \\ dG_4=0
    \end{gathered}
\end{equation}
where $G_7=*_{11}G_4$. In the absence of $I_8$, one may attempt to define a ``dual magnetic gauge field" by solving the first equation and introducing a trivialization of the electric field strength:
\begin{equation}
    G_7\equiv dB_6+\frac 12 C_3\wedge G_4
\end{equation}
However in order for such a trivialization to be gauge invariant we must modify the gauge transformations:
\begin{equation}
    \begin{gathered}
        C_3\to C_3+d\lambda_2 \\ B_6\to B_6-\frac 12 \lambda_2G_4+d\Lambda_5
    \end{gathered}
\end{equation}
The magnetic gauge field couples directly to the M5-brane, reproducing its anomalous gauge variation.  Of course we recognize this structure as precisely the expected transformation rules for gauge fields of $\frak G$. \\

This magnetic gauge field can not be treated on the same footing as the dual gauge field obtained from a duality transformation, as in Maxwell theory. This is due to the fact that the Chern-Simons term is cubic in the gauge field, and thus obstructs such a transformation. However there is a regime in which we \textit{can} perform such a duality. If we compactify the theory on a circle and take the low-energy limit then we arrive at the 10d type IIA supergravity effective action. The dimensional reduction of the M-theory 3-form gauge field gives rise to both a 3-form gauge field (which we also denote $C_3$) as well as a 2-form gauge field $B_2$. The resulting Chern-Simons term now includes the combination $B_2 G_4G_4$ which is \textit{linear} in the gauge field $B_2$. As a result the gauge field $B_2$ can be dualized following the usual S-duality transformation. However the Chern-Simons interaction enforces precisely the above modified gauge transformations on the dual magnetic gauge field $B_6$. Higher group symmetry structures from dimensional reductions were also studied in \cite{Gukov:2020btk, Gukov:2025dol, Sheckler:2025fql} \\ 

The pseudo-6-group features of the supergravity gauge fields are highly suggestive. However in this work we aim to primarily work with the fundamental M2 and M5-branes directly. The reason is that our interest lies in the fusion rules of the charges for $\mathbb G^{(6)}$ (and by extension $\frak G$). In 11d these charges are 3d and 6d defects, and while it may be useful to represent them as holonomies of spacetime fields, we will only ever treat features of the defects directly.

\section{Warmup: Charges of Finite Higher Groups}\label{sec3}

Before studying the charges of continuous higher groups, let us first get our footing in a more familiar and well-established setting: the symmetries and charges of \textit{finite} higher groups. We will first review some general formalism for how these symmetries and charges are defined, and what is the relationship between them. We will then zoom in and review a concrete familiar example which has appeared in different guises in the literature: the case of a finite $3$-group defined by the Pontryagin square $\frak P\in H^4(B^2\mathbb Z_N,\Gamma(\mathbb Z_N))$. This case happens to be particularly relevant to our eventual interests of continuous higher groups, so we will review it in detail. We will review the explicit construction of the non-invertible charge operators, which happen to include the familiar minimal abelian TQFTs $\mathcal A^{N,p}$ \cite{Hsin:2018vcg}. Having a grounded understanding of this example will pay dividends for our construction of continuous higher-group charges in the next section. \\

\subsection{Representations and Charges of Finite Higher Groups}\label{sec31}

Quantum field theories can admit global symmetries described by a finite higher group $\mathbb G^{(n)}$. As with any global symmetry, we are typically interested in studying two types of defect operators associated with it:
\begin{enumerate}
    \item The unitary topological defect operators which generate the symmetry, which we refer to as the symmetry defect operators (SDOs).
    \item The charges of $\mathbb G^{(n)}$, which may be supported along defect operators of varying dimension. 
\end{enumerate}

Let us begin with the first class of defects. As operators in the theory, the SDOs form a higher fusion category which captures the possible fusions representing the $\mathbb G^{(n)}$ symmetry. In a spacetime dimension $d$, this fusion category is in general given by a the $(d-1)$-category $(d-1)\text{-Vec}^\alpha(\mathbb G^{(n)})$ \cite{Freed:2022qnc, Bhardwaj:2022kot, Bhardwaj:2023ayw, Bartsch:2023wvv, Bartsch:2022ytj}. Here the $\alpha$ indicates the possibility for this symmetry to possess a 't Hooft anomaly. If we disregard anomalous data, then just as with ordinary group-like symmetries the algebraic structure of the fusion is completely determined by the algebraic data of $\mathbb G^{(n)} $ itself, which for a higher group is captured in the weak $n$-category defining $\mathbb G^{(n)}$.  \\

The second class of defects describes the possible charges of a higher-group symmetry $\mathbb G^{(n)}$. In principle, operators of our QFT which carry charge under the global symmetry need not be topological. Instead, what we mean by ``charge" is the specific \textit{representation} of the global symmetry $G$ that the operator transforms under. In fact, for group-like global symmetries, there is typically a different but \textit{related} theory in which the operators labeled by these representations $R\in \text{Rep}(G)$ are topological. For example, in the case of discrete 0-form symmetries, such a theory can be obtained by gauging $G$, in which case the resulting gauged theory has a dual ``quantum" symmetry given by $\text{Rep}(G)$, and generated by these topological operators. In this sense, there is a higher fusion category $(d-1)\text{-Rep}(\mathbb G^{(n)})$ composed of topological defects which capture the representations of the higher group \cite{Bartsch:2022mpm, Bartsch:2022ytj,Bhardwaj:2023wzd, Bhardwaj:2022lsg, Bhardwaj:2022yxj, Bhardwaj:2022kot, Bhardwaj:2022maz, Decoppet:2023bay, Bartsch:2023pzl, Delcamp:2023kew}.\footnote{For more on representation categories see e.g. \cite{ganter2007representationcharactertheory2categories, elgueta2005representationtheory2groupsfinite, Osorno_2010, etingof2009fusioncategorieshomotopytheory}}\\

An intuitive class of examples are Wilson lines for a flat connection of an ordinary finite non-anomalous $G$, which are labeled by representations $R \in \text{Rep}(G)$. These Wilson lines are defined using a gauge field for the original symmetry $G$, and thus belong to the gauged theory. The charged operators of the original theory have been projected out by the gauging, and their worldvolumes replaced by the topological Wilson line associated with their representation $R$.  \\ 

The category $(d-1)\text{-Rep}(\mathbb G^{(n)})$ is often defined as the $(d-1)$ category of higher functors between $\mathbb G^{(n)}\to (d-1)\text{-Vec}$ \cite{Bartsch:2023pzl}.  When the higher group is non-split, so that it has nontrivial Postnikov class, then the resulting representation category may non-invertible,\footnote{We note that this is neither a necessary nor sufficient condition for non-invertibility. See \cite{Bhardwaj:2022maz} for an example of non-invertible symmetry described by $\text{Rep}((\mathbb{Z}_2^2)^{(1)}\rtimes\mathbb{Z}_2^{(0)})$.   } so that the representing topological defects obey non-invertible fusion rules. In simple finite cases the structure of this category can be computed systematically, and the associated defects constructed. We will consider such an example in the following section. \\

As we have mentioned, there is a close relationship between a symmetry category $\text{Vec}(G)$ and its category of charges $\text{Rep}(G)$ under gauging. Indeed, after gauging a discrete invertible non-anomalous global symmetry $G$, the new gauged theory possesses a global symmetry described by the category of charges $\text{Rep}(G)$, sometimes referred to as the ``quantum symmetry", after the appearance of this phenomenon in orbfiolds \cite{Vafa:1986wx}. This is familiar from the case of finite abelian symmetries \cite{Tachikawa:2017gyf, Bhardwaj:2017xup} where gauging a symmetry $G$ yields a new theory with a dual quantum symmetry $\hat G=\text{Hom}(G,U(1))$. Moreover, in 2 dimensions gauging the new quantum symmetry $\text{Rep}(G)$ gives us back a theory with the original symmetry $\text{Vec}(G)$ \cite{Bhardwaj:2017xup}.  \\

\begin{table}[t]
\centering
\small
\renewcommand{\arraystretch}{1.4}
\setlength{\extrarowheight}{2pt}
\setlength{\tabcolsep}{2pt}

\begin{tabular}{
    |>{\raggedright\arraybackslash}m{0.315\textwidth}
    |>{\raggedright\arraybackslash}m{0.315\textwidth}
    |>{\raggedright\arraybackslash}m{0.315\textwidth}|
}
\hline
\cellcolor{BeforeBlue}
\textbf{Before gauging: \(\mathcal T\)}
&
\cellcolor{GaugingPurple}
\textbf{Gauging \(G^{(n)}\)}
&
\cellcolor{pink}
\textbf{After gauging: \(\mathcal T/G^{(n)}\)}
\\
\hline

Group-like symmetry defects organized by
\((D-1)\mathrm{Vec}_{G^{(n)}}\)
&
Sum over networks of symmetry defects
&
Dual topological defects organized by
\((D-1)\mathrm{Rep}\!\left(G^{(n)}\right)\)
\\

\rowcolor{black!4}
A \(q\)-dimensional operator \(\mathcal O_q\) carries a
\(q\)-charge \(\rho^{(q+1)}\)
&
Promote \(G^{(n)}\) from a global to a gauge symmetry
&
\(\rho^{(q+1)}\) labels a topological
\((q+1)\)-dimensional Wilson or condensation defect
\\

A genuine charged operator transforms nontrivially under
\(G^{(n)}\)
&
Impose gauge invariance
&
A gauge-invariant orbit, or a non-genuine operator attached
to a gauge defect
\\

\rowcolor{black!4}
A \(G^{(n)}\)-twisted operator is attached to a symmetry defect
&
Include the twisted sectors in the gauge sum
&
A dynamical gauge-flux defect
\\
\hline
\end{tabular}

\caption{
Schematic reorganization of symmetry and charge data under gauging
a finite higher-group symmetry \(G^{(n)}\). Before gauging, the
\(q\)-charges of \(q\)-dimensional operators are classified by
\((q+1)\)-representations of \(G^{(n)}\). After gauging, the
corresponding higher-representation data labels topological Wilson
and condensation defects. The table tracks symmetry and charge
labels rather than an objectwise map between individual operators
\cite{Bhardwaj:2023wzd,Bhardwaj:2023ayw,Bhardwaj:2022lsg,
Bartsch:2023pzl,Bartsch:2022mpm,Bartsch:2022ytj}.
}
\label{tab:higher-group-gauging}
\end{table}

The procedure of gauging a (non-anomalous) $\mathbb G^{(n)}$ global symmetry gives us a straightforward way to construct the defects of $\text{Rep}(\mathbb G^{(n)})$. It also elucidates why the resulting defects may be non-invertible. \\

\begin{tcolorbox}[breakable,colback=pink!30!white,colframe=black!75!black,
                    ]
    \textbf{Example: Finite 2-Group} \\ 

As an example, consider the case of a 2-group:
\begin{equation}
  H^{(1)}\to  \mathbb G^{(2)}\to G^{(0)}
\end{equation}
with trivial twist and non-trivial Postnikov class $\rho \in H^3(BG,H)$. If $\mathbb G^{(2)}$ is non-anomalous then its gauging can be achieved in a series of steps \cite{Tachikawa:2017gyf, Bhardwaj:2017xup, Kapustin:2014zva, Wang:2017loc, Villa:2026jmd, Bartsch:2022mpm, Bartsch:2022ytj}:
\begin{enumerate}
    \item We first gauge the abelian 1-form symmetry $H$ by introducing dynamical gauge fields $b_2\in H^2(M,H)$. The new theory will possess a new global $(d-3)$-form symmetry $\hat H^{(d-3)}$ where $\hat H=\text{Hom}(H,U(1))$ is the Pontryagin dual of $H$. This dual symmetry now has a mixed ‘t Hooft anomaly with the remaining global symmetry captured by $G^{(0)}$. Concretely, if we couple to background gauge fields $a_1$ for $G$ and $c_{d-2}$ for $\hat H$ then the anomaly cocycle describing this anomaly is defined by:
    \begin{equation} \label{mixedanom}
        \alpha_{d+1} = \exp\Big(i\int c_{d-2}\cup a^*\rho\Big)\in H^{d+1}(M,U(1))
    \end{equation}
    which is pulled back from the class in $H^{d+1}(BG\times B^{d-2}\hat H,U(1))$. 
    \item Now to complete the gauging of $\mathbb G^{(2)}$ we must promote $a_1$ to a dynamical gauge field and integrate over it. The consequence of the mixed anomaly is that the global $\hat H$ symmetry is naively broken by an “ABJ anomaly”, so that the original symmetry defect operators $\exp(i\int j_2)$ are no longer topological due to a lack of closure: $\delta j_2=a^*\rho$. It seems like the global symmetry is simply broken, since there is no topological operator to generate it. 
    \item In fact, it may still be possible to construct a topological operator by stacking the naive symmetry operator $\exp(i\int j_2)$ with a 2d worldvolume TQFT which cancels the anomalous divergence. As a result of this stacking, the resulting operator is a non-invertible topological symmetry defect operator belonging to $\text{Rep}(\mathbb G^{(2)})$. We will discuss a simple concrete example of stacking by such a TQFT in the next section.
\end{enumerate}
\end{tcolorbox}

From this procedure we can notice a general strategy for how to construct theories with either $\mathbb G^{(n)}$ global symmetry or $\text{Rep}(\mathbb G^{(n)})$ \cite{Cordova:2018cvg, Benini:2018reh, Tachikawa:2017gyf}. This can be achieved if we have a theory which begins at step 2 above: that is we have two global symmetries which share a mixed 't Hooft anomaly of the form \hyperref[mixedanom]{(3.2)}. Gauging the symmetry corresponding to the gauge field linear in the anomaly yields a theory with higher group $\mathbb G^{(n)}$ symmetry, while gauging the other symmetry yields a theory with non-invertible $\text{Rep}(\mathbb G^{(n)})$ symmetry. \\ 

This derivation of the $\text{Rep}(\mathbb G^{n})$ defects by gauging may seem rather ad-hoc. In fact there is typically a more universal procedure for studying these defects, as well as their relationship to the original $\text{Vec}(\mathbb G^{(n)})$ defects. The SymTFT provides an elegant framework which captures this relationship, and treats the defects for $\text{Vec}(\mathbb G^{(n)})$ and $\text{Rep}(\mathbb G^{(n)})$ on equal footing \cite{Gaiotto:2020iye, Freed:2022qnc, Bhardwaj:2023ayw, Kaidi:2022cpf, Thorngren:2019iar}. For a general symmetry category $\mathcal S$, the SymTFT is the $(d+1)$-dim TQFT defined by the Drinfeld center $\mathcal Z(\mathcal S)$. A simplifying property of the SymTFT is that it is preserved under Morita equivalence, so that symmetry categories exchanged by gauging are described by the same SymTFT. In our context, this means that we can study the topological defects for both $\text{Vec}(\mathbb G^{(n)})$ and $\text{Rep}(\mathbb G^{(n)})$ simultaneously.\footnote{We point out one subtlety in our use of the term ``SymTFT" throughout this paper. Typically the SymTFT of a $d$-dim theory $\mathcal T_d$ is a $(d+1)$-dim topological field theory $\mathcal S_{d+1}$, and the original theory is recovered by compactifying $\mathcal S_{d+1}$ on a slab with appropriate boundary conditions. Here we will typically only deal with the TFT $\mathcal S_{d+1}$ directly, and use it as an organizing tool for studying the topological defects of $\text{Rep}(G)$. We do not associate any specific lower-dim QFT $\mathcal T_d$ to this TFT. } In the 2-group example we just considered, the anomaly equation which resulted in the non-invertibility of the $\text{Rep}(\mathbb G^{(2)})$ amounts to an equation of motion in the SymTFT. The utility of this approach will be made manifest in our next example. Finally, we note that the full structure of $\text{Rep}(\mathbb{G})$ contains much more information than the genuine operators presented above, nonetheless this construction offers a heuristic approach to understanding the complicated $\text{Rep}(\mathbb{G})$ data.

\subsection{Finite 3-Group Symmetry and Charges}\label{sec32}

Let us consider the explicit example of a 3-group global symmetry $\mathbb G^{(3)}_N$. We define this as the untwisted extension of a 1-form symmetry $\mathbb Z_N^{(1)}$ by a 2-form symmetry $\Gamma(\mathbb Z_N)^{(2)}$, where $\Gamma(A)$ is the universal quadratic group of an abelian group $A$. For concreteness, we take the case where $N$ is even, so that $\Gamma(\mathbb Z_N)=\mathbb Z_{2N}$. Then we can write this 3-group as the extension:
\begin{equation}
    \mathbb Z_{2N}^{(2)}\to \mathbb G_N^{(3)}\to \mathbb Z_N^{(1)}
\end{equation}
We define this extension by the Postnikov class $\rho$, which we take to be the Pontryagin  square \footnote{See \cite{Kapustin:2013qsa, Benini:2018reh} for an introduction}
\begin{equation}
    \rho =\frak P \in H^4(B^2\mathbb Z_N, \mathbb Z_{2N})
\end{equation}
Explicitly, given a cocycle $a_2\in C^2(M,\mathbb Z_N)$ we can take an integer lift $\tilde a_2\in C^2(M,\mathbb Z)$ and express the Pontryagin square as:
\begin{equation}
    \frak P\tilde a_2=\tilde a_2\cup \tilde a_2-\tilde a_2\cup_1d\tilde a_2
\end{equation}
In fact more generally there exists a Pontryagin square $\frak P\in H^{2p}(B^p\mathbb Z_N, \mathbb Z_{2N})$ defined by a cohomology operation:
\begin{equation}
    \frak P: H^p(X,\mathbb Z_N)\to H^{2p}(X,\mathbb Z_{2N})
\end{equation}
which defines a quadratic refinement for the pairing on cocycles given by twice the cup product for even $p$:
\begin{equation}
    \frak P(a_p+b_p)-\frak P(a_p)-\frak P(b_p)=2a_p\cup b_p
\end{equation}
In fact this quadratic refinement law is a defining property of the Pontyagin square. We highlight here for later reference that the Postnikov class for this higher group defines a quadratic refinement of the natural cup product on cocycles. This bears a direct resemblance to our primary example of interest $\frak G$ that we introduced in \hyperref[sec2]{Section 2}. We will comment more on this connection soon. \\ 

As discussed in the previous section, a theory with $\mathbb G^{(3)}_N$ or $\text{Rep}(\mathbb G^{(3)}_N)$ global symmetries may be derived from any 4d theory with mixed 't Hooft anomaly whose 5d anomaly cocycle takes the form $\mathcal A_{5d}=\exp(i\int h_1\cup \frak P(a_2))$ by gauging either $h_1$ or $a_2$ respectively. 4-dimensional theories with such a mixed anomaly have been studied \cite{Kaidi:2021xfk}. \\

We would like to understand the topological defect operators belonging to $\text{Rep}(\mathbb G^{(3)}_N)$, which we refer to as the ``charges" of the 3-group $\mathbb G^{(3)}_N$. For this purpose, we find it convenient to use the 5d SymTFT as an organizing framework, since this captures the defects of both $\text{Vec}(\mathbb G^{(3)}_N)$ and $\text{Rep}(\mathbb G^{(3)}_N)$. A specific 4d theory which realizes one of these symmetries is then obtained by compactifying the 5d SymTFT on an interval, with topological and physical boundary conditions at respective ends. The choice of topological boundary condition determines which topological defects survive in the physical theory. \\ 

The 5d SymTFT for this theory describes the SymTFT for a 2-form $\mathbb Z_{N}$ and 3-form $\mathbb Z_{2N}$ gauge fields with a mixed anomaly term:
\begin{equation}
    S_{TFT}=\frac{2\pi}{ N }\int \Big\{ a_2\cup \delta f_2+\frac 12b_3\cup \delta h_1+\frac 12h_1\cup \frak P(a_2)\Big\}
\end{equation}
where the gauge fields are cochains $a_2, f_2 \in C^2(M,\mathbb Z_N)$, $b_3\in C^3(M,\mathbb Z_{2N})$ and $h_1\in C^1(M,\mathbb Z_{2N})$. The pair $(a_2, b_3)$ define the components of a flat $\mathbb G^{(3)}_N$ gauge field. Indeed, the equations of motion for $h_1$ give the condition for a flat higher group connection:
\begin{equation}\label{pontanom}
    \delta b_3=-\frak P(a_2)
\end{equation}
Moreover, one can see that the above topological action is naively not  invariant under a transformation $\tilde a_2\to \tilde a_2+N\lambda_2$ for $\lambda_2\in C^2(M,\mathbb Z)$, since it has variation $\delta\frak P(\tilde a_2)=N\delta(\tilde a_2\cup_1\lambda_2) \mod 2N$, which requires modified higher-group gauge transformations for $b_3$. \\ 

On the other hand, the pair $(h_1,f_2)$ are used to define the topological symmetry defect operators for a theory with $\mathbb G^{(3)}_N$ global symmetry. Such a 4d theory can be obtained from the SymTFT construction by giving Dirichlet boundary conditions to the gauge fields $(a_2,b_3)$ \cite{Bergman:2024its,Xue:2025enx,Kaidi:2023maf, Robbins:2025puq,Franco:2024mxa,Heckman:2022xgu}. We heuristically think of these fields as the global 1-form and 2-form currents that the gauge fields couple to, and indeed the $a_2$ equations of motion yield the familiar 3-group relation among the currents \cite{Cordova:2018cvg}:
\begin{equation}
    \delta f_2\sim  h_1\cup a_2
\end{equation}
A 4d theory with $\text{Rep}(\mathbb G^{(3)}_N)$ global symmetry is obtained by taking the topological boundary conditions to be Dirichlet for $h_1,f_2$. This projects out any topological defects constructed from these from the physical 4d theory. Therefore if we are interested in understanding the topological defects belonging to $\text{Rep}(\mathbb G^{(3)}_N)$ we should study those defects which can be constructed from the 3-group gauge fields $a_2,b_3$. \\ 

The easiest such defects are the $\mathbb Z_N$ Wilson surfaces:
\begin{equation}
    W_n[\Sigma_2]=\exp\Big(i n \int_{\Sigma_2}a_2\Big)
\end{equation}
for $n\in \mathbb Z_N$. On the other hand, the naive holonomy operators:
\begin{equation}
    V_p[\Sigma_3]=\exp\Big(ip \int_{\Sigma_3}b_3\Big)
\end{equation}
are neither topological nor gauge invariant, due to the modified gauge transformations and the anomalous equation of motion $\delta b_3=\frak P(a_2)$. However when $\gcd(p,N)=1$ then a genuine 3d topological defect can be defined by stacking this defect with a minimal abelian TQFT $\mathcal A^{N,p}[\Sigma_3,a_2]$  \cite{Hsin:2018vcg} which is coupled to $a_2$:
\begin{equation}
    \mathcal D_{\frac p N }[\Sigma_3]\equiv e^{i p \int b_3}\mathcal A^{N,p}[\Sigma_3, a_2]
\end{equation}
The resulting operator has non-invertible fusion, and hosts nontrivial topological order on its worldvolume. {The minimal abelian TQFTs $\mathcal A^{N,p}$ define the universal anyon content of any 2+1d abelian topological order with $\mathbb Z_N^{(1)}$ 1-form symmetry with anomaly labeled by $p$.   \\

From this point, the higher categorical data of $\text{Rep}(\mathbb G^{(3)}_N)$ can be computed in terms of the defect data on $\mathcal A^{N,p}$, including junctions between $W_n$ lines and between distinct $\mathcal D_{p/N}$ defects, which are interpreted as higher morphisms, as well as higher data such as the associator. The full structure of the category $\text{Rep}(\mathbb G^{(3)}_N)$ is much more detailed than just the Wilson operators defined above. \\ 

\begin{figure}[t]
    \centering

    \begin{tikzpicture}[
        scale=0.94,
        line join=round,
        font=\small
    ]

% ============================================================
% (a) Parallel fusion: before
% ============================================================
\begin{scope}[xshift=0cm,yshift=4.2cm]

    % Codimension-one operator D_p
    \path[interfaceA]
        (-2,-0.70)
        -- (1.40,-0.70)
        -- (2.10,0)
        -- (-1.30,0)
        -- cycle;

    % Invertible codimension-two operator above D_p
    \path[dpface]
        (-0.90,0.55)
        -- (0.70,0.55)
        -- (1.10,0.95)
        -- (-0.50,0.95)
        -- cycle;

    % Indicate that the two operators are initially separated
    \draw[
        -{Latex[length=2.2mm]},
        draw=DpBlue!80!black,
        line width=0.8pt
    ]
        (-0.35,0.48) -- (-0.35,0.08);

    \draw[
        -{Latex[length=2.2mm]},
        draw=DpBlue!80!black,
        line width=0.8pt
    ]
        (0.55,0.48) -- (0.55,0.08);

    \node[text=VBOrange!60!black]
        at (-2.55,-0.35)
        {$\mathcal{D}_p[\Sigma_3]$};

    \node[text=DpBlue!60!black]
        at (0.15,1.28)
        {$W_n[\Sigma_2]$};

    \node[morphismlabel]
        at (0,-1.10)
        {\textit{before fusion}};

\end{scope}

% Fusion arrow
\node[
    align=center,
    text=black
]
    at (3.15,4.15)
    {$\xrightarrow{\text{parallel fusion}}$};

% ============================================================
% (a) Parallel fusion: after
% ============================================================
\begin{scope}[xshift=6.3cm,yshift=4.2cm]

    % Codimension-one operator D_p
    \path[interfaceA]
        (-2,-0.70)
        -- (1.40,-0.70)
        -- (2.10,0)
        -- (-1.30,0)
        -- cycle;

    % W_n fused parallel onto D_p
    \path[dpface]
        (-0.75,-0.50)
        -- (0.75,-0.50)
        -- (1.12,-0.13)
        -- (-0.38,-0.13)
        -- cycle;

    \node[text=VBOrange!60!black]
        at (2.75,-0.35)
        {$\mathcal{D}_p[\Sigma_3]$};

    \node[text=DpBlue!60!black]
        at (0.25, 0.31)
        {$W_n|_{\mathcal{D}_p}$};

    \node[morphismlabel]
        at (0,-1.10)
        {\textit{after fusion}};

\end{scope}

\node[morphismlabel]
    at (3.15,2.55)
    {\textbf{(a) Parallel fusion onto \(\mathcal{D}_p\)}};

% ============================================================
% (b) Perpendicular termination
% ============================================================
% ============================================================
\begin{scope}[xshift=0cm,yshift=0cm]

    % Codimension-one operator D_p
    % Enlarged so that the endpoint line lies entirely inside it
    \path[interfaceA]
        (-2.40,-1.00)
        -- (1.80,-1.00)
        -- (2.80,0.10)
        -- (-1.40,0.10)
        -- cycle;

    % Codimension-two operator ending perpendicularly on D_p
    \path[dpface]
        (0,-0.80)
        -- (0.78,-0.02)
        -- (0.78,1.48)
        -- (0,0.70)
        -- cycle;

    % Endpoint line on D_p
    \draw[twistline]
        (0,-0.80) -- (0.78,-0.02);

    \node[text=VBOrange!60!black]
        at (-2.7,-0.45)
        {$\mathcal{D}_p[\Sigma_3]$};

    \node[text=DpBlue!60!black, align=center]
        at (0.38,1.75)
        {$W_{pk}[\Sigma_2]$};

    \node[text=SigmaRed, right]
        at (0.75,0.48)
        {$L^k=\partial W_{pk}$};

    \node[morphismlabel]
        at (0,-1.45)
        {\textbf{(b) Perpendicular termination}};

\end{scope}

% ============================================================
% (c) Trivalent junction of endpoint lines
% ============================================================
\begin{scope}[xshift=6.3cm,yshift=0cm]

    \coordinate (J)  at (0,-0.38);
    \coordinate (P1) at (-1.35,-0.38);
    \coordinate (P2) at (1.05,-0.65);
    \coordinate (P3) at (0.58,-0.04);

    % Codimension-one operator D_p
    % Drawn first so the vertical blue surfaces cover it
    \path[interfaceA]
        (-2,-0.70)
        -- (1.40,-0.7)
        -- (2.10,0)
        -- (-1.30,0)
        -- cycle;

    % Three codimension-two operators ending on D_p
    % Drawn after D_p so their blue shading lies on top
    \path[dpface]
        (P1) -- (J)
        -- ($(J)+(0,1.15)$)
        -- ($(P1)+(0,1.15)$)
        -- cycle;

    \path[dpface]
        (P2) -- (J)
        -- ($(J)+(0,1.15)$)
        -- ($(P2)+(0,1.15)$)
        -- cycle;

    \path[dpface]
        (J) -- (P3)
        -- ($(P3)+(0,1.15)$)
        -- ($(J)+(0,1.15)$)
        -- cycle;

    % Endpoint lines, drawn above both shaded regions
    \draw[
        twistline,
        -{Latex[length=2mm]}
    ]
        (P1) -- (J);

    \draw[
        twistline,
        -{Latex[length=2mm]}
    ]
        (P2) -- (J);

    \draw[
        twistline,
        -{Latex[length=2mm]}
    ]
        (J) -- (P3);

    % Pointlike trivalent junction
    \filldraw[
        fill=black,
        draw=white,
        line width=0.7pt
    ]
        (J) circle (3pt);

    \node[text=VBOrange!60!black]
        at (2.75,-0.63)
        {$\mathcal{D}_p[\Sigma_3]$};

    \node[text=DpBlue!60!black]
        at (-0.85,0.98)
        {$W_{pk_1}$};

    \node[text=DpBlue!60!black]
        at (1.18,0.73)
        {$W_{pk_2}$};

    \node[text=DpBlue!60!black]
        at (0.30,1.33)
        {$W_{pk_3}$};

    \node[text=SigmaRed, below]
        at (-0.72,-0.28)
        {$L^{k_1}$};

    \node[text=SigmaRed, below]
        at (0.62,-0.63)
        {$L^{k_2}$};

    \node[text=SigmaRed, above right]
        at (0.50,-0.08)
        {$L^{k_3}$};

    \node[above left=3pt]
        at (J)
        {$v_{k_1,k_2}^{k_3}$};

    \node[morphismlabel]
        at (0,-1.35)
        {\textbf{(c) Trivalent line junction}};

\end{scope}

    \end{tikzpicture}

    \caption{
        Physical configurations involving the codimension-one
        non-invertible operator
        \(
        \mathcal{D}_p[\Sigma_3]
        =
        \exp\!\left(ip\int_{\Sigma_3}b_3\right)
        \mathcal A^{N,p}[\Sigma_3,a_2]
        \)
        and the invertible codimension-two operators
        \(
        W_n[\Sigma_2]
        =
        \exp\!\left(in\int_{\Sigma_2}a_2\right)
        \).
        In (a), \(W_n\) is fused parallel onto \(\mathcal{D}_p\).
        In (b), \(W_{pk}\) terminates perpendicularly on \(\mathcal{D}_p\),
        leaving the line \(L^k=\partial W_{pk}\).
        In (c), three such endpoint lines meet at a pointlike
        junction \(v_{k_1,k_2}^{k_3}\), with
        \(k_1+k_2=k_3\) modulo \(N\) for the indicated orientations.
    }
    \label{fig:physical-higher-morphism-configurations}
\end{figure}
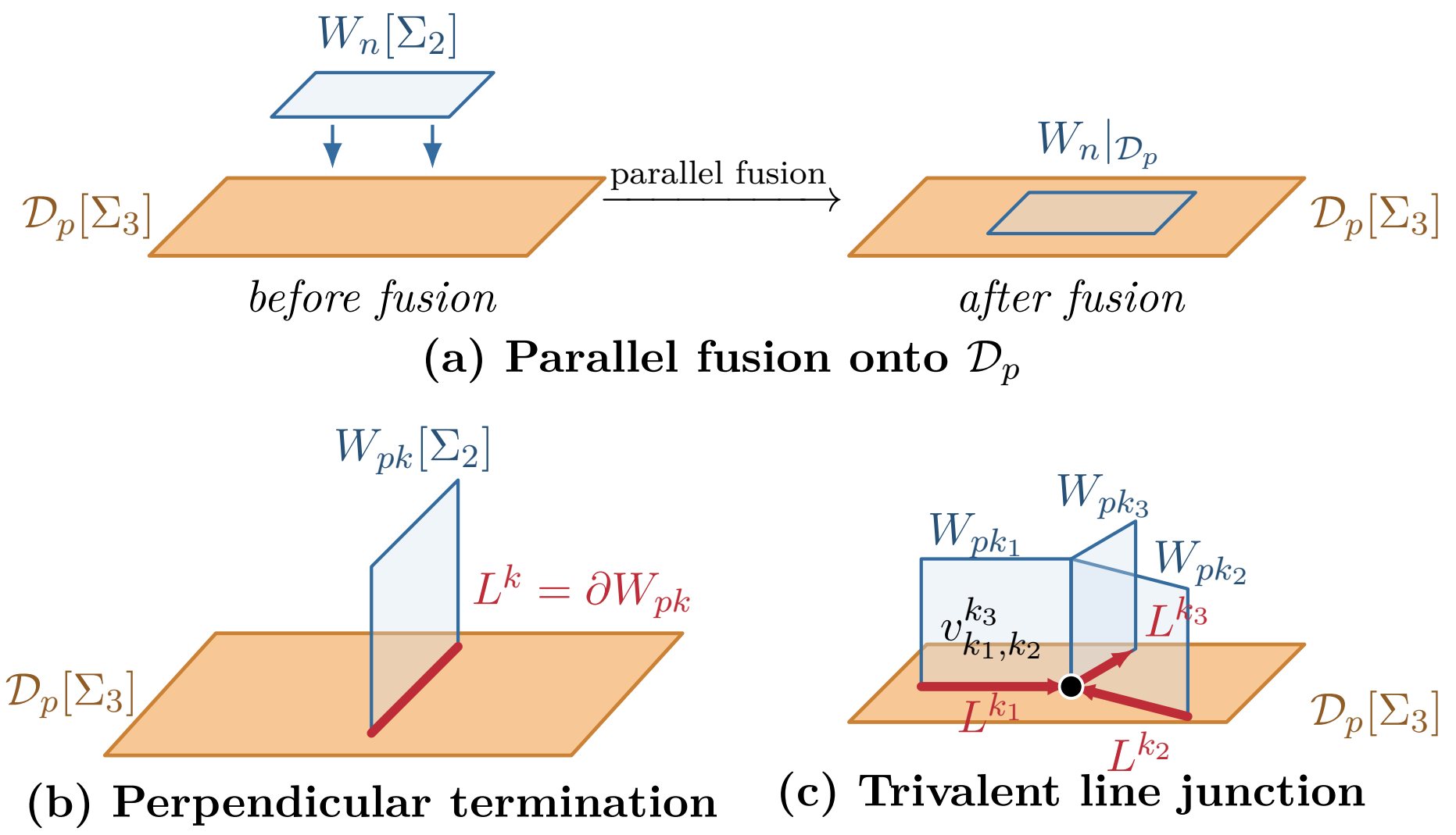

The category $\text{Rep}(\mathbb G^{(3)}_N)$ can also be embedded into a larger fusion category where we allow for all $N\in \mathbb N^+$. The non-invertible defects are then indexed by $\mathbb {Q/Z}$, , and thus all $\mathcal A^{N,p}$ are allowed for coprime pairs $p/N$. This higher fusion category appears as the non-invertible completion of the axial symmetry in 4d QED \cite{Choi:2022jqy, Cordova:2022ieu}, as well as many other theories whenever there is an ABJ anomaly of the form $dJ_{2p-1}= \frac{1}{4\pi} F_p\wedge F_p$. In the case of the non-invertible $\mathbb{Q/Z}$ chiral symmetry of 4d QED, a thorough study of the structure of this category was carried out in \cite{Copetti:2023mcq}, including higher morphisms and associators. 

\subsubsection*{Constructing the Non-Invertible Defects $\mathcal D_{p/N}$}

Let us give a quick review of the construction of the non-invertible operators $\mathcal D_{p/N}$ given in \cite{Choi:2022jqy}. Above we have defined these as stacking the original holonomy operator $\exp(ip\int_{\Sigma_3}b_3)$ with the minimal abelian TQFT $\mathcal A^{N,p}[\Sigma_3,a_2]$ coupled to the physical gauge field $a_2$. However this defect admits an illuminating definition as the \textit{boundary} theory of a 4d TQFT, which can be thought of as performing a ``half-space gauging" \cite{Choi:2021kmx, Hsin:2018vcg} of the $\mathbb Z_N$ symmetry generated by $a_2$ in the 4d bulk, along with an engineered choice of discrete torsion. This definition makes the calculation of the non-invertible fusion of these defects straightforward, since it translates the problem into manipulations of the bulk 4d action. \\ 

Specifically, consider a bounding 4-manifold $M_4$ such that $\Sigma_3=\partial M_4$, and introduce worldvolume $U(1)$ gauge fields $B_2$ and $C_1$ on $M_4$, with the total topological partition function:
\begin{equation}
    \mathcal T^{N,p}[a_2]=\int \mathcal DC_1\mathcal D B_2 \exp\Bigg(\frac{i}{2\pi}\int_{M_4}NdC_1B_2+\frac {kN}2B_2B_2+NB_2a_2\Bigg)
\end{equation}
where $k=p^{-1} \mod N$. The bulk gauge fields have modified gauge transformations in order to preserve gauge invariance:
\begin{equation}
    \begin{gathered}
        B_2\to B_2+d\Lambda_1 \\ C_1\to C_1+d\lambda_0-k \Lambda_1
    \end{gathered}
\end{equation}
The original gauge field $a_2$ is effectively treated as a background. The equation of motion for $C_1$ gives the constraint $NdB_2=0$, which effectively turns $B_2$ into a 2-form $\mathbb Z_N$ gauge field. Then the equations of motion for $B_2$ set $B_2+p a_2=0 \mod N$ which effectively identifies $B_2\cong -p a_2$ as $\mathbb Z_N$ gauge fields. The $B_2$ gauge field is given topological Dirichlet boundary conditions $B_2|_{\Sigma_3}=0$, which breaks the gauge symmetry and allows for genuine Wilson line operators $\exp(i\int C_1)$ on the boundary. Since $\gcd(k,N)=1$ then the bulk topological order is completely trivial and $B_2$ can be safely integrated out, leaving only the boundary Wilson lines and a Hall response SPT phase dependent on $a_2$. The result is the combination:
\begin{equation}
    \mathcal T^{N,p}[a_2]=\mathcal A^{N,p}[\Sigma_3,a_2]\otimes \exp\Big(-ip\int_{M_4} \frak P(a_2)\Big)
\end{equation}
Therefore we see that $\mathcal T^{N,p}$ effectively defines a ``disk operator" on $M_4$, with the minimal abelian TQFT supported on the boundary. The key is now to stack this topological disk operator with the gauge-invariant disk operator defined by the flat 3-group connection $\delta b_3+\frak P(a_2)=0$. The 4d bulk phases cancel and we are left with a genuine topological 3d non-invertible defect operator:
\begin{equation}
    \mathcal T^{N,p}[a_2]\otimes \exp\Big(ip\int_{\Sigma_3}b_3+ip\int_{M_4}\frak P(a_2)\Big)=e^{ip\int b_3}\otimes \mathcal A^{N,p}[\Sigma_3,a_2]\cong \mathcal D_{p/N}
\end{equation}
In fact,  there exists a convenient characterization of this definition using the discrete $SL(2,\mathbb Z_{2N})$ modular transformations for 4d theories with a $\mathbb Z_N^{(1)}$ 1-form symmetry \cite{Gaiotto:2014kfa}. Recall that, given a 4d partition function $Z[b_2]$ coupled to a background $b_2\in H^2(M_4,\mathbb Z_N)$, the following modular transformations can be defined:
\begin{equation} \label{CTST}
    S^{(k)} : Z[b]\to Z_{S^{(k)}}[B]=\frac{1}{|H^2(M,\mathbb Z_N)|^{1/2}}\sum_{b\in H^2(M,\mathbb Z_N)}Z[b]\exp\Big(ik \int b_2\cup B_2\Big) 
\end{equation}
\begin{equation}
    T^{(k)}: Z[b]\to Z_{T^{(k)}}[b]=Z[b]\times \exp\Big(i k \int_M\frak P(b)\Big)
\end{equation}
As well as a central transformation:
\begin{equation}
    C: Z[b]\to Z[-b]
\end{equation}
The $S$ trasnformation has the effect of gauging the $\mathbb Z_N^{(1)}$ symmetry by integrating over $b_2$, and subsequently coupling to the background $B_2$ for the dual $\mathbb Z_N^{(1)} $ symmetry in the gauged theory. The operation $T$ just shifts the overall background by an SPT phase $\exp(i\int \frak P(b))$. \\ 

Equivalently, given a Lagrangian $\mathcal L[B_2]$ for the theory in terms of a $U(1)$ lift of the gauge field $B_2$, we can express these as:
\begin{equation}
    \begin{gathered}
        S^{(k)}: \mathcal L[b] \to \mathcal L[b]+\frac{i N}{2\pi}b_2\wedge dC_1+\frac{i N k}{2\pi}b_2\wedge B_2 \\ T^{(k)}:\mathcal L[b]\to \mathcal L[b]+i\frac{Nk}{4\pi}B_2\wedge B_2
    \end{gathered}
\end{equation}
where it is implicitly assumed in $S^{(k)}$ that the field $b_2$ is subsequently gauged. From these operations, the above 4d bulk definition of $\mathcal T$ is equivalent to performing the combination of modular transformations $CTST$ in the 4d bulk region:
\begin{equation}\label{fCTST}
    \mathcal T^{N,1}\equiv  CTS T\circ e^{i\frac{N}{2\pi}B_2\wedge a_2}
\end{equation}
This framing in terms of the modular transformations makes the definition of the fusion straightforward. In particular, the parallel fusion of two defects is defined by simply performing this operation twice $(CTST)^2$ in the 4d bulk. The result for odd $N$ is \cite{Choi:2022jqy}:
\begin{equation}
    \mathcal D_{1/N}\otimes \mathcal D_{1/N}=\mathcal A^{N,2}[\Sigma_3,0] \otimes \mathcal D_{2/N}
\end{equation}
so that the result of the fusion yields a TQFT coefficient decoupled from $a_2$. The parallel non-invertible fusion of a defect and its conjugate (orientation reversal) is simply defined by implementing $CTST$ on a 4d slab between the defects and compactifying, which yields:
\begin{equation}\label{NIrepG}
    \mathcal D_{p/N}\otimes \mathcal D_{p/N}^\dag=\mathcal C(\Sigma_3)
\end{equation}
so that the fusion is non-invertible, and leaves behind a condensation defect \cite{Roumpedakis:2022aik, Choi:2022zal} where the $\mathbb Z_N$ symmetry generated by $a_2$ is higher-gauged on $\Sigma_3$. Further fusion rules and higher categorical data were computed in \cite{Copetti:2023mcq}. \\ 

\newpage

\section{Charges of Continuous Higher Groups}\label{sec4}

Armed with an understanding of the charges $\text{Rep}(\mathbb G_N)$ for finite higher groups $\mathbb G_N$, we can initiate our study of the case where our higher group $\mathbb G$ is continuous. In particular, we will be interested in the \hyperref[6groupex]{Examples 2.1, 2.2} discussed in section 2: namely the 2-group:
\begin{equation}\label{2groupH}
    U(1)^{(1)}\to \mathbb H^{(2)}\to U(1)^{(0)} \quad \text{with} \quad \rho \cong \frac{1}{2\pi}A_1dA_1\in H^3(BU(1),U(1))
\end{equation}
and the 6-group:\footnote{Here we have abused notation and used the isomorphism $H^3(BU(1),U(1))\cong H^4(BU(1),\mathbb Z)$ under which we identify $\frac{1}{2\pi}A_1dA_1\sim c_1^2\in H^4(BU(1),\mathbb Z)$.}
\begin{equation}
    U(1)^{(5)}\to \mathbb G^{(6)} \to U(1)^{(2)} \quad \text{with} \quad \rho \cong \frac{1}{2\pi}C_3dC_3\in H^7(B^3U(1),U(1))
\end{equation}
While the higher categorical framework governing finite symmetries $\text{Vec}( G_N)$ and their representation categories $\text{Rep}(G_N)$ is well-developed, the correct categorical technology for continuous symmetries is comparitively less mature.\footnote{In the case of $G=U(1)$ continuous symmetry, one interesting proposal for a categorical framework capturing the symmetry category and its Drinfeld center was given in \cite{Jia:2025vrj} building off of the work in \cite{Freed:2009qp}, which uses skyscraper sheaves to construct a natural generalization of the finite category $\text{Vec}(\mathbb Z_N)$.} As a result, our discussion will not follow any existing rigorous proposal, but will instead be guided by a direct generalization of the constructions in the finite case. We do not subscribe to any particular categorical framework which organizes the topological defects we study, but will attempt to interpret various features in categorical language. \\

We will follow the same strategy as the finite case: working directly with the Wilson surface holonomy operators for the higher group gauge fields. Let us recall the Wilson holonomy operators for the 6-group $\mathbb G^{(6)}$ from \hyperref[6charges]{(2.31)}:
\begin{equation}
    \begin{gathered}
        W_n(\gamma_3) = \exp\Big(in\int_{\gamma_3} C_3\Big) \\
        V_p(\Sigma_6)=\exp\Big(ip\int_{\Sigma_6}B_6\Big) \xrightarrow{\delta_{\lambda_2}} \exp\Big(ip\int_{\Sigma_6}\Big[B_6-\frac{1}{2\pi}\lambda_2G_4\Big]\Big)
    \end{gathered}
\end{equation}

We would like to identify these as charges of the 6-group $\mathbb G^{(6)}$.  The naive 6d defects $V_p[\Sigma_6]$ are not gauge invariant, but vary under the modified gauge transformation. Moreover, for a flat connection these defects are not topological due to the anomalous divergence: 
\begin{equation}
    dB_6=-\frac 1{2\pi}C_3dC_3
\end{equation}
Inspired by the example of the non-invertible charges for the finite 3-group $\mathcal D_{p/N}=e^{i\int b_3}\mathcal A^{N,p}$  in $\text{Rep}(\mathbb G^{(3)}_N)$, a solution presents itself: we can attempt to stack $e^{i\int B_6}$ with a gapped topological theory which has the opposite Hall response $-\frac{1}{2\pi}C_3dC_3$. This will be our undertaking in this section. \\

However, recall that the defects we are truly interested in are those which describe the gauge charges of the M2 and M5 branes:
\begin{equation}
    \begin{gathered}
        Z_{M2}\sim \exp\Big(i\int \check C_3\Big) \\ Z_{M5}\sim \exp\Big(i\int_{\Sigma_6}\mathcal B_6\Big) \xrightarrow{\delta_{\lambda_2}}\exp\Big(i\int_{\Sigma_6}\mathcal B_6-\delta_{\lambda_2} Q(\check C_3)\Big) \\ \qquad \qquad \qquad \qquad \qquad \qquad  \ \ \  \  \sim \exp\Big(i\int_{\Sigma_6}\mathcal B_6-\frac{1}{4\pi}\lambda_2(dC_3+\nu)\Big)
    \end{gathered}
\end{equation}
where recall $\check C_3\in \check H^4_{\frac 12\lambda}(M)$ is a shifted differential cocycle. As we discussed in \hyperref[sec23]{Section 2.3} these do not transform as the charges for any ordinary 6-group, but rather belong to the pseudo-6-group structure we denote $\frak G$. We comment on features and proposals for this structure later in \hyperref[sec55]{Section 5.5}. \\

As we have stated, our goal is to take inspiration from the finite higher group case and construct non-invertible defects corresponding to the above defects for $\mathbb G^{(6)}$. Surprisingly, it is in fact natural to first define non-invertible defects corresponding to the operators transforming as $\frak G$ charges. One reason for this is that the non-invertible defects for $\mathbb G^{(6)}$ require a Hall response of $\frac{1}{2\pi}C_3dC_3$ which is 7-form Chern-Simons theory at level 2. On the other hand, the defects of $\frak G$ require a Hall response of $Q(C_3)\sim \frac{1}{4\pi}C_3dC_3$ which is Chern-Simons theory at level 1. Thus the $\mathbb  G^{(6)}$ defects can naturally be obtained as the square of the $\frak G$ defects. Moreover Chern-Simons theory at level 1 is an invertible TQFT, while Chern-Simons theory at level 2 is non-invertible. We will soon see why this makes the charge defects with Hall response $Q(C_3)$ the natural ones to construct. We show later in \hyperref[sec414]{Section 4.1.4} that there also exists a simple bosonic construction for the $\mathbb G^{(6)}$ charges. However since our goal here is to model the M2 and M5 brane charges which transform as defects of $\frak G$ then this is the line we will follow.\footnote{We point out one subtlety of this definition is that, since the quadratic refinement depends on the spacetime Wu structure then so will the squared defects representing charges for $\mathbb{G}^{(6)}$, even though the Postnikov class itself does not require a Wu structure. For a purely bosonic construction which does not depend on a Wu structure see \hyperref[sec414]{Section 4.1.4}} \\

Before jumping to the case of the continuous 6-group $\mathbb G^{(6)}$ (or equivalently $\frak G$), recall that $\mathbb G^{(6)}$ is simply the higher-form version of the 2-group $\mathbb H^{(2)}$ defined by \hyperref[2groupH]{(4.1)}. Therefore, we can begin with the more comfortable case of studying the charges of $\mathbb H^{(2)}$ in 5 dimensions, and then straightforwardly generalize to charges of $\mathbb G^{(6)}$ in 11 dimensions. In the next section we study these charges, which we note might be interpreted as modeling the gauge charges of electric W-boson worldlines and magnetic 't Hooft surfaces in 5d Maxwell-Chern-Simons. \\

\subsection{2-Group Charges in 5d}\label{sec41}

We would like to construct the charges of the 2-group $\mathbb H^{(2)}$ from the Wilson surfaces for a flat 2-group gauge field, whose components are $(A_1,B_2)$. The naive charges are the holonomy operators:
\begin{equation}            
    W_n[\gamma_1]=\exp\Big(in\int_\gamma A_1\Big) 
\end{equation}
\begin{equation}
    V_p[\Sigma_2]=\exp\Big(ip\int_{\Sigma_2}B_2\Big)\xrightarrow{\delta_{\lambda_0}}\exp\Big(ip\int_{\Sigma_2}\Big[B_2-\frac{1}{2\pi}\lambda_0dA_1\Big]\Big)
\end{equation}

As shown, the surface operators $V_p[\Sigma_2]$ are not gauge invariant, and are also not topological due to the flat 2-connection relation $dB_2=\frac{1}{2\pi}A_1dA_1$. We could consider the gauge-invariant disk operator on $M_3$ with $\partial M_3=\Sigma_2$:
\begin{equation}
    D_p[M_3]=\exp\Big(ip\int_{\Sigma_2}B_2+\frac{ip}{2\pi}\int_{M_3}A_1dA_1\Big)
\end{equation}
but this does not define a genuine 2d surface operator. \\

Just as in the 6-group case where our true defects of interest belong to $\frak G$, it will be useful here to consider a refinement of the above defects whose variation coincides with the spin-Chern-Simons action $Q(\check A_1)$:
\begin{equation}\label{Hcharges}
    \begin{gathered}
        \exp\Big(in\int_\gamma \check A_1\Big) \\ 
        \exp\Big(ip\int_{\Sigma_2}B_2\Big)\xrightarrow{\delta_\lambda}\exp\Big(ip\int_{\Sigma_2}\Big[B_2-\delta_\lambda Q(\check A_1)\Big]\Big)\\
        \qquad \qquad \qquad \qquad \qquad \ \  \sim \exp\Big(ip\int_{\Sigma_2}\Big[B_2-\frac{1}{4\pi}\lambda_0(dA_1+\nu)\Big]\Big)
    \end{gathered}
\end{equation}
As with the case of $\frak G$, since $Q(A_1)$ does not define a valid Postnikov class in Eilenberg-MacLane cohomology then these do not correspond to holonomies of an ordinary 2-group. We likewise refer to such defects as belonging to an un-specified structure $\frak H$. The defects with these transformations will be our true charges of interest. \\

In fact, just as the defects of $\frak G$ were defined to model the electric/magnetic M2 and M5 branes of 11d supergravity, with Chern-Simons form $\exp\Big(2\pi i\int\frac 16C_3G_4G_4\Big)$, we can consider the defects of $\frak H$ as representing the electric/magnetic charges of 5d Maxwell-Chern-Simons theory (or alternatively 5d supergravity) \cite{Bekaert:2002eq}. The electric Wilson line holonomy operators $e^{i\int A_1}$ coincide with the charges of W-boson worldlines, while the magnetic surface operators $e^{i\int B_2}$ model the charge and anomaly of the 't Hooft monopole surface operators. \\

In 5D Maxwell-Chern-Simons, it is known that the 't Hooft surface operator must host chiral bosons/ fermions on the grounds of anomaly cancellation \cite{Garcia-Valdecasas:2024cqn,Bekaert:2002eq,Fukuda:2020imw,Boyarsky:2002ck}.  In our context, we are interested in singular Dirac-type magnetic surface operators, which admits UV completion by t' Hooft-Polyakov solutions. In 5D, one can have Wilson lines stretching between monopole worldsheets with the Wilson line ending on a chiral vertex operator on the worldsheets \cite{Boyarsky:2002ck}. These phenomena are analogous to the fact that the magnetic 5-branes in M-theory hosts chiral 2-forms, while the electric 2-branes can end vertically on magnetic branes and the boundary of the 2-brane are coupled to chiral 2-forms.  We will see that all these features can be reproduced by our model of the $\frak H$ charges. \\

\subsubsection*{2-Group SymTFT}

Just as for the case of the finite 3-group $\mathbb G^{(3)}_N$ and the dual non-invertible symmetry for the charges $\text{Rep}(\mathbb G^{(3)}_N)$, there exists a proposal for an organizing 5d TFT which captures both the topological defects for $\mathbb H^{(2)}$ global symmetry, as well as the above charges for $\mathbb H^{(2)}$. This 5d TFT is constructed using the proposed SymTFT for two global $U(1)$ symmetries with a mixed anomaly \cite{Brennan:2024fgj}, and has action \cite{Antinucci:2024zjp}:
\begin{equation}
    S = \frac 1{2\pi}\int_{M_5} d^5x \Bigg[ dA_1\wedge  f_3+dB_2\wedge h_2+ \frac{1}{2\pi}h_2\wedge A_1dA_1\Bigg]
\end{equation}
where $f_3$ and $h_2$ are $\mathbb R$-valued gauge fields. Notice that if we define the gauge-invariant 2-group curvature $H_3=dB_2+\frac{1}{2\pi}A_1dA_1$ then we can write this action as:
\begin{equation}
    S= \frac{1}{2\pi}\int d^5x \Bigg[dA_1\wedge f_3+H_3\wedge h_2\Bigg]
\end{equation}
The equations of motion for $f_3$ and $h_2$ therefore enforce the flatness conditions $\frac{1}{2\pi}dA_1=0$ and $dB_2+\frac{1}{2\pi}A_1dA_1=0$ respectively. Notice that, due to the global subtleties required in defining the spin-Chern-Simons theory $Q(A_1)\sim \frac{1}{4\pi}\int A_1dA_1$, then there is no obvious local topological action which captures the defects of $\frak H$.  In fact, we will stumble upon a possible construction when we study the 11d supergravity action in the presence of M2 and M5 branes later in section \hyperref[sec53]{Section 5.3}. \\ 

The gauge fields $f_3,h_2$ schematically play the role of the Noether currents which the 2-group gauge fields couple to. Indeed, when we want to define a 4d QFT with global 2-group symmetry then we perform the SymTFT compactification step while giving Dirichlet boundary conditions to $A_1,B_2$, then the surviving topological defects $e^{i\alpha\int f_3}$ and $e^{i\alpha\int h_2}$ generate the 2-group global symmetry. On the other hand, if we want to identify the dual non-invertible symmetry obtained from giving $A_1,B_2$ Neumann boundary conditions, then we are led to consider the topological defects constructed from $A_1,B_2$ directly. We will identify these defects as the charges of $\mathbb H^{(2)}$. We will not rely on the above SymTFT any longer, but instead focus directly on the defects themselves. 

\subsubsection{2-Group Non-Invertible Charges}

As we mentioned, it will first be easier to begin with the defects of $\frak H$ defined in \hyperref[Hcharges]{(4.9)}. The simplest defect is the Wilson line $W_n[\gamma_1]$. However the surface operators are neither topological nor gauge invariant:
\begin{equation}
    \begin{gathered}
        \exp\Big(ip\int_{\Sigma_2}B_2\Big)\xrightarrow{\delta_\lambda}\exp\Big(ip\int_{\Sigma_2}\Big[B_2-\delta_\lambda Q(\check A_1)\Big]\Big)\\
        \quad \quad\quad\quad\quad\quad\quad\quad\quad\quad\quad \sim \exp\Big(ip\int_{\Sigma_2}\Big[B_2-\frac{1}{4\pi}\lambda_0(dA_1+\nu)\Big]\Big) 
    \end{gathered}
\end{equation}
We would like to cure this by stacking with a 2d gapped TQFT which has an equivalent anomalous response. Note that well-known obstruction theorems prevent any such gapped phase from preserving the full symmetry. This is the point we will leverage: our gapped phase will be characterized as a Higgs phase which is explicitly charged under the symmetry. This is why it is essential that it be paired with $\exp({i\int B_2})$. \\

We will follow the same approach which was used to construct the defects $\mathcal D_{p/N}$ in \hyperref[sec32]{Section 3.2}. In this case, the correct TQFT was defined by a “half-space gauging” construction: by considering a bulk TQFT whose Hall response matched the anomaly of the bare defect. We will pursue this same strategy here. \\

We again use a half-space gauging.\footnote{In fact, since we are considering a 2d defect embedded in a 5d spacetime, this should really be considered a ``higher half-space gauging", as proposed in \cite{Choi:2022fgx}. Such a construction proceeds by first considering an invertible condensation defect on a manifold with boundary, and then defining the desired non-invertible defect on this boundary. We will see that our construction follows precisely such a procedure.} That is, we will consider a bounding 3d bulk manifold $M_3$ with boundary $\Sigma_2=\partial M_3$. In the 3d bulk we define a 3d TQFT coupled to $A_1$. We want this TQFT to admit gapped topological boundary conditions and have Hall response $\exp(-iQ(A_1))\sim \exp(-\frac{i}{4\pi}\int A_1dA_1)$. \\ 

We first present an overview of the construction of the non-invertible defect. We will then go back and analyze the construction in more detail, and comment on the motivation and virtues of this definition. In the 3d bulk we introduce two independent dynamical $U(1)$ gauge fields $a_1$ and $\bar a_1$. We will refer to the bulk $A_1$ gauge field as the ``physical" gauge field, which from the perspective of the worldvolume is effectively treated as a background. Locally we define the 3d action:
\begin{equation}
    \mathcal S\approx\int \mathcal Da_1 \mathcal D\bar a_1 \ \exp\Bigg(\frac{i}{2\pi}\int_{M_3} \Big[A_1da_1+\frac12 a_1da_1-\frac 12\bar a_1d\bar a_1\Big]\Bigg)
\end{equation}
Of course, since the action involves a 3d spin-Chern-Simons theory at level 1, the above action is not globally well-defined. We can use the more precise notation of differential cohomology in terms of the pairing and quadratic refinement defined in \hyperref[appa]{Appendix A}:
\begin{equation}
    \begin{gathered}
        (\check a_1,\check b_1)=\int \ \check a_1\star \check a_2 \sim \frac{1}{2\pi}\int a_1db_1 \\ 
        \tilde Q(\check a_1)=Q(\check a_1)-Q(0)\sim \frac{1}{4\pi}\int a_1(da_1-\nu)
    \end{gathered}
\end{equation}
Henceforth we will drop the $\sim$ on $\tilde Q$ and simply write $Q$. Then we can precisely define the 3d theory:
\begin{equation}
    \mathcal S\equiv \int \mathcal Da_1 \mathcal D\bar a_1 \ \exp i\Bigg[(A_1,a_1)+Q(a_1)-Q(\bar a_1)\Bigg]
\end{equation}
This theory admits topological boundary conditions for the gauge fields, so the boundary is manifestly topological. By using the quadratic refinement property of $Q$, we can complete the square of this action and see that it is in fact equivalent\footnote{There is a subtlety here in applying the quadratic refinement law which we will soon address. It amounts to the fact that on a manifold with boundary the Chern-Simons action $Q(a)$ can not be considered a $U(1)$ phase but rather defines a line  $q_\Sigma(x)$  over the boundary configuration of the gauge field $x=a|_\Sigma$. Therefore the quadratic refinement law does not define an equality of $\mathbb R/\mathbb Z$ phases but rather an isomorphism of lines $q_\Sigma(a+b)\cong q_\Sigma(a)\otimes q_\Sigma(b)\otimes(a,b)_\Sigma$.} to:
\begin{equation}
    \mathcal S\cong \exp\Big[-iQ(A_1)\Big]\times\int \mathcal D\tilde a_1 \mathcal D\bar a_1 \ \exp i\Big[Q(\tilde a_1)-Q(\bar a_1)\Big]
\end{equation}
where we have redefined the gauge field $\tilde a_1\equiv a_1+A_1$. The second factor defines a completely trivial invertible 3d spin Chern-Simons theory in the bulk. The topological boundary conditions have been carried through, which allows us to integrate out the dynamical 3d bulk gauge fields, leaving only a 2d topological theory $\mathcal W[\Sigma_2]$ on the boundary $\Sigma_2$. In total, we find that this defect $\mathcal S$ decomposes into a pure 3d SPT phase with Hall response $-Q(A_1)$ along with a boundary 2d TQFT:
\begin{equation}
    \tcboxmath[
        colback=pink!30!white,
        colframe=black!75,
        boxrule=0.6pt,
        arc=0pt,
        left=5pt,
        right=5pt,
        top=4pt,
        bottom=4pt
    ]{
        \mathcal S\cong \mathcal W_1[\Sigma_2,A_1]\otimes \exp\Big[-iQ(A_1)\Big]
    }
\end{equation}
We are then free to stack this theory with the gauge-invariant disk operator composed from the 2-group connection $\exp\Big(i\int_{\Sigma_2}B_2+iQ(A_1)\Big|_{M_3}\Big)$ which leaves us with the genuine 2d topological defect:
\begin{equation}
    \tcboxmath[
        colback=pink!30!white,
        colframe=black!75,
        boxrule=0.6pt,
        arc=0pt,
        left=5pt,
        right=5pt,
        top=4pt,
        bottom=4pt
    ]{
        \mathcal D_1\equiv \mathcal S\otimes \exp\Big(i\int_{\Sigma_2}B_2+iQ(A_1)\Big|_{M_3}\Big) = e^{i\int_{\Sigma_2 }B_2}\otimes \mathcal W_1[\Sigma_2, A_1] 
    }
\end{equation}

Let us now work through the details of this definition in order to understand this topological boundary theory $\mathcal W[\Sigma_2, A_1]$. We will argue that $\mathcal W[\Sigma_2,A_1]$ is effectively a \textit{Higgsing} defect: an almost-trivial topological phase whose vacuum is charged under the gauge symmetry of the bulk physical gauge field $A_1$, and which supplies local operators that the bulk Wilson lines $e^{i\int A_1}$ may end on. \\

The role of the gauge field $\bar a_1$ is purely as a spectator field, which does not couple to the physical gauge field $A_1$, and therefore does not change the desired Hall response. Its role is to supply the term $-Q(\bar a_1)$ in the action which cancels all (higher) chiral central charge \cite{Kaidi:2021gbs} of the 3d Chern-Simons theory. This is what allows for the gapped topological boundary conditions. What are these boundary conditions? \\

If we focus only the contributions from the dynamical gauge fields $a_1,\bar a_1$ for the moment, and disregard the coupling to $A_1$, then we identify the above Chern-Simons theory as $U(1)_1\times U(1)_{-1}$, which has K-matrix:
\begin{equation}
    K=\begin{pmatrix} 1 & 0 \\ 0 & -1\end{pmatrix}
\end{equation}
The classification of topological boundary conditions for a general Chern-Simons theory amounts to the identification of Lagrangian subgroups among the total set of anyons. Since $|\det K|=1$ then there is only a single such anyon, which indicates only one choice. In fact there is an extra choice when specifying the topological boundary. In the language of \cite{Kapustin:2010hk}, there is a choice of isotropic sublattice $\Lambda_0$ of the lattice $\Lambda\cong\mathbb Z\oplus\mathbb Z$ defining the gauge group, which results in a splitting:
\begin{equation}
    1\to \Lambda_0\to \Lambda\to \Lambda/\Lambda_0\to 1
\end{equation}
Two possible choices for such a sublattice are:
\begin{equation}
    \Lambda_0=\Lambda_{\pm}\equiv \{ (n,\pm n)  \ : \ n\in \mathbb Z\} 
\end{equation}
This determines which $U(1)$ subgroup of the overall $U(1)^2$ gauge group is preserved by the boundary. Since the theory is in fact a spin TQFT, there is also a transparent fermion line $\psi$ which may contribute to the boundary condition. In this case there are two resulting boundary theories exchanged by stacking the boundary with the 2d spin SPT phase $
\text{Arf}[\Sigma_2]\in \Omega^2_{spin}(\text{pt})$. \\

This can be physically understood in terms of the more familiar gapless boundary conditions for this $U(1)_1\times U(1)_{-1}$ Chern-Simons theory. To each Chern Simons term we can give holomorphic/antiholomorphic boundary conditions which define a pair of chiral/anitchiral modes on the boundary. The corresponding boundary theory is then a pair of left/right moving fermions. The gapped topological boundary condition can be thought of as introducing a gapping interaction which pairs these modes. The two possible choices for the gapped boundary condition correspond to the two possible gapping terms, which necessarily breaks one of the $U(1)$  symmetries. The two resulting gapped phases are  exchanged by stacking with $\text{Arf}[\Sigma_2]$ in the deep IR \cite{Okuda:2020fyl, Bhardwaj:2024ydc, Smith:2019jnh, Levin_2013}. \\

The boundary conditions for the chosen sublattices $\Lambda_\pm$ can be enforced by including a compact Lagrange multiplier constraint localized on the boundary \cite{Kapustin:2010hk, Kapustin:2010if} of the form:
\begin{equation}
    \frac{1}{2\pi}\int_{\Sigma_2} \varphi (da_1\pm d\bar a_1)
\end{equation}
where the Lagrange scalar $\varphi$ is charged under the surviving boundary gauge group. As a result, bulk Wilson lines belonging to $\Lambda/\Lambda_\pm$ may end on the vertex operators $\exp(in\varphi)$. Of course, the coupling to the background $\int a_1dA_1$ contributes an anomalous gauge variation, so in addition we must also add a boundary coupling: 
\begin{equation}
    \frac{1}{2\pi}\int_{\Sigma_2} \varphi (da_1\pm d\bar a_1+dA_1)
\end{equation}

Now we can carry out the change of variables $\tilde a_1\equiv a_1+A_1$ and apply the quadratic refinement law, which contributes an extra boundary term $\frac{1}{4\pi}\int_{\Sigma_2}A_1\wedge \tilde a_1$. In the differential cohomology formulation this extra term is absorbed into the definition of the Chern-Simons pairing as defining a nontrivial section of a line bundle associated to the boundary, so that the quadratic refinement law holds as an isomorphism between line bundles. We keep it here when working with a local trivialization to highlight its role. The total bulk+boundary action in a local trivialization is then:
\begin{equation}\label{bndyterm}
    S=-\frac{1}{4\pi}\int_{M_3} A_1dA_1+\frac 1{4\pi}\int_{M_3}\Big(\tilde a_1d\tilde a_1-\bar a_1d\bar a_1\Big)+\frac 1{2\pi}\int_{\Sigma_2}\Big[\varphi(d\tilde a_1-d\bar a_1)+\frac 12\tilde a_1\wedge A_1\Big]
\end{equation}
The phase $-\frac{1}{4\pi}\int A_1dA_1$ is canceled by stacking with the disk operator. What is left in the 3d bulk is a trivial phase, since $\varphi$ imposes the diagonal topological boundary conditions. However the topological boundary itself is not quite trivial since it must produce a response under an $A_1$ gauge transformation. The diagonal boundary conditions are twisted by the background $A_1$. To characterize the boundary let us suppress the $\bar a_1$ contribution since it does not couple to $A_1$. The key contribution to the boundary is then:
\begin{equation}
    \mathcal W[\Sigma_2,A_1]\sim \int \mathcal D\varphi \exp\Big(\frac{i}{2\pi}\int_{\Sigma_2}\tilde a_1\wedge (d\varphi +\frac 12 A_1)\Big)
\end{equation}
We see that $\tilde a_1$ schematically enforces the relation $d\varphi =-\frac 12A_1$.\footnote{This constraint should not be interpreted too literally. The proper statement is that the topological boundary condition imposed by $\varphi$ trivializes the total line defined by $Q(\tilde a_1)-Q(\bar a_1)-Q(A_1)$.} When we generalize to 6d this will be reminiscent of the required relation $dB_2=H_3\cong C_3$ between the worldvolume chiral 2-form $B_2$ of an M5-brane and the bulk $C_3$. \\

The interpretation of this theory can be illuminated by T-dualizing the scalar $\varphi$ into a dual scalar $\sigma$ which shifts under the gauge symmetry of $\tilde a_1-\bar a_1$, and thus in particular shifts under the gauge symmetry of the physical gauge field $A_1$. This yields the coupling:
\begin{equation}
         \exp\Big(-\frac{i}{4\pi}\int_{\Sigma}d\sigma\wedge A_1\Big) \quad \subset \quad \mathcal W[\Sigma_2,A_1]
\end{equation}
This manifestly realizes the required response $\delta_\lambda Q(A)\sim \frac{1}{4\pi}\int \lambda dA$ of the boundary theory under an $A_1$ gauge transformation. This allows us to interpret the theory $\mathcal W[\Sigma_2,A_1]$ as effectively describing a Higgsed phase where the vacuum is charged under the gauge symmetry of the bulk physical gauge field $A_1$ \cite{Maldacena:2001ss}. The dual vertex operators $e^{i\sigma}$ are local operators charged under this gauge symmetry. Thus the boundary topological theory is an almost trivial gapped phase: \textit{it’s role in life is to supply a local operator for the physical Wilson line to end on}. In categorical language, we interpret this junction between the Wilson line and the surface defect as a higher morphism. We will comment on the physical interpretation of this later in \hyperref[sec5]{Section 5}. \\

Notice that if we ignore the coupling to $A_1$ then the surface operators $\mathcal W[\Sigma_2]$ are precisely the interfaces studied in \cite{Kapustin:2010hk, Kapustin:2010if}. In particular $\mathcal W_1[\Sigma_2,0]$ is the invisible interface defined by the diagonal topological boundary conditions. However the coupling to the physical gauge field $A_1$ is what makes this surface operator nontrivial, as the vacuum is now charged under $A_1$. Condensation defects constructed from such surface operators were considered in \cite{Roumpedakis:2022aik}.  \\

\subsubsection{Modular Transformation Construction}

Recall that the 4d half-space gauging construction of the non-invertible charge defects for $\text{Rep}(\mathbb G^{(3)}_N)$ had a natural definition in terms of the $SL(2,\mathbb Z_{2N})$ modular transformations \hyperref[CTST]{(3.18)-(3.20)} for 4d theories with a $\mathbb Z_N^{(1)}$ symmetry. Now the non-invertible charge defects $\mathcal D_1$ we have just constructed are defined using a bulk 3d abelian Chern-Simons theory. In this case there also exists a natural set of modular $SL(2,\mathbb Z)$ transformations introduced by Witten \cite{Witten:2003ya, Kapustin:1999ha}. Given a partition function $Z[A_1]$ coupled to a background $U(1)$ gauge field $A_1$ they are defined by:
\begin{equation}
    \begin{gathered}
         S : Z[A_1]\to Z_S[a_1]\equiv \int \mathcal D A_1 \ Z[A_1] \exp\Bigg(\frac{i}{2\pi}\int_{M_3}a_1dA_1\Bigg)\\
        T: Z[A_1]\to Z_T[A_1]\equiv Z[A_1]\exp\Bigg(\frac{i}{4\pi}\int A_1dA_1\Bigg)\\ 
        C : Z[A_1]\to Z[-A_1]
    \end{gathered}
\end{equation}
The transformation $S$ promotes the background $A_1$ to a dynamical gauge field and subsequently introduces a new background for the resulting topological symmetry, while the transformation $T$ shifts by a background spin Chern-Simons phase. We can also express these transformations in a globally-well-defined way using the pairing and quadratic refinement on differential cohomology:
\begin{equation}
    \begin{gathered}
         S : Z[\check A_1]\to Z_S[\check a_1]\equiv \int \mathcal D \check A_1 \ Z[\check A_1] \exp\Big[i(\check a_1,\check A_1)\Big]\\
        T: Z[\check A_1]\to Z_T[\check A_1]\equiv Z[\check A_1]\exp\Big[iQ(\check A_1)\Big]
    \end{gathered}
\end{equation}
Of course, as we’ve argued above, for each modular transformation we also want to add a decoupled spectator sector, which implements precisely the \textit{opposite} modular transformation for a background spectator. We therefore define the \textit{doubled} modular transformations:
\begin{equation}\label{dmod}
    \begin{gathered}
        \mathbf S: Z[A,\bar A]\to Z_{\mathbf S}[a,\bar a]=\int \mathcal DA\mathcal D\bar A \ Z[A,\bar A] \ \exp\Big[i(a,A)-i(\bar a,\bar A)\Big] \\ \mathbf T:Z[A,\bar A]\to Z_{\mathbf T}[A,\bar A]= Z[A,\bar A]\exp\Big[iQ(A)-iQ(\bar A)\Big] \\ \mathbf C:Z[A,\bar A]\to Z[-A,-\bar A]
    \end{gathered}
\end{equation}
Using these modular transformations, we can express the half-space gauging definition of our non-invertible defect $\mathcal S$ as applying the combination $\mathbf{CTST}$ in the 3d half-space:
\begin{equation}
    \mathcal S=\mathbf{CTST}\circ e^{i (a,A)}
\end{equation}
Notice that this is precisely the same combination of modular transformations which defined the non-invertible defects \hyperref[fCTST]{(3.22)} for the finite case $\text{Rep}(\mathbb G^{(3)}_N)$, which was constructed in \cite{Choi:2022jqy}. In our presentation of \hyperref[fCTST]{(3.22)} we have followed their convention of highlighting only the dynamical fields contributed by the $\mathbf{ST}$, while suppressing the background field manipulations of $\mathbf{CT}$. Nevertheless, these are necessary in order to reproduce the same coupling of the physical gauge field $A_1$ to the background. While the combination $\mathbf{CTST}$ certainly gives us the desired properties we want for our defect, we will argue further below why it is in fact the most natural definition. \\

\subsubsection{Higher-Charge Defects}

We have successfully defined the minimal non-invertible charge: 
\begin{equation}
    \mathcal D_1= e^{i\int_{\Sigma_2} B_2}\mathcal \otimes \mathcal W[\Sigma_2,A_1]
\end{equation}
Constructed from the 3d half-space gauging defect $\mathcal S$ which had Hall response $-Q(A_1)\sim -\frac{1}{4\pi}A_1dA_1$. This reflects precisely the expected charge for defects of $\frak H$. What about defects with higher Hall response $-k\mathcal Q(A_1)\sim -\frac{k}{4\pi}A_1dA_1$? For example, to describe the defects of the genuine 2-group $\mathbb H^{(2)}$ we need a defect with twice the Hall response, which corresponds to the level-2 Chern-Simons term $-\frac{1}{2\pi}A_1dA_1$. \\

We will first give a simple but incorrect guess for the definition, and explain its fatal features. The simplest approach would be a straightforward generalization of the above definition for the minimal defect:
\begin{equation}
    \mathcal S_k \stackrel{?}{=}\mathbf{C}\mathbf{T}_k\mathbf{S}_k\mathbf{T}_k=\int\mathcal Da_1 \mathcal D\bar a_1 \ \exp i\Bigg[k(A_1,a_1)+kQ(a_1)-kQ(\bar a_1)\Bigg]
\end{equation}
 Indeed we can complete the square using the same field redefinition as in the minimal defect, and find the desired Hall response:
 \begin{equation}
     \mathcal S\cong \exp\Big[-ikQ(A_1)\Big]\times\int \mathcal D\tilde a_1 \mathcal D\bar a_1 \ \exp i\Big[kQ(\tilde a_1)-kQ(\bar a_1)\Big]
 \end{equation}

The leftover dynamical theory in the 3d bulk is $U(1)_k\times U(1)_{-k}$ Chern-Simons theory. However there are several maladies with this definition. The first is that the bulk Chern-Simons theory is now a fully non-invertible TQFT, which means it is not so innocent to attempt to integrate out the bulk fields and obtain a genuine 2d boundary theory. The bulk path integral will instead prepare a nontrivial Hilbert space of states on the boundary, so that the boundary is a relative theory \cite{Freed:2012bs}. One might think that this is actually a desirable feature, since we expect the 6d $\mathcal N=(2,0)$ SCFT embedded in the worldvolume theory of a stack of $k$ M5-branes to be a relative theory defined by some partition vector. We will comment more on this interpretation later in \hyperref[sec5]{Section 5}. \\

However there is a second issue which is more troubling. Recall that our construction of the 2-group charge surfaces requires stacking the above theory with the disk operator $\exp\Big(ik\int_{\Sigma_2} B_2+ik Q(A_1)\big|_{M_3}\Big)$ to obtain:
\begin{equation}
    \mathcal D_k \sim e^{ik\int B_2}\otimes \mathcal W_k[\Sigma_2,A_1]
\end{equation}
Moreover, we want to be able to perform parallel fusion of these defects $\mathcal D_k\otimes \mathcal D_q$, from which we see the holonomy contributions add linearly: $e^{i(k+q)\int B_2}$. Therefore, we require that the parallel fusion $\mathcal S_k\otimes \mathcal S_q$ have total Hall response $\exp\Big(-i(k+q)Q(A_1)\Big)$. However if we use the above definition and define the fusion:
\begin{equation}
    \begin{gathered}
        \mathcal S_k\otimes \mathcal S_q\stackrel{?}{=}(\mathbf{C}\mathbf{T}_q\mathbf{S}_q\mathbf{T}_q)(\mathbf{C}\mathbf{T}_k\mathbf{S}_k\mathbf{T}_k) \\
        =\int\mathcal Da_1 \mathcal D\bar a_1\mathcal D b_1\mathcal D\bar b_1 \ \exp i\Bigg[k(A_1,a_1)+\Big(kQ(a_1)-k(a_1,b_1)+(k+q)Q(b_1)\Big) \\
        -\Big(kQ(\bar a_1)-k(\bar a_1,\bar b_1)+(k+q)Q(\bar b_1)\Big)\Bigg]
    \end{gathered}
\end{equation}
Then in fact this total theory does not have the required Hall response. This implies that the fusion $\mathcal D_k\otimes \mathcal D_q$ does not yield a genuine topological 2d defect. This is a key property we need to preserve with any definition of the topological defects: not only should the genuine defect $\mathcal D_k$ itself be topological, but the result of its parallel fusion with any other $\mathcal D_q$ should yield a genuine topological defect. A necessary requirement for this is for the Hall responses to add linearly. \\

In fact this is a key reason we have defined the minimal defect $\mathcal S_1$ as the combination of modular transformations $\mathbf{CTST}$. Not only does this definition have the desired Hall response, but its parallel fusion with itself $\mathcal S_1^{\otimes k}\sim (\mathbf{CTST})^k$ has total Hall response $\exp(-ik Q(A_1))$. \\

We will therefore define the \textit{defect of charge k} as the iterated parallel fusion of $k$ minimal defects:
\begin{equation}
    \tcboxmath[
        colback=pink!30!white,
        colframe=black!75,
        boxrule=0.6pt,
        arc=0pt,
        left=5pt,
        right=5pt,
        top=4pt,
        bottom=4pt
    ]{
        \mathcal S_k
        \equiv
        \mathcal S_1^{\otimes k}
        =
        \mathcal S_1\otimes\mathcal S_1\otimes\dots\otimes\mathcal S_1
        =
        (\mathbf{CTST})^k
    }
\end{equation}
\hspace{0em}Let us study the defect obtained from this iterated parallel fusion. If we ignore the coupling to the physical gauge field $A_1$ and focus only on the dynamical 3d worldvolume gauge fields $a_1^{(1)},a_1^{(2)},a_1^{(3)},\dots$ and their spectators $\bar a_1^{(1)},\bar a_1^{(2)}, \bar a_1^{(3)},\dots$ then the total K-matrix for the 3d Chern-Simons theory takes the form:
\begin{equation}
    \mathbb K=\begin{pmatrix} K & 0 \\ 0 & -K \end{pmatrix} 
\end{equation}
where:
\begin{equation}\label{Kmatrix}
K=
\left(
\begin{array}{c|cccccc}
1      & -1     & 0      & 0      & 0      & \cdots & 0      \\ \hline
-1     & 2      & -1     & 0      & 0      & \cdots & 0      \\
0      & -1     & 2      & -1     & 0      & \cdots & 0      \\
0      & 0      & -1     & 2      & -1     & \cdots & 0      \\
0      & 0      & 0      & -1     & 2      & \ddots & \vdots \\
\vdots & \vdots & \vdots & \vdots & \ddots & \ddots & -1     \\
0      & 0      & 0      & 0      & \cdots & -1     & 2
\end{array}
\right)
=
\left(
\begin{array}{c|ccc}
    1      & -1 & 0 & \dots \\ \hline
    -1     &     &   &       \\
    0      &     &
    \smash{\mathclap{
        \scalebox{1.6}{ \ $\mathbb A_{k-1}$}
    }}
    &       \\
    \vdots &     &   &
\end{array}
\right).
\end{equation}
Here $\mathbb A_{k-1}$ is the Cartan matrix for the Lie algebra $\frak a_{k-1}$. The appearance of this Cartan matrix is remarkable, as we will discuss when we relate the stack of fused 6d non-invertible defects to a stack of $k$ M5 branes in Sec \ref{sec5}.\\

In fact it is simple to see that this 3d Chern-Simons theory has total Hall response  $\exp(-ik Q(A_1))$. To do this, we recall that we must also include the coupling of the physical gauge field to the first dynamical gauge field $(a_1,A_1)=\frac{1}{2\pi}\int a_1dA_1$. If we include this coupling, we write the total K-matrix as:   
\begin{equation}
\overline{\mathbb K}
=
\vcenter{\hbox{%
\begin{tikzpicture}[
    x=0.8cm,
    y=0.8cm,
    baseline=(current bounding box.center)
]

% Background region bounded by the parentheses
\begin{scope}
    \clip
        (-2.05,2.05)
        -- (2.05,2.05)
        .. controls (2.30,1.40) and (2.30,-1.40)
        .. (2.05,-2.05)
        -- (-2.05,-2.05)
        .. controls (-2.30,-1.40) and (-2.30,1.40)
        .. (-2.05,2.05)
        -- cycle;

    % Gray first block row and first block column
    \fill[blue!15!white]
        (-2.30,-2.05) rectangle (2.30,2.05);

    % Red lower-right block
    \fill[red!18]
        (-1.00,-2.05) rectangle (2.30,1.00);

    % Dashed vertical block partitions
    \draw[
        black!70,
        line width=0.6pt,
        dash pattern=on 3pt off 2pt
    ]
        (-1.00,-2.10) -- (-1.00,2.10);

    \draw[
        black!70,
        line width=0.6pt,
        dash pattern=on 3pt off 2pt
    ]
        (1.00,-2.10) -- (1.00,2.10);

    % Dashed horizontal block partitions
    \draw[
        black!70,
        line width=0.6pt,
        dash pattern=on 3pt off 2pt
    ]
        (-2.30,1.00) -- (2.30,1.00);

    \draw[
        black!70,
        line width=0.6pt,
        dash pattern=on 3pt off 2pt
    ]
        (-2.30,-1.00) -- (2.30,-1.00);
\end{scope}

% Left parenthesis
\draw[line width=0.75pt]
    (-2.05,2.05)
    .. controls (-2.30,1.40) and (-2.30,-1.40)
    .. (-2.05,-2.05);

% Right parenthesis
\draw[line width=0.75pt]
    (2.05,2.05)
    .. controls (2.30,1.40) and (2.30,-1.40)
    .. (2.05,-2.05);

% First block row
\node at (-1.50, 1.50) {$0$};
\node at (-0.50, 1.50) {$1$};
\node at ( 0.50, 1.50) {$\mathbf{0}^{\mathsf T}_{k-1}$};
\node at ( 1.50, 1.50) {$\mathbf{0}^{\mathsf T}_k$};

% First block column
\node at (-1.50, 0.50) {$1$};
\node at (-1.50,-0.50) {$\mathbf{0}_{k-1}$};
\node at (-1.50,-1.50) {$\mathbf{0}_k$};

% Central k-by-k block
\node at (0,0)
    {\scalebox{1.1}{\ensuremath{K}}};

% Upper-right k-by-k zero block
\node at (1.70,0)
    {$\mathbf{0}_{k\times k}$};

% Lower-middle k-by-k zero block
\node at (0,-1.50)
    {$\mathbf{0}_{k\times k}$};

% Bottom-right k-by-k block
\node at (1.50,-1.50)
    {\scalebox{1.1}{\ensuremath{-K}}};

\end{tikzpicture}%
}}.
\end{equation}
where the entries with blue background denote couplings to the physical gauge field $A_1$, and the entries with red background denote dynamical gauge fields on the worldvolume of the higher half-space gauging defect. \\

In fact this matrix can be diagonalized by an $SL(N,\mathbb Z)$ transformation where $N=2k+1$. This is achieved by noticing that $K$ can be diagonalized to the identity matrix using:
\begin{equation}\label{diagU}
    K=U \mathbf{1}_k U^T \quad \text{where} \quad U=\begin{pmatrix} -e_1 \\ \alpha_1 \\ \alpha_2 \\ \vdots \\ \alpha_{k-1} \end{pmatrix}
\end{equation}
where $e_1=(1,0, 0, \dots, 0)$ and $\alpha_i=e_i-e_{i+1}$ are the generators of the root lattice for $\frak a_{k-1}$. Thus the full matrix $\overline{\mathbb K}$ can be diagonalized by the change of variables $\vec b= -U^T \vec a + \mathbf 1 A$ and $\vec{\bar b}=-U^T \vec {\bar a}$, or more explicitly:
\begin{equation}\label{cov}
\begin{alignedat}{2}
       & b_1^{(k)} \equiv a_1^{(k)}+A_1, \qquad\qquad          &b_1^{(i)}\equiv a_1^{(i)}-a_1^{(i+1)}+A_1 \quad \text{for }  \ 1\leq i \leq k-1   \\
        &\bar b_1^{(k)} \equiv \bar a_1^{(k)},\qquad \qquad  &\bar b_1^{(i)}\equiv \bar a_1^{(i)}-\bar a_1^{(i+1)} \quad \text{for }  \ 1\leq i \leq k-1 
\end{alignedat}
\end{equation}
The result is the diagonal:
\begin{equation}
\overline{\mathbb K}
\sim
\vcenter{\hbox{%
\begin{tikzpicture}[
    x=1.1cm,
    y=1.1cm,
    baseline=(current bounding box.center)
]

% Background region bounded by the parentheses
\begin{scope}
    \clip
        (-1.48,1.55)
        -- (1.48,1.55)
        .. controls (1.72,1.05) and (1.72,-1.05)
        .. (1.48,-1.55)
        -- (-1.48,-1.55)
        .. controls (-1.72,-1.05) and (-1.72,1.05)
        .. (-1.48,1.55)
        -- cycle;

    % Gray first block row and first block column
    \fill[blue!15!white]
        (-1.75,-1.55) rectangle (1.75,1.55);

    % Red lower-right 2-by-2 block
    \fill[red!18]
        (-0.50,-1.55) rectangle (1.75,0.50);

    % Dashed vertical block partitions
    \draw[
        black!70,
        line width=0.6pt,
        dash pattern=on 3pt off 2pt
    ]
        (-0.50,-1.60) -- (-0.50,1.60);

    \draw[
        black!70,
        line width=0.6pt,
        dash pattern=on 3pt off 2pt
    ]
        (0.50,-1.60) -- (0.50,1.60);

    % Dashed horizontal block partitions
    \draw[
        black!70,
        line width=0.6pt,
        dash pattern=on 3pt off 2pt
    ]
        (-1.75,0.50) -- (1.75,0.50);

    \draw[
        black!70,
        line width=0.6pt,
        dash pattern=on 3pt off 2pt
    ]
        (-1.75,-0.50) -- (1.75,-0.50);
\end{scope}

% Left parenthesis
\draw[line width=0.75pt]
    (-1.48,1.55)
    .. controls (-1.72,1.05) and (-1.72,-1.05)
    .. (-1.48,-1.55);

% Right parenthesis
\draw[line width=0.75pt]
    (1.48,1.55)
    .. controls (1.72,1.05) and (1.72,-1.05)
    .. (1.48,-1.55);

% First block row
\node at (-1,1) {$-k$};
\node at ( 0,1) {$\mathbf{0}_k^{\mathsf T}$};
\node at ( 1,1) {$\mathbf{0}_k^{\mathsf T}$};

% First block column
\node at (-1, 0) {$\mathbf{0}_k$};
\node at (-1,-1) {$\mathbf{0}_k$};

% Lower-right 2-by-2 block
\node at (0,0) {$\mathbf{1}_k$};

\node at (1,0)
    {$\mathbf{0}_{k\times k}$};

\node at (0,-1)
    {$\mathbf{0}_{k\times k}$};

\node at (1,-1)
    {$-\mathbf{1}_k$};

\end{tikzpicture}%
}}.
\end{equation}
so we can write the full 3d partition function:
\begin{equation}\label{factored}
    \mathcal S_k =\exp\Big[-ikQ(A_1)\Big]\times\int \prod_i^k \mathcal D b_1^{(i)}\mathcal D \bar b_1^{(i)} \ \exp\Bigg(i\sum_i^kQ(b_1^{(i)})-Q(\bar b_1^{(i)})\Bigg)
\end{equation}
The dynamical fields thus form $U(1)_1^{k}\times U(1)_{-1}^{k}$ Chern-Simons theory. Notice in particular that $|\det K| =1$ so the dynamical K-matrix is unimodular, and therefore the dynamical bulk 3d Chern-Simons theory is invertible. \\

As in the case of the minimal defect, we can therefore give topological boundary conditions to the bulk dynamical gauge fields, and safely integrate out the bulk invertible theory. We are left with only a 2d boundary TQFT and the bulk Hall response SPT phase:
\begin{equation}
    \mathcal S_k = \exp\Big[-ikQ(A_1)\Big]\otimes \mathcal W_k[\Sigma_2,A_1]
\end{equation}
We then stack this with the gauge-invariant disk operator $\exp\Big(ik\int_{\Sigma_2}B_2+ikQ(A_1)\Big|_{M_3}\Big)$ which leaves us with the genuine 2d topological defect:
\begin{equation}
    \mathcal D_k \equiv e^{ik\int_{\Sigma_2} B_2} \otimes \mathcal W_k[\Sigma_2,A_1]
\end{equation}
Since the overall theory is still fermionic and the SPT phase $\text{Arf}[\Sigma_2]\in \Omega^2_{spin}$ has order 2 then there is still an overall possibility of stacking the boundary theory with this invertible phase. Once again we can enforce diagonal boundary conditions relating the fields $b_1^{(i)}$ and $\bar{b}_1^{(i)}$ along the boundary $\Sigma_6$. Locally the boundary conditions by adding Lagrange multipliers:
\begin{equation}
    \frac{1}{2\pi}\sum_i^k\varphi_i(db_1^{(i)}-d\bar b_1^{(i)})
\end{equation}
If we establish these boundary conditions first and then apply the quadratic refinement law to yield \hyperref[factored]{(4.43)} then this manifests the additional boundary terms:
\begin{equation}
    \frac{1}{2\pi}\sum_i^k\Big[\varphi_i(db_1^{(i)}-d\bar b_1^{(i)})+\frac 12b_1^{(i)}\wedge A_1\Big]
\end{equation}
Each term of the sum takes the same form as for the single defect \hyperref[bndyterm]{(4.24)}, so the same characterization of the boundary theory holds for $\mathcal W_k[\Sigma_2,A_1]$. This includes a $k$-dimensional cubic lattice of local vertex operators charged under the gauge symmetry of the physical gauge field $A_1$, which the bulk physical Wilson lines may end on.  We will remark on the physical interpretation of this when we compare to the symmetries and charges of the 6d $\mathcal N=(2,0)$ SCFTs in \hyperref[sec5]{Section 5}. \\ 

\subsubsection{Bosonic 2-Group Charges }\label{sec414}

Our analysis has prioritized the construction of charge defects which correspond to the pseudo-2-group $\frak H$, since these are the charges we will eventually generalize to 11d in order to model the charges of M2 and M5 branes. We noted that this also provides a construction of the charges of the ordinary 2-group $\mathbb H^{(2)}$, since we may simply take the squared defect $\mathcal S_1^2$ which has the required level-2 Hall response. However one drawback of this construction is that the definition of $\mathcal S_1$ requires the specification of a Wu-structure, while there is no hint of such a requirement in the pure 2-group $\mathbb H^{(2)}$ data or its gauge fields. It would seem that there should be some simple construction which avoids this requirement. \\

Indeed there does exist such a construction which mirrors that of the half-space gauging defect $\mathcal S_1$. Once again we consider a 3d half-space $M_3$ bounding our 2d surface of interest $\Sigma_2 = \partial M_3$, and introduce worldvolume $U(1)$ gauge fields $a_1, b_1$. The bulk partition function is taken to be:
\begin{equation}
    \begin{gathered}
        \mathcal T \equiv \int \mathcal Da_1 \mathcal D b_1 \ \exp\Bigg(i(a_1,b_1)+i(a_1+b_1,A_1)\Bigg) \\
        \cong\int \mathcal Da_1\mathcal D b_1\exp\Bigg(\frac{i}{2\pi}
\int a_1db_1+a_1dA_1+b_1dA_1\Bigg)        
    \end{gathered}
\end{equation}
One can see that this has the required Hall response $\exp(-i(A_1,A_1))\cong \exp(-\frac{i}{2\pi}\int A_1dA_1)$ for the 2-group defect $\mathbb H^{(2)}$ if we perform the change of variables $\tilde a_1=a_1+A_1$ and $\tilde b_1=b_1+A_1$ then this becomes:
\begin{equation}
    \int \mathcal Da_1 \mathcal D b_1 \ \exp\Bigg(i(a'_1,b'_1)-i(A_1,A_1)\Bigg)
\end{equation}
This is just a trivial BF theory, which admits manifestly topological Dirichlet boundary conditions for either $a_1'$ or $b_1'$. One can then carry out the same construction by stacking this with the 2-group disk operator:
\begin{equation}
    \exp\Big(i\int_{\Sigma_2}B_2+\frac i{2\pi}\int_{M_3}A_1dA_1\Big)
\end{equation}
to cancel the bulk Hall responses and obtain a genuine 2d defect. \\

Since our ultimate interest is to construct the charges for the pseudo-6-group $\frak G$ by generalizing the charges of the pseudo-2-group $\frak H$ then we will not elaborate on this construction. However we note that the steps we have carried out in analyzing the fusion rules could be applied to these 2-group charge defects as well.

%{\color{red} I think the move should instead show the defect lifting off of the worldsheet and remaining linked}

\subsection{6-Group Charges in 11d}\label{sec42}

We have successfully constructed non-invertible charges for our 2-group $\mathbb H^{(2)}$ (and correspondingly the defects of $\frak H$) in 5d. We may now straightforwardly extend these results to our main interest: the higher-form 6-group $\mathbb G^{(6)}$ (and correspondingly $\frak G$). The overall construction will be (almost) precisely the same, with minor variations and subtleties that we will point out. Nevertheless we state the same results explicitly for the 6-group case, since these are the defects we will be comparing to M2 and M5 branes in the next section. \\

We point out that for the genuine 6-group $\mathbb G^{(6)}$ there exists an 11d SymTFT which organizes the defects:
\begin{equation}
    \begin{gathered}
        S = \frac 1{2\pi}\int_{M_{11}} d^{11}x \Bigg[ C_3\wedge  df_7+B_6\wedge dh_4+ \frac{1}{2\pi}h_4\wedge C_3dC_3\Bigg] \\
        =\frac{1}{2\pi}\int_{M_{11}}d^{11}x\Bigg[C_3\wedge df_7+h_4\wedge H_7\Bigg]
    \end{gathered}
\end{equation}
Once again, due to the global subtlety in defining $Q(\check C_3)$ then there is no obvious candidate for a TFT capturing the defects of $\frak G$, but we will stumble upon a possible construction in \hyperref[sec53]{Section 5.3} when we study 11d supergravity directly. As before, we will not work with this SymTFT, but instead focus directly on the defects. \\ 

In particular, we begin with the $\frak G$ defects. The simplest is the Wilson line:
\begin{equation}
    W_n[\gamma_3]=\exp\Big(in\int_{\gamma_3}C_3\Big)
\end{equation}
while the 6d defects:
\begin{equation}
    \begin{gathered}
        V_p[\Sigma_6]= \exp\Big(ip\int_{\Sigma_6}B_6\Big) \xrightarrow{\delta_{\lambda_2}}\exp\Big(ip\int_{\Sigma_6} B_6-\delta_{\lambda_2} Q(\check C_3)\Big) \\ 
    \qquad \qquad \qquad \qquad \qquad \qquad \  \ \ \ \  \  \sim \exp\Big(ip\int_{\Sigma_6}\Big[ B_6-\frac{1}{4\pi}\lambda_2(dC_3+\nu)\Big]\Big)
    \end{gathered}
\end{equation}
are neither gauge invariant nor topological. We will fix this by the same method as the 2-group: we extend to a bounding 7-manifold $M_7$ with boundary $\partial M_7=\Sigma_6$. In the bulk we introduce dynamical 3-form gauge fields $a_3,\bar a_3$ and define the bulk 7d Wu-Chern-Simons theory with action:
\begin{equation}
    \begin{gathered}
        \mathcal S_1  \equiv\int \mathcal Da_3\mathcal D\bar a_3 \ \exp\Big(i\Big[ (C_3,a_3)+Q(a_3)-Q(\bar a_3)\Big]\Big)\\
         \ \ \ \quad \quad \  \quad  \sim \int \mathcal Da_3 \mathcal D\bar a_3 \ \exp\Bigg(\frac{i}{2\pi}\int_{M_7}\Big[ C_3da_3+\frac12 a_3da_3-\frac 12\bar a_3d\bar a_3 \Big]\Bigg)
    \end{gathered}
\end{equation}
We isolate the Hall response by performing the change of variables $\tilde a_3=a_3+C_3$: 
\begin{equation}
    \mathcal S_1\cong \exp\Big[-iQ(C_3)\Big]\times \int\mathcal Da_3\mathcal D\bar a_3 \ \exp\Big(iQ(\tilde a_3)-iQ(\bar a_3)\Big)
\end{equation}
Choosing topological boundary conditions and integrating out the bulk invertible phase leaves a TQFT $\mathcal W_1[\Sigma_6,C_3]$ on the boundary. Stacking this with the gauge-invariant disk operator yields our non-invertible defect:
\begin{equation}
    \mathcal D_1\equiv \mathcal S_1\otimes \exp\Bigg(i\int_{\Sigma_6}B_6+Q(C_3)\Big|_{M_7}\Bigg)=e^{i\int_{\Sigma_6} B_6}\otimes \mathcal W_1[\Sigma_6,C_3]
\end{equation}
In order to define $\mathcal W_1[\Sigma_6]$ we must specify topological boundary conditions. Now the 7d TQFT is $U(1)_1\times U(1)_{-1}$ Wu-Chern-Simons theory. We work under the assumption that the classification of topological boundary conditions follows the same structure as the 3d case developed in \cite{Kapustin:2010hk}. We again make the same choices of isotropic sublattices:
\begin{equation}\label{tbc}
    \Lambda_0=\Lambda_{\pm}\equiv \{ (n,\pm n)  \ : \ n\in \mathbb Z\}  
\end{equation}
However now the relation between the 6d boundaries defined by these two choices is less well-understood. More broadly, while in the 2d/3d case we could stack the boundary with an SPT phase classified by the Anderson dual of spin cobordism $\Omega^2_{spin}$, we might expect a generalization in the 6d case. A natural candidate is the Arf-Kervaire invariant \cite{Kerv, browder, brown} considered in this context in \cite{Monnier:2016jlo, Gukov:2020btk}. By analogy with the 2d case, the dependence of the theory on a Wu structure suggests that the allowed SPT phases should belong to the Anderson dual of Wu cobordism $\Omega^{6}_{Wu}$. We do not include a calculation of this group or a detailed analysis of the associated phases in this work, but we hope to report on it in the near future. \\

We can enforce these boundary conditions with a Lagrange multiplier 2-form $\beta_2$ interaction on the boundary:
\begin{equation}\label{6dtbc}
    \frac{1}{2\pi}\int_{\Sigma_6}\beta_2\wedge (da_3-d\bar a_3+dC_3)
\end{equation}
The interpretation of the topological boundary theory $\mathcal W_1[\Sigma_6,C_3]$ follows the same analysis as the 2d theory in the previous section. In particular, we interpret the boundary as a gapped symmetry-breaking Higgsed phase which is charged under the gauge symmetry of the bulk physical $C_3$ field. It includes a coupling:
\begin{equation}\label{betacoupling}
    -\frac{1}{4\pi}\int_{\Sigma_6}d\alpha_2\wedge C_3 \quad \subset \quad \mathcal W_1[\Sigma_6,C_3]
\end{equation}
so the charged operators on $\Sigma_6$ are now ``vertex" surface operators $e^{i\int_{\Gamma_2}\alpha_2}$. Moreover, when we integrate out the bulk invertible theory, leaving only the above coupling, then since $\alpha_2$ is charged under the gauge symmetry of the physical gauge field $C_3$ then the bulk physical Wilson lines $\exp(i\int C_3)$ may end on these vertex operators. \\

We may straightforwardly generalize the (doubled) Witten modular transformations of \hyperref[dmod]{(4.28)} to the case of 7d Wu-Chern-Simons theories:
\begin{equation}
    \begin{gathered}
        \mathbf S: Z[A_3,\bar A_3]\to Z_{\mathbf S}[a_3,\bar a_3]=\int \mathcal DA_3\mathcal D\bar A_3 \ Z[A_3,\bar A_3] \ \exp\Big[i(a_3,A_3)-i(\bar a_3,\bar A_3)\Big] \\ \mathbf T:Z[A_3,\bar A_3]\to Z_{\mathbf T}[A_3,\bar A_3]= Z[A_3,\bar A_3]\exp\Big[iQ(A_3)-iQ(\bar A_3)\Big] \\ \mathbf C:Z[A_3,\bar A_3]\to Z[-A_3,-\bar A_3]
    \end{gathered}
\end{equation}
From these we can define the above bulk 7d TQFT by the same combination $\mathbf{CTST}$ of modular transformations:
\begin{equation}
    \mathcal S_1=\mathbf{CTST}\circ e^{i(a_3,C_3)}
\end{equation}

\subsubsection*{Higher-Charge Defects}

We can construct non-invertible defects with higher charge by following the same construction as in the 2-group case (up to an interesting subtlety we point out shortly). That is, we may define the defect of charge $k$ using the iterated parallel fusion of the minimal defect:
\begin{equation}
    \mathcal S_k \equiv \mathcal S_1^{\otimes k} =(\mathbf{CTST)^k}
\end{equation}
 This defines a bulk 7d Chern-Simons theory with overall K-matrix (including the coupling to $C_3$): 
\begin{equation}\label{bigK}
\overline{\mathbb K}
=
\vcenter{\hbox{%
\begin{tikzpicture}[
    x=0.8cm,
    y=0.8cm,
    baseline=(current bounding box.center)
]

% Background region bounded by the parentheses
\begin{scope}
    \clip
        (-2.05,2.05)
        -- (2.05,2.05)
        .. controls (2.30,1.40) and (2.30,-1.40)
        .. (2.05,-2.05)
        -- (-2.05,-2.05)
        .. controls (-2.30,-1.40) and (-2.30,1.40)
        .. (-2.05,2.05)
        -- cycle;

    % Gray first block row and first block column
    \fill[blue!15!white]
        (-2.30,-2.05) rectangle (2.30,2.05);

    % Red lower-right block
    \fill[red!18]
        (-1.00,-2.05) rectangle (2.30,1.00);

    % Dashed block partitions
    \draw[
        black!70,
        line width=0.6pt,
        dash pattern=on 3pt off 2pt
    ]
        (-1.00,-2.10) -- (-1.00,2.10);

    \draw[
        black!70,
        line width=0.6pt,
        dash pattern=on 3pt off 2pt
    ]
        (1.00,-2.10) -- (1.00,2.10);

    \draw[
        black!70,
        line width=0.6pt,
        dash pattern=on 3pt off 2pt
    ]
        (-2.30,1.00) -- (2.30,1.00);

    \draw[
        black!70,
        line width=0.6pt,
        dash pattern=on 3pt off 2pt
    ]
        (-2.30,-1.00) -- (2.30,-1.00);
\end{scope}

% Left parenthesis
\draw[line width=0.75pt]
    (-2.05,2.05)
    .. controls (-2.30,1.40) and (-2.30,-1.40)
    .. (-2.05,-2.05);

% Right parenthesis
\draw[line width=0.75pt]
    (2.05,2.05)
    .. controls (2.30,1.40) and (2.30,-1.40)
    .. (2.05,-2.05);

% First block row
\node at (-1.50, 1.50) {$0$};
\node at (-0.50, 1.50) {$1$};
\node at ( 0.50, 1.50)
    {$\mathbf{0}_{k-1}^{\mathsf T}$};
\node at ( 1.50, 1.50)
    {$\mathbf{0}_{k}^{\mathsf T}$};

% First block column
\node at (-1.50, 0.50) {$1$};
\node at (-1.50,-0.50)
    {$\mathbf{0}_{k-1}$};
\node at (-1.50,-1.50)
    {$\mathbf{0}_{k}$};

% Central k-by-k block
\node at (0,0)
    {\scalebox{1.1}{\ensuremath{K}}};

% Upper-right k-by-k zero block
\node at (1.50,0)
    {\scalebox{1}{
        \ \  \ensuremath{\mathbf{0}_{k\times k}}
    }};

% Lower-middle k-by-k zero block
\node at (0,-1.50)
    {\scalebox{1}{
        \ensuremath{\mathbf{0}_{k\times k}}
    }};

% Bottom-right k-by-k block
\node at (1.50,-1.50)
    {\scalebox{0.8}{\ensuremath{-K}}};

\end{tikzpicture}%
}}
\qquad \text{where} \qquad
K=
\left(
\begin{array}{c|ccc}
    1      & -1 & 0 & \dots \\ \hline
    -1     &     &   &       \\
    0      &     &
    \smash{\mathclap{
        \scalebox{1.6}{$\mathbb A_{k-1}$}
    }}
    &       \\
    \vdots &     &   &
\end{array}
\right).
\end{equation}
In order to isolate the Hall response, we would like to diagonalize this matrix by performing the change of variables \hyperref[cov]{(4.41)}. However there is a subtlety in this procedure: in order for the result to yield a diagonal K-matrix we must have that $Q(-A_3)=Q(A_3)$. If we consider the definitions of these quadratic refinements given in \hyperref[appa]{Appendix A} then we see that this is automatically true for the 3d spin-Chern-Simons theory $Q(-A_1)=Q(A_1)$, but it is not necessarily true for the 7d Wu-Chern-Simons theory. However notice that this property \textit{does} hold if we require the 3-form fields to have the shifted quantization $dC_3+\frac 14 p_1(TM)$. This is precisely the shifted quantization of M-theory, which we required of defects in $\frak G$! Thus the shifted quantization required of the M-theory $C$ field is precisely compatible with achieving the correct Hall response in the non-invertible fusion. \\

Assuming this shifted quantization, we may then use the change of variables \hyperref[cov]{(4.41)}:
\begin{equation}\label{6cov}
    \begin{gathered}
        C_3=C_3  \\
        b_3^{(i)}\equiv a_3^{(i)}-a_3^{(i+1)}+C_3 \quad \text{for }  \ 1\leq i \leq k-1   \\
        b_3^{(k)} \equiv a_3^{(k)}+C_3 \\
        \bar b_3^{(i)}\equiv \bar a_3^{(i)}-\bar a_3^{(i+1)} \quad \text{for }  \ 1\leq i \leq k-1   \\
        \bar b_3^{(k)} \equiv \bar a_3^{(k)} \\
    \end{gathered}
\end{equation}
and express the total 7d partition function as:\
\begin{equation}
    \mathcal S_k \cong \exp\Big[-ikQ(C_3)\Big]\times\int \prod_i^k \mathcal D b_3^{(i)}\mathcal D \bar b_3^{(i)} \ \exp\Bigg(i\sum_i^kQ(b_3^{(i)})-Q(\bar b_3^{(i)})\Bigg)
\end{equation}
We give the topological boundary conditions \hyperref[tbc]{(4.57)} to each pair $b_3^{(i)}, \bar b_3^{(i)}$ and integrate out the bulk invertible phase. We then stack $\mathcal S_k$ with the charge-$k$ disk operator to obtain our non-invertible charge defects:
\begin{equation}
     \mathcal D_k\equiv \mathcal S_k\otimes \exp\Bigg(ik\int_{\Sigma_6}B_6+kQ(C_3)\Big|_{M_7}\Bigg)=e^{ik\int_{\Sigma_6} B_6}\otimes \mathcal W_k[\Sigma_6,C_3]
\end{equation}

Once again the boundary can be characterized by the inclusion of couplings:
\begin{equation}
    \exp\Big(-\frac{i}{4\pi}\int_{\Sigma_6}d\alpha_2^{(i)}\wedge C_3\Big) \quad \subset \quad \mathcal W_k[\Sigma_6,C_3]
\end{equation}
Thus bulk Wilson surfaces $W_1[\gamma_3]=e^{i\int C_3}$ may end on the vertex surface operators $e^{i\int_{\Gamma_2}\alpha_2^{(i)}}$. This defines the lattice of string charges which in the next section we will interpret as the junctions of M2-branes on a stack of M5-branes. 

\section{Comparison to M-Theory} \label{sec5}

We have successfully constructed 3d and 6d topological defects which capture the desired features of the $\frak G$ charges:
\begin{equation}
    \begin{gathered}
        W_n(\gamma_3)=\exp\Big(in\int_{\gamma_3}C_3\Big) \\ \mathcal D_k[\Sigma_6]=e^{ik\int_{\Sigma_6}B_6}\otimes \mathcal W_k[\Sigma_6,C_3] \
    \end{gathered}
\end{equation}
This brings us to our main objective, which is to compare these defects with M2 and M5 branes of M-theory:
\begin{equation}
    \begin{gathered}
    Z_{M2}[\gamma_3]\sim W_1[\gamma_3]\\
    \mathcal Z_{M5}[\Sigma_6]\sim \mathcal D_1[\Sigma_6]
    \end{gathered}
\end{equation}
What features of these branes can we hope to capture? We have argued that the properties \hyperref[Mcharges]{(2.32)} represent the relevant coupling of the 3-form gauge field $C_3$ to the M2 and M5 branes. The hope is that these topological defects and their fusions act as a model for the charges and fusions of M2 and M5 branes.  We therefore study the possible fusion rules for these topological defects and compare them to those features of M2 and M5 branes which derive from the gauge coupling. Such features include fusion and higher-codimension junctions between branes, and operators living on the worldvolume whose charges derive from the $C_3$ gauge symmetry. This comparison will be the content of the next subsection. \\

We would also like to better understand the relationship between the higher-group-like algebra $\frak G$ and other aspects of M-theory. We will discuss further hints of this higher-algebraic structure in the following subsections. \\

\subsection{Fusion Rules}

We begin by comparing the various fusion rules of our topological defects to the corresponding configurations of M2 and M5 branes.

\subsubsection{Stacks of M5 Branes}

The first configuration we would like to analyze is a stack of $N$ M5-branes, which become coincident on a single worldvolume $\Sigma_6$. Let us begin by recalling some known features of this system.\footnote{See \cite{Heckman:2018jxk, Moore_FK} for a review.} In particular, we will be interested in the local QFT living on the worldvolume which is isolated in the decoupling limit. This theory can be studied concretely, for example in AdS/CFT, where the theory becomes dual to M-theory on $AdS_7\times S^4$ with $N$ units of magnetic $G_4$ flux threaded through $S^4$ \cite{Maldacena:1997re}. The worldvolume theory is the famous maximal 6d $\mathcal N=(2,0)$ superconformal field theory of type $\frak a_{N-1}$, along with an overall 6d superconformal tensor multiplet corresponding to the center of mass of the brane stack, typically called the singleton. In general there is an ADE classification of 6d $\mathcal N=(2,0)$ SCFTs by the simply-laced Lie algebras from the type IIB construction, however the construction of $N$ coincident nonsingular M5 branes realizes only the $\frak a_{N-1
}$ type. We will comment on the $\frak d$-type later in \hyperref[sec6]{Section 6}. \\

This 6d $\mathcal N=(2,0)$ SCFT is a highly intriguing and mysterious theory, and still evades direct description. However many features of it have been deduced from various different approaches, including top-down studies from M-theory \cite{Witten:1996hc, Witten:1995em, Strominger:1995ac} and type II string theory \cite{Witten:1995zh}, as well as bottom-up approaches such as the conformal bootstrap \cite{Beem:2015aoa}, or uplifting from lower-dimensional theories \cite{Aharony:1997an, Arkani-Hamed:2001wsh, Douglas:2010iu, Papageorgakis:2014foa, Cordova:2015vwa}. \\

An illuminating approach is to first consider the case of $N$ separated M5-branes, and then take the strong-coupling limit where they all become coincident. The worldvolume theory of a single M5-brane is described by a single 6d (2,0) tensor supermultiplet, consisting of a 2-form gauge field $b_2$ with self dual field strength $db_2=H_3=*_g H_3$, five scalars $\Phi_i$ parametrizing the spacetime embedding in the 5 transverse directions of 11d, and the 8 fermionic superpartners. From the perspective of the 6d SCFT, separating the \(N\) M5-branes corresponds to moving onto the tensor branch of its moduli space of supersymmetric vacua.. The tensor branch for a 6d (2,0) theory of type $\frak g$ is described as $(\mathbb R^5)^{r(\frak g)}/W(\frak g)$ where $r(\frak g)$ is the rank of $\frak g$ and $W(\frak g)$ is the Weyl group. The resulting theory at a generic point of the moduli space is then described by $N$ decoupled tensor multiplets, and the moduli space can be parametrized by the vevs of the scalars $\Phi_i$ determining the relative separation of the branes.  \\ 

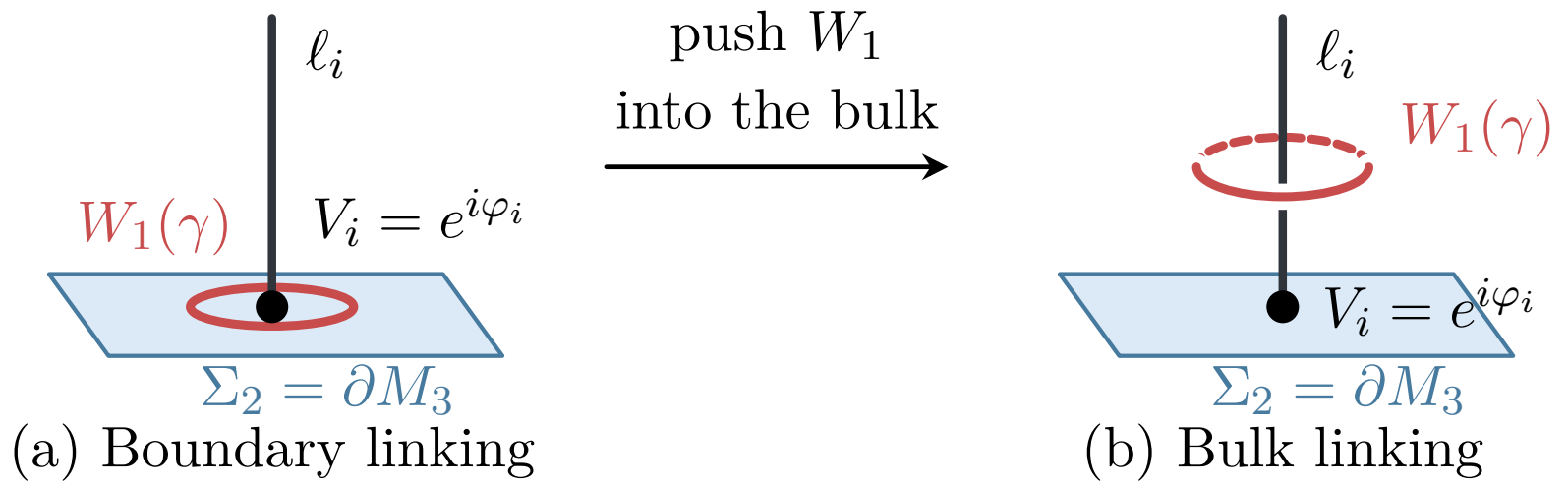
\begin{figure}[t]
    \centering

    \begin{tikzpicture}[
        font=\sffamily,
        line cap=round,
        line join=round
    ]

    % =========================================================
    % Left panel: boundary linking
    % =========================================================

    \begin{scope}[xshift=-3.4cm]

        % Boundary surface Sigma_2
        \draw[boundary plane]
            (-1.50,0)
            -- (1.15,0)
            -- (1.55,-0.55)
            -- (-1.10,-0.55)
            -- cycle;

        \node[
            text=BoundaryEdge,
            font=\small,
            anchor=north east
        ] at (1.38,-0.47)
            {$\Sigma_2=\partial M_3$};

        % Wilson loop lying on the boundary and surrounding
        % the endpoint of the bulk Wilson line
        \draw[probe Wilson line]
            (0,-0.22)
            ellipse[
                x radius=0.55,
                y radius=0.13
            ];

        \node[
            text=ProbeRed,
            font=\small,
            anchor=east
        ] at (-0.12,0.32)
            {$W_1(\gamma)$};

        % Drawn after the loop so the black line covers it
        \draw[bulk Wilson line]
            (0,-0.22) -- (0,1.72);

        \node[
            font=\small,
            anchor=west
        ] at (0.08,1.48)
            {$\ell_i$};

        % Boundary vertex operator
        \node[vertex operator] at (0,-0.22) {};

        \node[
            font=\small,
            anchor=west
        ] at (0.12,0.38)
            {$V_i=e^{i\varphi_i}$};

        \node[
            font=\small,
            anchor=north
        ] at (0,-0.86)
            {(a) Boundary linking};

    \end{scope}

    % =========================================================
    % Deformation into the bulk
    % =========================================================

    \draw[braid arrow]
        (-1.15,0.72) -- (1.15,0.72)
        node[
            midway,
            above=3pt,
            align=center,
            font=\small
        ]
        {push \(W_1\)\\into the bulk};

    \node[
        font=\small
    ] at (0,0.18)
        {};

    % =========================================================
    % Right panel: bulk linking
    % =========================================================

    \begin{scope}[xshift=3.4cm]

        % Boundary surface Sigma_2
        \draw[boundary plane]
            (-1.50,0)
            -- (1.15,0)
            -- (1.55,-0.55)
            -- (-1.10,-0.55)
            -- cycle;

        \node[
            text=BoundaryEdge,
            font=\small,
            anchor=north east
        ] at (1.38,-0.47)
            {$\Sigma_2=\partial M_3$};

        % Back half of the linked probe loop
        \draw[
            probe Wilson line,
            densely dashed
        ]
            (0.58,0.72)
            arc[
                start angle=0,
                end angle=180,
                x radius=0.58,
                y radius=0.20
            ];

        % Bulk Wilson line
        \draw[bulk Wilson line]
            (0,-0.22) -- (0,1.72);

        \node[
            font=\small,
            anchor=west
        ] at (0.08,1.48)
            {$\ell_i$};

        % Front half of the linked probe loop
        \draw[
            probe Wilson line,
            preaction={
                draw=white,
                line width=5pt
            }
        ]
            (-0.58,0.72)
            arc[
                start angle=180,
                end angle=360,
                x radius=0.58,
                y radius=0.20
            ];

        \node[
            text=ProbeRed,
            font=\small,
            anchor=west
        ] at (0.64,0.98)
            {$W_1(\gamma)$};

        % Boundary vertex operator
        \node[vertex operator] at (0,-0.22) {};

        \node[
            font=\small,
            anchor=west
        ] at (0.12,-0.22)
            {$V_i=e^{i\varphi_i}$};

        \node[
            font=\small,
            anchor=north
        ] at (0,-0.86)
            {(b) Bulk linking};

    \end{scope}

    \end{tikzpicture}

    \caption{
        Braiding move detecting the charge of the boundary vertex
        operator. A bulk Wilson line \(\ell_i\) terminates
        on \(\Sigma_2=\partial M_3\) at
        \(V_i=e^{i\varphi_i}\). The configuration remains linked in the 3D bulk.
    }
    \label{fig:boundary-vertex-braiding}

\end{figure}

Importantly, the tensor branch also admits a class of $\frac 12$-BPS string-like solitonic defects. From the bulk perspective, these correspond to the 2d junctions of an M2-brane ending on an M5-brane. From the perspective of a single M5-brane, these strings are both electrically and magnetically charged under $H_3$. More generally, on the tensor branch these M2-branes may stretch between adjacent M5-branes. As a result, the full spectrum of string states carry charges in the root lattice of $\frak a_{N-1}$.  The tension of these string solutions is proportional to the separation distance of the branes. Therefore when we take the strongly-coupled limit of coincident M5-branes their tension effectively goes to zero. This perspective of the local 6d SCFT fixed point as the tensionless limit of string states is one of the many engaging mysteries of this theory. \\ 

A particularly accessible approach to the 6d $\mathcal N=(2,0)$ is the study of its symmetries and anomalies \cite{Witten:1996hc, Freed:1998tg, Harvey:1998bx, Intriligator:2000eq, Ganor:1998ve, Ohmori:2014kda, Monnier:2017klz, Monnier:2014txa, Monnier:2013rpa, Sheckler:2025lfv}. In particular, on the tensor branch the string-like solitons are charged under the 2-form symmetry associated with the $b_2^{(i)}$ gauge fields of the tensor multiplets \cite{Gaiotto:2014kfa}. Naively, the $N$ tensor multiplets indicate that the overall 2-form symmetry should be $[U(1)^N]^{(2)}$. However the string charges in the $\frak a_{N-1}$ root lattice break this down to an overall center $\mathbb Z_N^{(2)}$ symmetry. For theories of general ADE type \(\mathfrak g\), the corresponding discrete symmetry can be determined from the defect group using the type IIB construction\cite{DelZotto:2015isa}. \\

Holography provides a relevant perspective on this symmetry. The topological term in the 7d bulk action is the level $N$ Chern-Simons term:
\begin{equation}
    \frac{N}{4\pi}\int_{M_7}C_3dC_3
\end{equation}
This is a non-invertible 7d TQFT, so the boundary 6d CFT is a relative theory, giving rise to a nontrivial Hilbert space formed from the conformal blocks on the boundary, whose dimension is $|H^3(\Sigma_6,\mathbb Z_N)|^{1/2}$. The 6d (2,0) SCFT associated to a general flux sector is therefore described by a partition vector, while the specification of a genuine absolute theory with a partition function requires a choice of  polarization \cite{Tachikawa:2013hya, Gukov:2020btk, Gukov:2025dol}. The self-dual strings are realized as the 2d ends of the 3d anyons $e^{i\oint C_3}$ ending on the boundary, which derive from M2-branes in the asymptotic limit. Their charge under the boundary 2-form symmetry can be measured by fusing a 3d anyon with the boundary and linking with the string. The $\mathbb Z_N$ charge can then be deduced from the bulk braiding:
\begin{equation}
    \braket{e^{i\int_{\Gamma_3} C_3 }e^{i\int_{\Gamma'_3}C_3}}=\exp\Big(\frac{2\pi i}N\text{Link}(\Gamma_3,\Gamma'_3)\Big)
\end{equation}

\subsubsection*{M2 Junctions}

A key relationship between M2 and M5 branes which we would like to model with our topological charge defects is the fact that an M2 brane can end on an M5 brane. This requires a 2d string operator on the non invertible defect which can act as a junction. In categorical language this junction is a higher morphism. \\

In fact we have already seen that the 6d non-invertible defects have this property. Recall that our characterization of the topological theory $\mathcal W_1[\Sigma_6,C_3]$ was as a gapped Higgs phase with string vertex operators $e^{i\int_{\gamma_2}\alpha_2}$ charged under the gauge symmetry of the bulk physical gauge field $C_3$, with the schematic coupling
\hyperref[betacoupling]{(4.59)}:
\begin{equation}
    -\frac{1}{4\pi}\int d\alpha_2\wedge C_3
\end{equation}
Thus a bulk Wilson line can end on the 6d non-invertible defects with the vertex string operator $e^{i\int_{\Gamma_2}\alpha_2}$ supplying the junction. We see that this vertex operator naturally plays the role of the self-dual string of an M5-brane. 

\subsubsection*{Parallel Fusion and Singleton Decomposition}

We would like to study the charge carried by a stack of $N$ M5-branes using our model for the single M5-brane charge as the non-invertible defect $\mathcal D_{1}[\Sigma_6]$. The natural candidate for this stack is the parallel fusion of $N$ minimal defects:
\begin{equation}
    \mathcal D_N[\Sigma_6]=\bigotimes_{i=1}^N \mathcal D_1[\Sigma_6]=e^{iN\int_{\Sigma_6}B_6}\otimes \mathcal W_N[\Sigma_6,C_3]
\end{equation}
We constructed this defect using the bulk 7d TQFT $\mathcal S_1^{\otimes N} $. It was found that there exists a cubic lattice of vertex string operators $\exp({i\int\alpha^{(i)}_2)}$ which carry the same charges as the redefined gauge fields $b_3^{(i)}$, which themselves were defined by \hyperref[6cov]{(4.64)}:
\begin{equation}
    b^{(i)}_3=-U^T_{ij}a^{(j)}_3+C_3
\end{equation}
Recall in particular that $U$ is the $SL(N,\mathbb Z)$ matrix which diagonalizes the 7d bulk $U(1)^N$ K-matrix \hyperref[diagU]{(4.38)}:
\begin{equation}
K=
\left(
\begin{array}{c|cccccc}
1      & -1     & 0      & 0      & 0      & \cdots & 0      \\ \hline
-1     & 2      & -1     & 0      & 0      & \cdots & 0      \\
0      & -1     & 2      & -1     & 0      & \cdots & 0      \\
0      & 0      & -1     & 2      & -1     & \cdots & 0      \\
0      & 0      & 0      & -1     & 2      & \ddots & \vdots \\
\vdots & \vdots & \vdots & \vdots & \ddots & \ddots & -1     \\
0      & 0      & 0      & 0      & \cdots & -1     & 2
\end{array}
\right)
=
\left(
\begin{array}{c|ccc}
    1      & -1 & 0 & \dots \\ \hline
    -1     &     &   &       \\
    0      &     &
    \smash{\mathclap{
        \scalebox{1.6}{  $\mathbb A_{N-1}$}
    }}
    &       \\
    \vdots &     &   &
\end{array}
\right).
\end{equation}
to the purely identity K-matrix:

\begin{equation}
    K=U \mathbf{1}_N U^T \quad \text{where} \quad U=\begin{pmatrix} -e_1 \\ \alpha_1 \\ \alpha_2 \\ \vdots \\ \alpha_{N-1} \end{pmatrix}
\end{equation}
 where $\vec \alpha_i$ are the generators of the $\frak a_{N-1}$ root lattice. If we consider the K-matrix of an abelian Chern-Simons theory as the Gram matrix for lattice $\Lambda$, then we see that this Chern-Simons theory with K-matrix $\mathbf 1_N$ is defined by the canonical cubic $N\times N$ lattice $\mathbb Z^N$. $U$ provides a decomposition of the cubic lattice into the sublattice $A_{N-1}$ defining the root lattice of $\frak a_{N-1}$ and its complement. We obtain $N-1$ charges taking value in the root lattice $\frak a_{N-1}$ and one leftover charge in the complementary direction $e_1$. \\

As we showed, the bulk Wilson lines of the 7d Chern-Simons theory may end on string vertex operators of the boundary theory $\mathcal W_N[\Sigma_6,C_3]$. We interpret the generic string vertex operators:
 \begin{equation}
     \exp\Bigg(i\int \sum_i c_i\alpha_2^{(i)}\Bigg) \quad \text{for} \quad \vec c\in \mathbb Z^N
 \end{equation}
as modeling the charge of a 2d self-dual strings on the worldvolume of a stack of $N$ M5-branes. In particular, each individual string $\exp(i\int \alpha_2^{(i)})$ carries unit charge under the gauge symmetry of the bulk $C_3$ gauge field. Therefore a bulk Wilson line $e^{i\int C_3}$ representing an M2-brane may end on such a charge. However there is a distinguished \textit{sublattice} of charges which are gauge neutral, so that $\sum_ic_i=0$. This is precisely the sublattice $A_{N-1}$ defining the root lattice of $\frak a_{N-1}$. In the limit in which the stack of M5-branes become coincident, one might consider these as describing the strings resulting from arbitrary configurations of M2-branes stretching between adjacent M5-branes. From the perspective of the stack worldvolume theory these are \textit{genuine} string operators, which do not require an M2-brane coming from the bulk to end on them.  \\

The orthogonal complement of the sublattice $A_{N-1}$ is the completely diagonal vector $v = (1,1,1,\dots,1)$ which has length $|v|^2=N$. This manifestly represents the charge of the singleton degree of freedom, which parametrizes the center of mass mode of the M5-brane stack, and couples to each brane equally. In total the orthogonal combination: 
\begin{equation}
     A_{N-1}\oplus \mathbb Zv \subset \mathbb Z^N
\end{equation}
defines a degree $N$ sublattice of the total cubic lattice $\mathbb Z^N$.\footnote{This decomposition was used to study the anomaly field theory of the 6d $\mathcal N=(2,0)$ theory in \cite{Monnier:2017klz, Monnier:2014txa}, however in a different setting. The cubic lattice was associated only to the case of \textit{separated} M5-branes.} This reproduces the defect group \cite{Gaiotto:2014kfa, DelZotto:2015isa, Gukov:2020btk, Gukov:2025dol} which is equivalent to the $\mathbb Z_N^{(2)}$ 2-form symmetry of the worldvolume theory:
\begin{equation}
    \mathbb Z^N/(A_{N-1}\oplus \mathbb Zv)\cong \mathbb Z_N
\end{equation}
In holography, the singleton mode \cite{Aharony:1998qu, Witten:1998wy} is typically removed by a bulk gauge transformation in order to isolate the 6d (2,0) theory \cite{Aharony:1999ti, Maldacena:2001ss}. However the $A_{N-1} $ sub-block alone does not define an invertible 7d Chern-Simons theory, but instead has $|\det {\mathbb A_{N-1}}|=N$. If we give the 7d CS theory the usual gapless chiral boundary conditions corresponding to a chiral 2-form CFT on the boundary, then the this prepares a Hilbert space of states on the conformal boundary spanned by multiple conformal blocks. This is the familiar statement that the 6d (2,0) theory is a relative theory \cite{witten2008conformalfieldtheorydimensions, Freed:2012bs}. Projecting out the singleton destroys invertibility of the theory. \\

Nevertheless, these conformal blocks of the isolated $A_{N-1}$ can be consistently combined with those of the singleton $U(1)_N$ theory in order to recover the original invertible theory. Indeed, the original partition function for the $\mathbb Z^N$ lattice is invertible, and so in the end there should only be one conformal character on the boundary. The decomposition of the cubic lattice allows us to cleanly combine the conformal characters of the $A_{N-1}$ boundary theory along with those of the $U(1)_N$ singleton theory back into this single character. This is level-rank dual to the familiar decomposition:
\begin{equation}
    U(N)_1\cong \frac{SU(N)_1\times U(1)_N}{\mathbb Z_N}
\end{equation}
Let us review this decomposition of the $\mathbb Z^N$ character. For simplicity we review the 2d/3d case following the analysis \cite{Dijkgraaf:2007fe}, but this can be straightforwardly generalized to 6d/7d as was considered in \cite{Henningson:2000gu}. Since the original $U(1)^N$ cubic theory is an invertible Chern-Simons theory, then when we consider it on a manifold with boundary and give the ordinary gapless chiral boundary conditions realizing $N$ chiral scalars, then there is only a single conformal block and thus a single character defining the partition function:
\begin{equation}
    \begin{gathered}
        \chi_{\frak u(1)^N}(z_i,q)=\chi_{\frak u(N)_1}(z_i,q)=\text{Tr}\Big(e^{2\pi i \sum_iz_iJ_i }q^{L_0-\frac{N}{24}}\Big) \\
        =\frac{1}{\eta(q)^N}\sum_{\vec p\in \mathbb Z^N}q^{\frac 12\vec p^2}e^{2\pi i \vec z\cdot \vec p}  \equiv \frac{\Theta_{\mathbb Z^N}(z_i,q)}{\eta(q)^N}
    \end{gathered}
\end{equation}
Now if consider the rank $N$ sublattice $ A_{N-1}\oplus \mathbb Zv \subset \mathbb Z^N$, then we can recover the original $\mathbb Z^N$ lattice by supplying the $N$ missing cosets, which we index by $r\in \mathbb Z_N$. Explicitly, we can express any element $\vec p \in \mathbb Z^N$ of the cubic lattice using a specific value of $r$ as:
\begin{equation}
    \vec p=\begin{pmatrix}n+n_1+r \\ n-n_1+n_2 \\ n-n_2+n_3 \\ \vdots \\ n-n_{N-2}+n_{N-1} \\n-n_{N-1} \end{pmatrix} 
\end{equation}
Then we can re-organize the above character according to this decomposition as:
\begin{equation}
    \begin{gathered}
        \chi_{\frak u(1)^N}(z_i,q)=\frac{1}{\eta(q)^N}\sum_{r=0}^{N-1}\sum_{n }q^{\frac N2(n+r/N)^2}x^{n+r/N}\sum_{\{n_i\}}q^{\sum_i(n_i^2-n_in_{i+1})+n_1r+\frac {r^2}2\frac{N-1}N}\prod_{i=1}^{N-1}x_i^{n_i+\frac{N-i}Nr}
    \end{gathered}
\end{equation}
where we have defined $x\equiv e^{2\pi i \sum_iz_i}   $and $  x_i\equiv e^{2\pi i (z_i-z_{i+1})}$. We can then recognize this decomposition in terms of the characters for $\chi_r^{\frak u(1)_N}$ and $\chi^{\frak{su}(N)_1}_r=\chi^{\frak a_{N-1}}_r$ as:
\begin{equation}
    \chi_{\frak u(1)^N}(z_i,q)=\sum_{r=0}^{N-1}\chi_r^{\frak u(1)_N}(x)\chi_r^{\frak{su}(N)_1}(x_i)
\end{equation}
In each factor, the $r$ is interpreted as the $\frak u(1)$ charge and $\frak{su}(N)_1$ weight respectively. As noted in \cite{Moore:2004jv, Belov:2004ht, Gukov:2004id}, this decomposition takes precisely the same form as the decomposition of the bulk holographic partition function of AdS/CFT (or in Hamiltonian quantization the boundary groundstate wavefunction) in terms of the contributions for the singleton and boundary gauge theory contributions, where the sum is over ‘t Hooft flux sectors:
\begin{equation}
    Z=\sum_\beta Z^{\text{singleton}}_\beta Z_{SU(N)/\mathbb Z_N}^\beta
\end{equation}

%{\color{gray}

%We can also see that the 2-form symmetry of the worldvolume theory is reproduced. Indeed, those anyons of the bulk 7d TFT with K-matrix $\overline{\mathbb K}$ given by \hyperref[bigK]{(4.61)} which do not belong to the isotropic sublattice $\Lambda_0$ defining the topological boundary conditions may be fused with the boundary and remain as topological 3d operators. They may then link with the charged string vertex operators which are the terminal points of bulk anyons. The braiding phase is evaluated as the same braiding phase from the bulk linking. Since $|\det \overline{\mathbb K}|=N$ then generically the linking phase will be $\exp(2\pi i p/N)$, indicating a worldvolume $\mathbb Z_N^{(2)}$ 2-form symmetry. This matches the expected 2-form symmetry of the M5-brane worldvolume theory \cite{Gaiotto:2014kfa, DelZotto:2015isa, Gukov:2020btk, Gukov:2025dol}.}\\

\subsubsection{Non-Invertible Fusion}

Another feature of M5-branes that we might hope to reproduce with our model are the charges of M5/anti-M5 states \cite{Yi:1999hd, Houart:1999bi, Bergman:2000xf}. The obvious candidate for this system is the result of the parallel fusion between a minimal defect $\mathcal D_1$ and its orientation reversal $\mathcal D_1^\dag$, representing the charge of an M5-brane and anti-M5-brane respectively:
\begin{equation}
    \mathcal D_1\otimes \mathcal D_1^\dag
\end{equation}
The interpretation of D-brane/anti-D-brane states as the fusion of non-invertible defects has been considered previously in holographic settings \cite{Apruzzi:2022rei, GarciaEtxebarria:2022vzq, Heckman:2022muc, Bonetti:2024etn, Antinucci:2022vyk, Bah:2023ymy, Apruzzi:2023uma, Damia:2022bcd, Calvo:2025usj, Yu:2023nyn, Franco:2024mxa, Gutperle:2024vyp, Heckman:2024obe, Calvo:2025kjh}, as we discuss in \hyperref[sec52]{Section 5.2}. \\

This fusion can be straightforwardly computed from our half-space gauging construction, as in the case of the non-invertible fusion \hyperref[NIrepG]{(3.24)} for the $\text{Rep}(\mathbb G^{(3)}_N)$ defects. We consider a system where the defects are parallel but separated by a finite distance $L$. The defects are defined by inserting the 7d TQFT $\mathcal S$ inside the 7d slab between them, and the resulting parallel fusion is obtained by compactifying the slab on the interval $[0,L]$:
\begin{equation}
    \mathcal D_1\otimes \mathcal D_1^\dag=\lim_{L\to 0}\int \mathcal D a_3\mathcal D\bar a_3 \exp\Big(\frac{i}{2\pi}\int_{\Sigma_6\times [0,L]}\Big[a_3dC_3+\frac 12a_3da_3-\frac 12\bar a_3d\bar a_3 \Big]\Big)
\end{equation}
The result from compactifying a Chern-Simons theory on a slab in this way is easily computed following \cite{Kapustin:2010hk}. The result is the BF theory:
\begin{equation}
    \int \mathcal D \beta_2\mathcal D\alpha_2\mathcal D\bar\alpha_2\mathcal Da_3^{||}\mathcal D\bar a_3^{||} \exp\Big(\frac{i}{2\pi}\int_{\Sigma_6}\Big[ \alpha_2 dC_3+\alpha_2da_3^{||}-\bar \alpha_2d\bar a_3^{||}+\beta_2(da_3-d\bar a_3) \Big]\Big)
\end{equation}
here $\alpha_2\sim\int_{[0,L]}a_3$ is the surviving mode of the bulk gauge field transverse to $\Sigma_6$, while $a_3^{||}$ is the surviving longitudinal component. Successively integrating out the various dynamical fields leaves a single pairing $\frac{i}{2\pi}\int d\gamma_2\wedge C_3$ where $\gamma_2$ is a specific combination of the 2-forms. Notice that this takes nearly the same form as the coupling in the original non-invertible defect $\mathcal W_1[\Sigma_6,C_3]$, which matches our interpretation of this defect as a projection operator. The only difference is that here $\gamma_2$ is not charged under the same gauge symmetry of $C_3$. \\

This precisely matches the expected coupling of M5/anti-M5 systems to the bulk supergravity $C_3$ field \cite{Yi:1999hd}. The chiral 2-form of the M5-brane combines with the anti-chiral 2-form of the anti-M5-brane to form a non-chiral 2-form $b_2$ with non-self-dual field strength $H_3$, which couples to the bulk $C_3$ field as:
\begin{equation}
    \frac{1}{2\pi}\int C_3\wedge H_3
\end{equation}
This coupling captures the charge of the tachyonic string for this brane/anti-brane system, which is the uplift of the familiar tachyonic modes from open ended strings of type IIA brane/anti-brane states up to M-theory \cite{Yi:1999hd}. \\ 

\subsubsection{Hanany-Witten Moves}

Another feature of M5-brane charges which is captured by our 6d non-invertible defects are Hanany-Witten moves between M5 branes \cite{Hanany:1996ie}. In particular, we consider a configuration of two flat M5 branes supported on the coordinate planes $\Sigma_6=x^{1,2,3,4,5,6}$ and $\Sigma_6'=x^{1,2,7,8,9,10}$ in a flat background. We then pass the first M5 brane through the second, so that they momentarily meet along the 2d locus $C_2=x^{1,2}$. Charge conservation requires that after these branes pass through each other there is an M2 brane stretched between them. \\

We can see this same move using the half-space gauging construction of our non-invertible defects.\footnote{The relationship between Hanany-Witten moves of branes and holographically-dual non-invertible defects was also studied in AdS/CFT in \cite{Apruzzi:2022rei, Apruzzi:2023uma, Bonetti:2024etn}.} We work with flat M5 branes in a flat background. We place the first defect along the 6d plane $\Sigma_6$ spanned by the $x^{1,2,3,4,5,6}$ axes. We take the 7d bulk $M_7=\Sigma_6\times \{x^{11}\geq 0\}$ to extend along the positive $x_{11}$ axis: 
\begin{equation}
    \mathcal S_1[\Sigma_6]=\int \mathcal Da_3\mathcal D\bar a_3\exp\Big(\frac i{2\pi}\int_{M_7}\Big[a_3dC_3+\Big(\frac 12a_3da_3-B_3da_3+\frac 12B_3dB_3 \Big)-c.c.\Big]\Big)
\end{equation}
where we have included the coupling to the background gauge field $B_3$ defined by the full $\mathbf{CTST}$ operation. We use the convention that all background fields are represented by capital letters, while dynamical fields are lower-case. The $c.c.$ refers to the spectator fields.  \\ 

\begin{figure}[h]
    \centering

    \begin{tikzpicture}[
        >=Latex,
        every node/.style={font=\small}
    ]

% ============================================================
% Coordinate directions: leftmost block
% ============================================================
\begin{scope}[shift={(0,-0.20)}]

    \coordinate (O) at (0,0);

    % x^{11}: horizontal to the right
    \draw[->,thick]
        (O) -- (1.25,0)
        node[right] {$x^{11}$};

    % x^{3,4,5,6}: vertically upward
    \draw[->,thick]
        (O) -- (0,1.25)
        node[above] {$x^{3,4,5,6}$};

    % x^{7,8,9,10}: diagonally into the third quadrant
    \draw[->,thick]
        (O) -- (-0.78,-0.78)
        node[below right,align=center]
        {$x^{7,8,9,10}$};

\end{scope}

% ============================================================
% Middle figure: turn on the vertical red brane
% ============================================================
\begin{scope}[xshift=1.70cm]

    % Extended red CTST half-space
    \path[
        draw=braneC!80!black,
        fill=braneC!16,
        line width=0.7pt
    ]
        (1,-1.50)
        -- (5.20,-1.50)
        -- (5.20,1.50)
        -- (1,1.50)
        -- cycle;

    % Physical vertical red brane
    \draw[brane,braneC]
        (1,-1.50) -- (1,1.50);

    % Label
    \node[
        braneC,
        above
    ]
        at (2.00,-1.35)
        {$S[a_3,C_3]$};

\end{scope}

% ============================================================
% Right figure: turn on the slanted blue brane
% ============================================================
\begin{scope}[xshift=9.20cm]

    % Points on the slanted blue M5-brane
    \coordinate (BA) at (-0.35, 0.75);
    \coordinate (BC) at (-1.00, 0.00);
    \coordinate (BB) at (-1.65,-0.75);

    % Extended points translated along x^{11}
    \coordinate (BAp) at (5.85, 0.75);
    \coordinate (BCp) at (5.20, 0.00);
    \coordinate (BBp) at (4.55,-0.75);

    % Entire extended blue CTST plane
    \path[
        fill=braneB!18,
        draw=none
    ]
        (BA) -- (BAp) -- (BBp) -- (BB) -- cycle;

    % Exterior boundary of the extended blue parallelogram
    \draw[
        braneB!80!black,
        line width=0.7pt
    ]
        (BA) -- (BAp) -- (BBp) -- (BB) -- cycle;

    % Extended red CTST plane
    \path[
        draw=braneC!80!black,
        fill=braneC!16,
        line width=0.7pt
    ]
        (1,-1.50)
        -- (5.20,-1.50)
        -- (5.20,1.50)
        -- (1,1.50)
        -- cycle;

    % Front portion of the extended blue CTST plane
    \path[
        fill=braneB!18,
        draw=none
    ]
        (BC) -- (BCp) -- (BBp) -- (BB) -- cycle;
    
    % Redraw only the visible exterior boundary

        % Bottom-right and bottom boundaries of the blue parallelogram
    \draw[
        draw=braneB!80!black,
        line width=0.7pt
    ]
        (BCp) -- (BBp) -- (BB);

    \node[
        text=green!55!black,
        above,
        yshift=2pt
    ]
        at (3,0.05)
        {$L$};
    % Green intersection locus
    \draw[
        line width=2.2pt,
        draw=green!55!black
    ]
        (1,0) -- (BCp);

    % Physical slanted blue brane
    \draw[brane,braneB]
        (BB) -- (BA);

    % Physical vertical red brane
    % The portion behind the front blue plane is omitted.
    \draw[brane,braneC]
        (1,-1.50) -- (1,-0.75);

    \draw[brane,braneC]
        (1,0) -- (1,1.50);

    % Red intersection point, drawn last so it remains visible
    \fill[braneC]
        (1,0) circle (3.2pt);

    % Labels
    \node[
        braneB,
        anchor=south east
    ]
        at (0.67,-0.45)
        {$S[b_3,C_3]$};

    \node[
        braneC,
        above
    ]
        at (2.00,-1.35)
        {$S[a_3,C_3]$};

    % Exponential annotation
    \node[
        text=black!75
    ] (alabel)
        at (2.90,1.05)
        {$e^{-\frac{i}{2\pi}
        \int_{M_7}\delta(L)b_3\wedge da_3}$};

    % Curved dashed arrow pointing to the intersection term
    \draw[
        ->,
        dashed,
        draw=black!75,
        line width=0.9pt
    ]
        (alabel.south west)
        to[out=-160,in=75]
        (1.55,0);

\end{scope}

    \end{tikzpicture}

    \caption{
        Sequential construction of the intersecting \(CTST\)
        configuration. First, the vertical red brane and its
        \(CTST\) half-space are introduced. The slanted blue brane
        and its \(CTST\) half-space are then turned on. The
        \(x_{1,2}\) directions are suppressed; the horizontal,
        vertical, and diagonal directions represent \(x_{11}\),
        \(x_{3,4,5,6}\), and \(x_{7,8,9,10}\), respectively.
    }
    \label{fig:sequential-CTST-construction}

\end{figure}
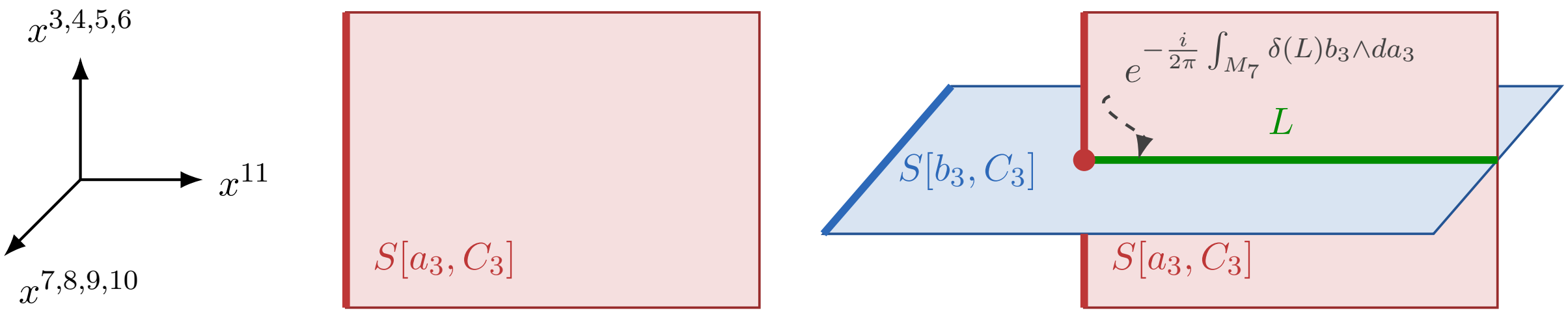

% ===========================
% ===========================

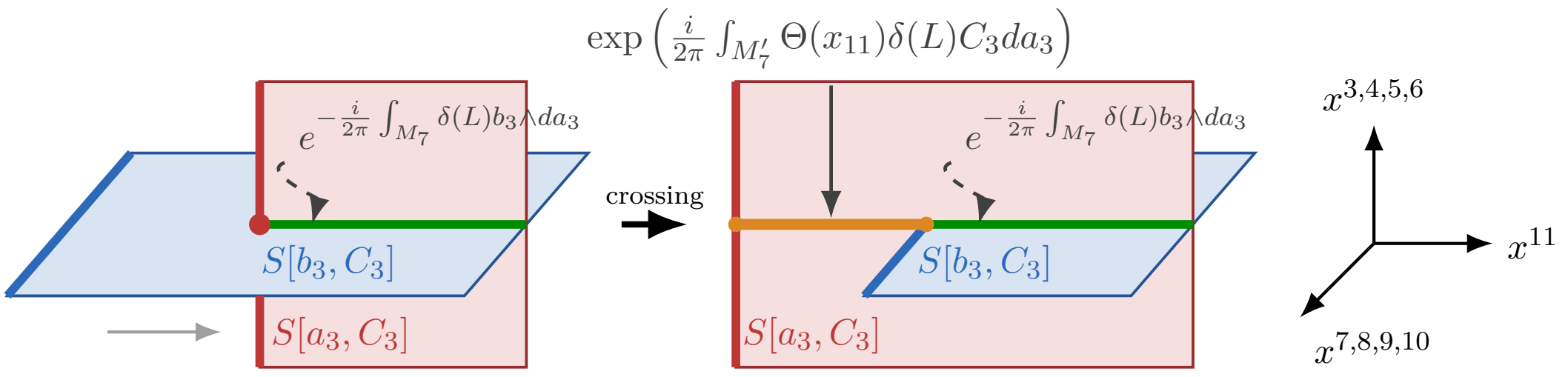
\begin{figure}[h]
    \centering

    \begin{tikzpicture}[
        scale=0.85,
        transform shape,
        >=Latex,
        every node/.style={font=\small}
    ]

% ============================================================
% Before
% ============================================================
\begin{scope}[xshift=0cm]

    % Points on the slanted blue M5-brane
    \coordinate (BA) at (-0.35, 0.75);
    \coordinate (BC) at (-1.00, 0.00);
    \coordinate (BB) at (-1.65,-0.75);

    % Points translated along x^{11}
    \coordinate (BAp) at (4.45, 0.75);
    \coordinate (BCp) at (3.80, 0.00);
    \coordinate (BBp) at (3.15,-0.75);

    % Entire blue CTST plane
    \path[
        fill=braneB!18,
        draw=none
    ]
        (BA) -- (BAp) -- (BBp) -- (BB) -- cycle;

    % Exterior boundary of the blue parallelogram
    \draw[
        braneB!80!black,
        line width=0.7pt
    ]
        (BA) -- (BAp) -- (BBp) -- (BB) -- cycle;

    % Red CTST plane
    \path[
        draw=braneC!80!black,
        fill=braneC!16,
        line width=0.7pt
    ]
        (1,-1.50)
        -- (3.80,-1.50)
        -- (3.80,1.50)
        -- (1,1.50)
        -- cycle;

    % Front portion of the blue CTST plane
    \path[
        fill=braneB!18,
        draw=none
    ]
        (BC) -- (BCp) -- (BBp) -- (BB) -- cycle;

    % Visible exterior boundary
    \draw[
        braneB!80!black,
        line width=0.7pt
    ]
        (BCp) -- (BBp) -- (BB);

    % Green intersection locus
    \draw[
        line width=2.2pt,
        draw=green!55!black
    ]
        (1,0) -- (BCp);

    % Physical blue M5-brane
    \draw[brane,braneB]
        (BB) -- (BA);

    % Physical red M5-brane
    \draw[brane,braneC]
        (1,-1.50) -- (1,-0.75);

    \draw[brane,braneC]
        (1,0) -- (1,1.50);

    % Red intersection point
    \fill[braneC]
        (1,0) circle (3.2pt);

    % Labels
    \node[
        braneB,
        anchor=south west
    ]
        at (0.87,-0.75)
        {$S[b_3,C_3]$};

    \node[
        braneC,
        above
    ]
        at (1.85,-1.50)
        {$S[a_3,C_3]$};

    % Direction of motion
    \draw[motion]
        (-0.60,-1.13) -- (0.60,-1.13);

    % Exponential annotation
    \node[
        text=black!75
    ] (beforelabel)
        at (2.90,1.05)
        {$e^{-\frac{i}{2\pi}
        \int_{M_7}\delta(L)b_3\wedge da_3}$};

    \draw[
        ->,
        dashed,
        draw=black!75,
        line width=0.9pt
    ]
        (beforelabel.south west)
        to[out=-160,in=75]
        (1.55,0);

\end{scope}

% ============================================================
% Transition arrow
% ============================================================
\draw[->,ultra thick]
    (4.80,0)
    -- node[above,font=\scriptsize] {crossing}
    (5.50,0);

% ============================================================
% After
% ============================================================
\begin{scope}[xshift=7.00cm]

    % Points on the displaced blue M5-brane
    \coordinate (AA) at (1.65, 0.75);
    \coordinate (AC) at (1.00, 0.00);
    \coordinate (AB) at (0.35,-0.75);

    % Points translated along x^{11}
    \coordinate (AAp) at (4.45, 0.75);
    \coordinate (ACp) at (3.80, 0.00);
    \coordinate (ABp) at (3.15,-0.75);

    % Back part of the blue CTST plane
    \path[
        draw=braneB!80!black,
        fill=braneB!18,
        line width=0.7pt
    ]
        (AA) -- (AAp) -- (ACp) -- (AC) -- cycle;

    % Back portion of the blue M5-brane
    \draw[brane,braneB]
        (AC) -- (AA);

    % Red CTST plane
    \path[
        draw=braneC!80!black,
        fill=braneC!16,
        line width=0.7pt
    ]
        (-1,-1.50)
        -- (3.80,-1.50)
        -- (3.80,1.50)
        -- (-1,1.50)
        -- cycle;

    % Front part of the blue CTST plane
    \path[
        draw=braneB!80!black,
        fill=braneB!18,
        line width=0.7pt
    ]
        (AC) -- (ACp) -- (ABp) -- (AB) -- cycle;

    % Front portion of the blue M5-brane
    \draw[brane,braneB]
        (AB) -- (AC);

    % Green intersection locus
    \draw[
        line width=2.2pt,
        draw=green!55!black
    ]
        (AC) -- (ACp);

    % Physical red M5-brane
    \draw[brane,braneC]
        (-1,-1.50) -- (-1,1.50);

    % Created M2-brane
    \draw[createdbrane]
        (-1,0) -- (AC);

    % Orange M2 endpoints
    \fill[created]
        (-1,0) circle (2.3pt);

    % This remains orange at the junction with the green locus
    \fill[created]
        (AC) circle (2.3pt);

    % Vertical annotation for the created M2-brane
    \node[
        text=black!75
    ] (m2label)
        at (0,1.95)
        {$\exp\Big(\frac{i}{2\pi}\int_{M_7'}\Theta(x_{11})\delta(L)C_3da_3\Big)$};

    \draw[
        ->,
        draw=black!75,
        line width=0.9pt
    ]
        (m2label.south) -- (0,0.06);

    % Labels
    \node[
        braneC,
        above
    ]
        at (-0.2,-1.50)
        {$S[a_3,C_3]$};

    \node[
        braneB,
        anchor=south west
    ]
        at (0.76,-0.75)
        {$S[b_3,C_3]$};

    % Exponential annotation
    \node[
        text=black!75
    ] (afterlabel)
        at (2.90,1.05)
        {$e^{-\frac{i}{2\pi}
        \int_{M_7}\delta(L)b_3\wedge da_3}$};

    \draw[
        ->,
        dashed,
        draw=black!75,
        line width=0.9pt
    ]
        (afterlabel.south west)
        to[out=-160,in=75]
        (1.55,0);

\end{scope}

% ============================================================
% Coordinate directions
% ============================================================
\begin{scope}[shift={(12.70,-0.20)}]

    \coordinate (O) at (0,0);

    \draw[->,thick]
        (O) -- (1.25,0)
        node[right] {$x^{11}$};

    \draw[->,thick]
        (O) -- (0,1.25)
        node[above] {$x^{3,4,5,6}$};

    \draw[->,thick]
        (O) -- (-0.78,-0.78)
        node[below right,align=center]
        {$x^{7,8,9,10}$};

\end{scope}

    \end{tikzpicture}

    \caption{
        Illustration of the Hanany--Witten move. The \(x^{1,2}\)
        directions are suppressed. The horizontal, vertical, and
        diagonal directions represent \(x^{11}\),
        \(x^{3,4,5,6}\), and \(x^{7,8,9,10}\), respectively.
    }
    \label{fig:hw-unit-charge-3d}

\end{figure}

We now would like to insert the other defect. We take $\Sigma_6'$ to extend along the $x_{1,2,7,8,9,10}$ plane, but we place it at $x_{11}=-\epsilon$, so that the extended 7d volume defining the half-space gauging is $M_7'=\Sigma_6'\times \{x_{11}\geq -\epsilon\}$. This means that the intersection of the two bulks is the 3d half-plane $M_7\cap M_7'\equiv L_3=\mathbb R_{12}\times \{x_{11}\geq 0\}$. The second defect is defined by applying the half-space gauging operation $\mathbf{CTST}$ in $M_7'$. In particular, these operations act on the existing background gauge field in this region. Away from $L$, this background simply couples directly to $dC_3$. However the region $L\subset M_7$ already supports the background $B_3$ above.  Concretely, if we restrict to $L$, the gauge field is the 3-form $ \delta(L) B_3(x_1,x_2,x_{11})dx_1\wedge dx_2\wedge dx_{11}$, where $\delta(L)$ is the delta function in the transverse coordinates which localizes to $L$. The $\mathbf S$ operation promotes this to a dynamical gauge field $b_3$ on $M_7'$. Then we can write the total combined partition function (suppressing backgrounds) as:
\begin{equation}
    \begin{gathered}
        \mathcal S_1[\Sigma_6]\oplus \mathcal S_1'[\Sigma_6']=\int\mathcal Da_3\mathcal Db_3\dots \exp \Big(\frac{i}{2\pi}\int_{M_7}\Big[ a_3dC_3+\Big(\frac 12 a_3da_3-\delta(L)b_3da_3 \Big)-c.c.\Big]\Big) \\ \times \exp\Big(\frac{i}{2\pi}\int_{M_7'}\Big[b_3dC_3+\frac 12b_3db_3-c.c.\Big]\Big)
    \end{gathered}
\end{equation}
Notice that the term containing $\delta(L)$ may be integrated over the transverse directions, and so is supported only on $L$. We can thus choose to include it in either factor, and so are free to move it from the action of the first defect to the second:
\begin{equation}
    \begin{gathered}
        \mathcal S_1[\Sigma_6]\oplus \mathcal S_1'[\Sigma_6']=\int\mathcal Da_3\mathcal Db_3\dots \exp \Big(\frac{i}{2\pi}\int_{M_7} \Big[ a_3dC_3+\frac 12 a_3da_3-c.c. \Big] \Big) \\ \times \exp\Big(\frac{i}{2\pi}\int_{M_7'}\Big[b_3dC_3+\Big(\frac 12b_3db_3-\Theta(x_{11})\delta(L)b_3da_3\Big)-c.c.\Big]\Big)
    \end{gathered}
\end{equation}

We now would like to move the defect on $\Sigma_6'$ from $x_{11}=-\epsilon$ to $x_{11}=+\epsilon$. This can be achieved by performing the path integral for $b_3$ over the slab $\Sigma_6'\times [-\epsilon,\epsilon]$. Performing the quadratic path integral and subtracting the usual Hall response $\exp(-\frac{i}{4\pi}\int C_3dC_3)$ leaves us with the cross-term:
\begin{equation}
    \exp\Big(\frac{i}{2\pi}\int_{M_7'}\Theta(x_{11})\delta(L)C_3da_3\Big)
\end{equation}
Since this term is localized on $L$ it can be moved back into the action on $M_7$. We see that the effective change can be viewed as shifting the support of $C_3$ by adding a contribution along the segment of $L_3$ stretching from $x_{11}\in [0,\epsilon]$.  We can interpret this as the insertion of a $C_3$ Wilson line along this segment, which is the expected outcome from a Hanany-Witten move. 

\subsubsection{Codimension-2 Defects}

Our half-space gauging construction of the non-invertible defects $\mathcal D_1$ may also be used to analyze junctions of other dimensions, corresponding to higher morphisms. As an illustration of such an analysis, we can consider the albeit almost trivial example of two individual intersecting non-invertible defects along a 4d locus. \\

Such a configuration is expected to have relevance for the worldvolume theory of the M5-branes. Besides the self-dual strings, M5-brane worldvolume theories also possess another species of $\frac 12$-BPS defect. These are 4d defects, and are often simply referred to as codimension-2 defects \cite{Gaiotto:2009we, Alday:2010vg, Tachikawa:2011dz, Chacaltana:2012zy, Gukov:2006jk, Balasubramanian:2014jca}. They can be realized in M-theory as the intersections of M5-branes along a 4d locus $V_4$. These defects are familiar in the context of 4d $\mathcal N=2$ SCFTs, where in the construction of class-S theories from dimensional reduction of the 6d $\mathcal N=(2,0)$ theories on punctured Riemann surfaces they correspond to the punctures. For a 6d (2,0) theory of type $\frak g$ regular punctures are labeled in a compactified theory by nilpotent orbits of $\frak g$, or equivalently homomorphisms $\rho : \frak{su}(2)\to \frak g$ which define boundary conditions of the adjoint scalars at the defect. \\

We can realize this junction in our model of the M5-brane charges as non-invertible defects $\mathcal D_1$, and compare the operators supported at the locus $V_4$ with those which have support away from it. We consider a configuration similar to the previous subsection, where two flat $\mathcal D_1$ defects intersect along $V_4$. We insert the first defect on the $\Sigma_6=x^{1,2,3,4,5,6}$ plane, with the 7th bulk direction taken to be $x_9\geq 0$ so that $M_7=\Sigma_6\times\{x_9\geq 0\}$:
\begin{equation}
    \begin{gathered}
        \mathcal S_1[\Sigma_6]=\int Da_3D\bar a_3\exp\Big(\frac{i}{2\pi}\int_{M_7}\Big[a_3dC_3+\Big(\frac 12a_3da_3-B_3da_3+\frac 12B_3dB_3\Big)-c.c.\Big]\Big)
    \end{gathered}
\end{equation}
where we have included the coupling to the background $B_3$. We now insert another defect along $\Sigma_6'$ which we take to be the $x^{1,2,3,4,7,8}$ plane, and again take the bulk $M_7'$ along the positive 9-axis. This means that $V_4=\Sigma_6\cap \Sigma_6'$ is the $x^{1,2,3,4}$ plane, and $W_5=M_7\cap M_7'=V\times\{x_9\geq 0\}$. The insertion of this defect requires performing $\mathbf{CTST}$ in $M_7'$. The full bulk action for both defects can then be written:
\begin{equation}
    \begin{gathered}
        \int \mathcal Da_3\mathcal D b_3\dots \exp\Big(\frac{i}{2\pi}\int_{M_7}\Big[ a_3dC_3+\Big(\frac 12a_3da_3-B_3da_3\Big)-c.c.\Big]\Big) \\ 
        \times\exp\Big(\frac{i}{2\pi}\int_{M_7'}\Big[b_3dC_3+\frac 12 b_3db_3-c.c.\Big]\Big) 
    \end{gathered}
\end{equation}
We must identify $B_3\cong b_3$ along $W_5$, which is why we have dropped the $B_3dB_3$ term. However it will be more convenient to leave $B_3$ generic for the time being and impose this constraint later. By our construction, we are instructed to perform the change of variables $\tilde a_3=a_3+C_3-B_3$ in order to isolate the Hall response. Applying this we obtain:   
\begin{equation}
    \begin{gathered}
        \int \mathcal D\tilde a_3\mathcal D b_3\dots \exp\Big(\frac{i}{2\pi}\int_{M_7}\Big[\frac 12\tilde a_3d\tilde a_3-c.c.\Big]-Q(B_3+C_3)\Big) \exp\Big(\frac{i}{2\pi}\int_{\Sigma_6}\beta_2(d\tilde a_3-d\bar a_3)\Big) \\ 
        \times\exp\Big(\frac{i}{2\pi}\int_{M_7'}\Big[b_3dC_3+\frac 12 b_3db_3-c.c.\Big]\Big)\exp\Big(\frac{i}{2\pi}\int_{\Sigma_6'}\gamma_2(db_3-d\bar b_3)\Big) 
    \end{gathered}
\end{equation}
where we have included the Lagrange multiplier interactions enforcing the topological boundary conditions. From our analysis we know that the boundaries support string vertex operators $e^{i\int \alpha_2}$ and $e^{i\int \eta_2}$ dual to $\beta_2,\gamma_2$ respectively, which are charged under the gauge symmetry of $\tilde a_3$ and $b_3$. However we know that on $V_4$ we must identify $B_3=b_3|_{V_4}$ which is a genuine gauge field. Since $\tilde a_3=a_3-B_3+C_3$ then we see that the string operator $e^{i\int \alpha_2}$ picks up an extra gauge redundancy corresponding to that of $b_3$ when supported on $V_4$, and as a result bulk anyons on $M_7'$ may end on it. However away from $V_4$ then only the Wilson lines of $\tilde a_3$ may end on it. Moreover, along $V_4$ we may consider the novel relative string $e^{i\int \alpha_2-\eta_2}$. We may interpret this as an enhancement of this junction operator $e^{i\int \alpha_2}$ when it itself is supported at the junction $V_4$. \\

There are expected to be interesting trapped operators for more complex configurations of junctions.  In \cite{Bhardwaj:2021mzl} it was argued that irregular punctures may possess their own classes of ``trapped" higher-form symmetries: for example defect operators which belong only to the locus of the intersection and remain mutually nonlocal. Such examples require more sophisticated configurations of M5-branes than those considered here, but it would be interesting to reproduce such trapped higher symmetries using these non-invertible charge defects. Moreover, we point out that this calculation might be used to study exotic intersections of $\mathbb{Q}/\mathbb{Z}$ type non-invertible symmetry defects in regular quantum field theories. We leave such an analysis to future work.  \\ 

\subsection{M5-Branes and Non-Invertible Defects}\label{sec52}

We have argued that the charge of the M5-brane may be modeled by the 6d non-invertible defect $\mathcal D_1$, which is defined using a higher half-space gauging procedure as the boundary of a 7d abelian Wu-Chern-Simons theory. We have seen that this model reproduces many desired features of the M5-brane, but it would be better to understand the precedent for such a model. How is our definition of such a non-invertible defect related to the definition of the M5-brane? \\

A more familiar setting in which a relationship between branes and non-invertible topological defects has been established is holography. In AdS/CFT, the boundary field theory may host generalized global symmetries which are generated by extended topological operators. There has been much work on investigating the bulk AdS duals of these operators \cite{Apruzzi:2022rei, GarciaEtxebarria:2022vzq, Heckman:2022muc, Bonetti:2024etn, Antinucci:2022vyk, Heckman:2022xgu, Bah:2023ymy, Apruzzi:2023uma, Etheredge:2023ler, Damia:2022bcd, Heckman:2024oot, Bergman:2022otk, Bergman:2024aly, Calvo:2025usj, Cvetic:2023plv, Yu:2023nyn, Franco:2024mxa, Gutperle:2024vyp, Heckman:2024obe, Waddleton:2024iiv, Calvo:2025kjh, Heckman:2025isn, Bah:2025vfu, Bah:2026gcf, Apruzzi:2021nmk}, especially in the case where the boundary symmetry is non-invertible. In several proposals, the bulk dual of a boundary topological symmetry defect operator is naturally identified with a brane in the bulk gravitational theory, perhaps wrapping suitable cycles of the bulk geometry. In this correspondence, a limit is taken where the brane approaches the asymptotic conformal boundary, allowing for an expansion of the brane worldvolume action. The gauge sector of this action is then argued to be dominated by a topological term, while the kinetic contributions to the DBI action is subleading. The non-topological modes are thus “frozen out”. Correlators of the resulting brane configurations in this limit can then by analyzed by considering the Bianchi identities and Gauss laws. Under suitable boundary conditions, branes which host nontrivial topological terms may then be identified with the non-invertible topological defects of the boundary theory. Indeed, the desired features of the field theory topological defects are reproducible in the bulk, such as non-invertible fusion. A particularly relevant example was considered in \cite{Bonetti:2024etn} who studied the holographic duality between the 6d $\mathcal N=(2,0)$ SCFTs of type-D and M-theory on $AdS_7\times \mathbb{RP}^4$. In this correspondence, the non-invertible symmetry defect operators of the boundary were identified precisely with those deriving from M5-branes of the bulk theory wrapped on cycles of the bulk geometry. This included a bulk derivation of the TQFTs stacked on the topological defect worldvolumes which led to their non-invertibility. \\

While a correspondence across holography is suggestive, it is distinct from our proposal here. We do not treat the non-invertible defects $\mathcal D_k$ as the holographic duals of M5-branes in AdS/CFT, but rather note that they appear to capture precisely the charge information relevant to how the M-theory $C_3$ field couples to the M5-brane. \\ 

In fact there is a stronger precedent for the defect $\mathcal D_1$ as representing the M5-brane charge. Recall that we have defined this defect as arising on the boundary of a 7d  Chern-Simons theory:
\begin{equation}
    \mathcal S_1=\int \mathcal D a_3\mathcal D\bar a_3\exp\Big(i\int_{M_7}\Big[\frac{1}{2\pi}a_3dC_3+\frac{1}{4\pi}a_3da_3-\frac{1}{4\pi}\bar a_3d\bar a_3\Big]\Big)
\end{equation}
The relevant feature of the M5-brane we need to capture is the chiral 2-form $B_2$ and its coupling to the bulk gauge field $C_3$.  It was argued in \cite{Witten:1996hc, Belov:2006jd}  that in fact the natural definition of the partition function of a chiral 2-form is to use the bulk partition function for a 7d level-1 Wu Chern-Simons theory:
\begin{equation}
    \int \mathcal DH_3\exp\Big(\frac{i}{4\pi}\int_{M_7}H_3dH_3\Big)
\end{equation}
More precisely, when considered on a manifold with boundary $\partial M_7=\Sigma_6$ then this partition function defines a section of a line bundle over the moduli space $J_3\equiv H^3(\Sigma_6, \mathbb R)/H^3(\Sigma_6,\mathbb Z)$ of boundary gauge field configurations. The chirality of the boundary theory is enforced by taking this to be the unique holomorphic section of this line bundle. The partition function of the boundary chiral 2-form may then be defined as this holomorphic section. Notice this definition is completely local to the boundary. As noted in \cite{Bonetti:2024etn}, adding the coupling to the bulk physical gauge field
\begin{equation}
    \int \mathcal DH_3\exp\Big(i\int_{M_7}\Big[\frac{1}{2\pi}H_3dC_3+\frac{1}{4\pi}H_3dH_3\Big]\Big)
\end{equation}
(and subsequently redefining $H_3$ by a sign) enforces the required Bianchi identity of the worldvolume field strength: $dH_3=G_4$ \cite{Witten:1996hc, Townsend:1995af}. \\

Of course, the holomorphic boundary conditions define a gapless chiral CFT on the boundary. For our purposes, we care only about capturing the charge information of the worldvolume theory in a topological defect hosting a gapped worldvolume. This is the purpose of including the spectators, which allow for a gapped boundary while preserving the crucial coupling $dH_3=G_4$. \\ 

There is also another perspective which illuminates the non-invertible nature of the M5-brane charge when it is considered as a heavy magnetic defect in the low-energy supergravity theory, which of course contains the Chern-Simons interaction. It was noted in \cite{Hsin:2022heo} that whenever a magnetic defect is inserted in a theory whose action contains a topological action then the magnetic defect may undergo flux attachment, which is a generalization of the Witten effect. As a result, if the attached flux induces an anomalous variation on the magnetic worldvolume then the resulting magnetic defect may be non-invertible. In our case, this flux attachment is precisely equivalent to the anomalous variation that the 11d Chern-Simons term induces on the M5-brane worldvolume. This is the key input which was used to define the non-invertible defect $\mathcal D_1$.  \\

\subsection{Appearance of the 6-group Action}\label{sec53}

We have remarked on the fact that the 6-group SymTFT has been a useful tool which allows us to study the symmetries and charges of the 6-group $\mathbb G^{(6)}$. We have also noted that, due to the subtlety in defining the quadratic refinement $Q(C_3)$, there does not seem to exist a natural way to extend this local action to one which captures the charges $\frak G$. One might expect a local action to take the heuristic form:
\begin{equation}\label{pseudotft}
S\sim \frac{1}{2\pi}\int \Big[ C_3\wedge df_7+ B_6\wedge dh_4+h_4\wedge \Big(\frac 12C_3dC_3-T_7\Big)\Big]
\end{equation}
where $C_3$ is a 3-form abelian gauge field with the shifted quantization law and $dT_7=I_8$. However if we want to make these quantities globally well-defined by using shifted differential cohomology $\check C_3\in \check H^4_{\lambda/2}(M)$ then it is unclear how to define the pairing between $h_4$ and the quadratic refinement $Q(\check C_3)$. \\

The above action is a useful heuristic tool, but does it have any connection to M-theory? In fact we will see that the general form of the above heuristic action does make a suggestive appearance in the 11d supergravity action when we couple to M2 and M5 branes. As we will see, the interpretation of this action we will derive is quite different from the one above, it does not define the action of an 11d TFT by any means. However its appearance is illuminating nonetheless. \\

To see it, let us start with the gauge sector of the low-energy 11d supergravity action:
\begin{equation}
    S_{sugra}\sim \int d^{11}x \ \Bigg[ G_4 *G_4 + \frac 16 C_3G_4G_4-C_3I_8\Bigg]
\end{equation}
If we want to express this in a globally-defined way, then we should write:
\begin{equation}
    Z_{sugra}=\int\mathcal D C_3\dots \exp\Bigg(i\int d^{11}x  \ G_4*G_4\Bigg) \Phi(\check C)
\end{equation}
where $\Phi(\check C)\sim \exp\Big(2\pi i\int \frac 16 C_3G_4G_4-C_3I_8\Big)$ is the globally-well defined cubic refinement defined in \hyperref[appa]{Appendix A}. We would like to consider how this low-energy effective action changes upon inserting an M5-brane along a 6d submanifold $\Sigma_6\subset M_{11}$. We consider the limit in which interactions between supergravity fields and the M5-brane are suppressed, so that the total effective action can be decomposed as:
\begin{equation}
    S\approx S_{sugra}+S_{M5}[\Sigma_6]
\end{equation}
Then the insertion of a (non-singular) M5-brane requires two steps:
\begin{enumerate}
    \item Inserting the M5-brane action $S_{M5}[\Sigma_6]$ along the worldvolume submanifold.
    \item Enforcing the magnetic flux condition on the 11d gauge field $C_3$ by performing a shift $G_4\to \tilde G_4+e_4$ where $e_4$ has unit magnetic flux around the generating 4-cycle linking $\Sigma_6$, and $\tilde G_4$ encodes the rest of the field strength which has no magnetic flux around this cycle.  
\end{enumerate}

Let us first express how these two steps modify the local action in terms of only the local trivializations of the gauge fields. We will then revisit the argument in the language of differential cohomology and attempt to make these steps more precise. First we imagine that the relevant portion of the M5 effective action which couples to the bulk $C_3$ gauge field admits a description as some local density which can be integrated over: 
\begin{equation}
    S_{M5}[\Sigma_6]\supset \int_{\Sigma_6}\mathcal B_6
\end{equation}
Morally we may think of $\mathcal B_6$ as playing the role of the “dual magnetic gauge field” which trivializes the modified Gauss law operator $d\mathcal B_6=*G_4+\frac 12 C_3G_4$. In this case, we may write the gauge sector of the total action as:
\begin{equation}
    S= \int_{M_{11}}d^{11}x \Big[\frac{1}{l^3}G_4*G_4+\frac 16 C_3G_4G_4-C_3I_8+J_5\wedge \mathcal B_6\Big]
\end{equation}
where $J_5(x)$ is the current density localized to $\Sigma_6$. For a singular configuration then we take this simply as the $\delta$-function $J_5=\delta^{(5)}(\Sigma_6\hookrightarrow M_{11})$. However in the spirit of treating the M5-brane action truly as an effective action, we can instead excise some finite tubular neighborhood $D_{\epsilon}(\Sigma_6)$ which is the total space of some disk bundle of radius $\epsilon$ over $\Sigma_6$. The boundary $N_{10}=\partial D_{\epsilon}(\Sigma_6)$ is some $S^4$ bundle over $\Sigma_6$. We may then smooth out the source using a bump form $\rho(r)$ so that the current takes the form of the Thom class for the normal bundle \cite{Freed:1998tg, Harvey:1998bx}:
\begin{equation}
    J_5=d\rho \wedge e_4/2
\end{equation}
where $e_4/2$ is the global angular form normalized to integrate to 1 over any 4-sphere in the fiber of $N_{10}$. \\

Now we implement the second step of the insertion, which is to include the magnetic backreaction of the M5-brane on the bulk $C_3$ field. This can be achieved by shifting the bulk gauge field by adding a component which captures the magnetic flux:
\begin{equation}
    C_3\to \tilde C_3+c_3
\end{equation}
where $G_4=\tilde G_4+h_4$ so that $\int_{Z_4} \tilde G_4=0$ around any 4-cycle linking $\Sigma_6$, while we require that near $D_{\epsilon}(\Sigma_6)$ we have $h_4=\rho e_4/2+d\sigma_3$. Notice that the resulting $C_3$ gauge field is no longer a closed gauge field on $D_{\epsilon}(\Sigma_6)$, since $dG_4=J_5$. \\

Finally we can also couple to an M2-brane by adding an M2-brane action $S_{M2}$ supported on $\gamma_3$. We keep only the coupling to the gauge field $\int_{\gamma_3}C_3 = \int_{M_{11}}\mathcal J_8\wedge C_3$ where $\mathcal J_8=\delta^{(8)}(\gamma_3\hookrightarrow  M_{11})$. In total, the resulting action is: 
\begin{equation}
    S= \int_{M_{11}}d^{11}x \Bigg\{G_4*G_4+\frac 16 (\tilde C_3+c_3)(\tilde G_4+h_4)(\tilde G_4+h_4)-(\tilde C_3+c_3)I_8+J_5\wedge \mathcal B_6+\mathcal J_8\wedge C_3\Bigg\}
\end{equation}
We can then take account of the total change to the action from inserting the M2 and M5 branes. Dropping the kinetic terms we get:
\begin{equation}
    \Delta S=\int \Big[ df_7\wedge C_3+dh_4\wedge \mathcal B_6+\Big\{\text{CS}_{11}( G_4+h_4)-\text{CS}_{11}(G_4)\Big\}\Big]
\end{equation}
where we have also chosen to trivialize $df_7\equiv \mathcal J_8$. As we show in \hyperref[appa]{Appendix A}, if we define the last quantity:
\begin{equation}
    \tilde Q_h(a_4)=\text{CS}_{11}( a_4+h_4)-\text{CS}_{11}(a_4)
\end{equation}
then this defines a quadratic refinement for the bilinear pairing:
\begin{equation}
    \braket{a_4,b_4}_h\equiv(a_4,b_4,h_4)=\tilde Q_h(a_4+b_4)-\tilde Q_h(a_4)-\tilde Q_h(b_4)+\tilde Q_h(0)
\end{equation}
In particular, we have:
\begin{equation}
    \braket{G_4,G_4}_{h_4}=(G_4,G_4,h_4)\cong \frac{1}{2\pi}\int h_4\wedge C_3dC_3
\end{equation}
The purpose of this exercise has been to notice the resemblance of this change:
\begin{equation}
    \Delta S=\int \Big[df_7\wedge C_3+dh_4\wedge \mathcal B_6+\tilde Q_h(G_4)\Big]
\end{equation}
with the form of the anticipated pseudo-6-group SymTFT \hyperref[pseudotft]{(5.33)}. Of course these are completely different objects: the above is obviously not a valid TQFT, the forms $h_4,f_7,\mathcal B_6$ are not gauge fields on spacetime, and this is only the action obtained from inserting a single M2 and M5 brane. Nevertheless it is amusing and suggestive to notice this resemblance. \\

We can make the above more precise using the language of differential cohomology. Following \cite{Freed:2000ta}, we should promote the magnetic current $J_5$ to a differential cocycle $\check J_5\in \check Z^5[M_{11}]$. We would also like to promote $\check h_4$ to a differential cocycle, which we think of as parametrizing the embedding of the M5-brane flux in spacetime. However as noted in \cite{Freed:2000ta} this object should naturally be considered as a differential cochain $\check h_4\in \check C^4[M_{11}]$ which trivializes the magnetic current $d\check h_4=\check J_5$. We point out that this appears related to the problem discussed in \cite{Antinucci:2024zjp} that the $U(1)$ SymTFT is not preserved under gauging, but must use the replacements $dh_4 \leftrightarrow f_5$. On the other hand, there is nothing wrong with promoting the electric current to a differential cocycle $\check J_8\in \check Z^8[M_{11}]$, so that we may use the pairing $(\check G_4,\check J_8)=\int df_7\wedge C_3$. \\

\subsection{Electric Symmetry and Charge}\label{sec54}

We noted in \hyperref[sec24]{Section 2.4} that hints of a pseudo-6-group $\frak G$ are present in the equations of motion for the 11d supergravity $C_3$ field. In particular, the 11d Chern-Simons term modifies the equation for the electric flux to take the form:
\begin{equation}
    dG_7=\frac 12G_4G_4-I_8
\end{equation}
This evidently requires a departure from the usual definition of electric flux and electric charge that we are familiar with in Maxwell theory. There has been thorough investigation into precisely how the definition of the electric charge \cite{Marolf:2000cb} should be modified in this case. In this section we collect some of these perspectives and compare them to features of our proposed charge algebra $\frak G$. In the next section we will review more formal approaches to the characterization of the charge algebra. \\

\subsection*{Electric Symmetry}

We begin by considering the electric higher form global symmetry in the gauge sector of pure 11d supergravity, before we have included any brane charges. If we were working in pure 3-form Maxwell theory then the electric flux $G_7=*G_4$ would define a 3-form symmetry generated by the topological symmetry defect operators:
\begin{equation}
    U_\alpha[\Sigma_7]=\exp\Big(i\alpha\int_{\Sigma_7}G_7\Big)
\end{equation}

which measure the electric charges of 3d Wilson surfaces $W_n[\gamma_3]=\exp(in\int_{\gamma_3}C_3)$ by linking with them. However the above modified equation of motion for supergravity spoils the $U(1)^{(3)}$ symmetry since these operators are no longer topological. Nevertheless, it was pointed out in \cite{Garcia-Valdecasas:2023mis} that if we ignore the gravitational term a non-invertible $\mathbb{Q/Z}^{(3)}$ symmetry may be preserved by stacking these with 7d minimal abelian TQFTs:
\begin{equation}
    \mathcal D_{p/N}\equiv e^{i\frac p N \int_{\Sigma_7}G_7}\otimes \mathcal A^{N,p}[\Sigma_7,G_4]
\end{equation}
Therefore a non-invertible 3-form electric symmetry is preserved in 11d supergravity. Of course this symmetry is expected to be explicitly broken once we include dynamical M2 branes, since it can be unlinked from a Wilson line which ends on an M2-brane. \\

Recall from \hyperref[sec3]{Section 3} that the operators $\mathcal D_{p/N}$ may naturally identified as belonging to the representation fusion category for a finite 7-group $\text{Rep}(\mathbb G^{(7)}_N)$ defined by:
\begin{equation}
    1 \to \Gamma(\mathbb Z_N)^{(6)}\to \mathbb G_N^{(7)}\to \mathbb Z_N^{(3)}\to 1
\end{equation}
with Postnikov class $\frak P\in H^8(K(\mathbb Z_N,4),\Gamma(\mathbb Z_N))$, which is a quadratic refinement for the cup product pairing on cocycles. We note that this group qualitatively seems related to the pseudo-6-group structure $\frak G$ which describes the gauge charges of the M2 and M5 branes, since recall that the form playing the role of the Postnikov class defines a quadratic refinement on differential cocycles, and thus its global characteristic class must define a quadratic refinement of the cup product on integral cohomology. \\

Recall also from our review in \hyperref[sec3]{Section 3} that gauging an ordinary global symmetry $G$ yields a theory with a dual global symmetry, sometimes referred to as the quantum symmetry. For discrete symmetries this dual global symmetry is $\text{Rep}(G)$, while for $U(1)$ global symmetries it is given by a dual $U(1)$ higher-form symmetry (for example the magnetic 1-form symmetry of 4d Maxwell). It is thus tempting to interpret the $\mathbb {Q/Z}^{(3)}\sim \text{Rep}(\mathbb G^{(7)})$ symmetry as the global quantum symmetry obtained from a $\frak G$ gauge redundancy. \\

\begin{figure}[h]
    \centering

    \begin{tikzpicture}[
        >=Latex,
        every node/.style={font=\small}
    ]

% ============================================================
% Before
% ============================================================
\begin{scope}[xshift=0cm]

    %\node[font=\bfseries] at (0.55,2.15) {Before};

    % Points on the slanted blue brane
    \coordinate (BA) at (-0.09, 1.05);
    \coordinate (BC) at (-1.00, 0.00);
    \coordinate (BB) at (-1.91,-1.05);

    % Endpoint of the intersection line along x^8
    \coordinate (BCp) at (2.20,0);

    % Red CTST half-space
    \path[
        draw=braneC!80!black,
        fill=braneC!16,
        line width=0.7pt
    ]
        (1,-1.50)
        -- (2.20,-1.50)
        -- (2.20,1.50)
        -- (1,1.50)
        -- cycle;

    % Physical blue brane, parallel to x^{9,10,11}
    \draw[brane,braneB]
        (BB) -- (BA);

    % Physical red brane, parallel to x^{1,2,3,4}
    \draw[brane,braneC]
        (1,-1.50) -- (1,1.50);

    % Labels
    \node[
        braneB,
        anchor=south east
    ]
        at (-1.17,0.45)
        {$mM5$};

    \node[
        braneC,
        above
    ]
        at (1,1.50)
        {$\mathcal{D}_{p/N}$};

    % Direction of motion
    \draw[motion]
        (-0.60,-1.13) -- (0.60,-1.13);

\end{scope}

% ============================================================
% Transition arrow
% ============================================================
\draw[->,ultra thick]
    (3.05,0)
    -- node[above,font=\scriptsize] {crossing}
    (3.75,0);

% ============================================================
% After
% ============================================================
% ============================================================
% After
% ============================================================
\begin{scope}[xshift=5.70cm]

    % Points on the displaced blue brane
    \coordinate (AA) at (1.91, 1.05);
    \coordinate (AC) at (1.00, 0.00);
    \coordinate (AB) at (0.09,-1.05);

    % Upper part of the blue brane:
    % drawn first so that it lies behind the red CTST plane
    \draw[brane,braneB]
        (AC) -- (AA);

    % Red CTST half-space, extended further to the right
    \path[
        draw=braneC!80!black,
        fill=braneC!16,
        line width=0.7pt
    ]
        (-1,-1.50)
        -- (3.30,-1.50)
        -- (3.30,1.50)
        -- (-1,1.50)
        -- cycle;

    % Lower part of the blue brane:
    % drawn after the red plane so that it lies in front
    \draw[brane,braneB]
        (AB) -- (AC);

    % Physical red brane
    \draw[brane,braneC]
        (-1,-1.50) -- (-1,1.50);

    % Created brane
    \draw[createdbrane]
        (-1,0)
        -- node[above right,created]
        {$\exp\!\left(
            2\pi i\frac{pm}{N}\int_L G_4
        \right)$}
        (AC);
    \draw[createdbrane]
        (-1,0)
        -- node[above left,created]
        {$L$}
        (AC);

    % Endpoints of the created brane
    \fill[created]
        (-1,0) circle (2.3pt);

    \fill[created]
        (AC) circle (2.3pt);

    % Labels
    \node[
        braneC,
        above
    ]
        at (-1,1.50)
        {$\mathcal{D}_{p/N}$};

    \node[
        braneB,
        anchor=south west
    ]
        at (0.8,-0.77)
        {$mM5$};

\end{scope}

% ============================================================
% Coordinate directions
% ============================================================
\begin{scope}[shift={(10.20,-0.20)}]

    \coordinate (O) at (0,0);

    % x^8: horizontal to the right
    \draw[->,thick]
        (O) -- (1.25,0)
        node[right] {$x^{8}$};

    % x^{1,2,3,4}: vertically upward
    \draw[->,thick]
        (O) -- (0,1.25)
        node[above] {$x^{1,2,3,4}$};

    % x^{9,10,11}: diagonally into the third quadrant
    \draw[->,thick]
        (O) -- (-0.95,-0.95)
        node[below right,align=center]
        {$x^{9,10,11}$};

\end{scope}

    \end{tikzpicture}

    \caption{
        The symmetry action of $\mathcal{D}_{p/N}$ on 5-brane stack,
        with the \(x^{5,6,7}\) directions suppressed to a point.
        The horizontal, vertical, and diagonal directions represent
        \(x^8\), \(x^{1,2,3,4}\), and \(x^{9,10,11}\),
        respectively. The blue brane is represented by a slanted
        line, with no associated blue shaded region. After the
        crossing, a suspended brane is created between the red and
        blue branes.
    }
    \label{fig:hw-move-3d-second-projection}

\end{figure}
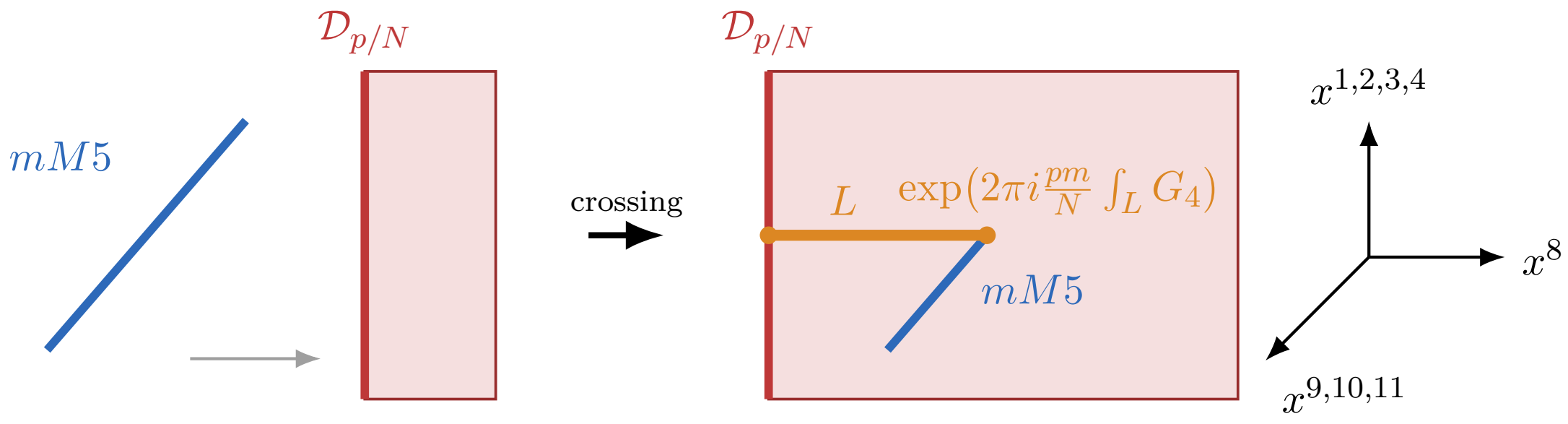

The topological operators also have an interesting effect on M5-branes which was shown in \cite{Garcia-Valdecasas:2023mis}. If we consider a 6d worldvolume $H_m[\Sigma_6]$ supporting $m$ units of magnetic flux, then passing the non-invertible operator $\mathcal D_{p/N}$ transversely through the magnetic defect so that they briefly intersect along a 3d locus results in a stretched sheet $\exp(2\pi i\frac{pm}N\int G_4)$ between the defects. In fact we point out that an equivalent way to see this may be at the level of the equations of motion. The above equation of motion for $G_7$ was derived by varying the $C_3$ field in the absence of any brane charges. Of course if we include an electric M2-brane supported on worldvolume $\gamma_3$ then this contributes a source:
\begin{equation}
    dG_7=\frac 12G_4G_4-I_8+\delta^{(8)}(\gamma_3)
\end{equation}
However if we insert an M5-brane then we know that the $C_3$ field also couples to the M5-brane, and thus we must also take into account its variation. The resulting contribution to the equation of motion is \cite{Heidenreich:2020pkc}:
\begin{equation}
    dG_7=\frac 12G_4G_4-I_8+H_3\delta^{(5)}(\Sigma_6)
\end{equation}
where recall $dH_3=G_4$. This can for example be argued using our original requirement that the M5-brane partition function cancel any gauge variation from the bulk action. If we vary the $C_3$ field in order to obtain its equation motion by $C_3\to C_3+\delta C_3$ then the 11d Chern-Simons term will also vary, and thus the anomalous variation from a gauge transformation on the M5-brane worldvolume will also change. Thus the M5-brane variation must respond with an opposite variation in order to cancel all gauge anomalies. The resulting contribution to the equations of motion for this variation is as above. We see that this manifests the additional junction between the $\mathcal A^{N,p}[M_7,G_4]$ and the M5-brane since $dG_4=\delta^{(5)}(M_6)$. \\

Another point about this contribution to the electric defect can be made if we trivialize the Thom class $J^{(5)}(\Sigma_6)=dh_4\sim d\rho\wedge e_4/2$. Integrating by parts then allows us to write this contribution schematically as:
\begin{equation}
    dG_7\sim  h_4\wedge G_4+\dots
\end{equation}
We recognize this as taking precisely the same form as the equations relating 6-group currents. For example, in the SymTFT for our 6-group gauge theory (really the $\frak G$ theory) considered in \hyperref[sec4]{Section 4}, the equation of motion for $c_3$ gave:
\begin{equation}
    df_7= h_4\wedge dc_3
\end{equation}
An important consequence of the above action of the non-invertible symmetry defect $\mathcal A^{N,p}$ on the magnetic defect $H_m[\Sigma_6]$ which was highlighted in \cite{Garcia-Valdecasas:2023mis} following \cite{Choi:2022fgx} is that if the total magnetic flux $m$ is not a multiple of $N$ then in fact the symmetry defect annihilates the magnetic source. This can be seen by shrinking the defect to a point after the move is made, which results in a topological boundary for the sheet $\exp(2\pi i \frac{pm}N\int F_4)$, which is impossible since the magnetic symmetry acts faithfully. \\

We note that this is the same conditions as the constraint on the Page charge operators $U[\omega_3]=\exp(iP[\omega_3])$ where $P[\omega_3]=\int_{N_{10}}\omega_3\wedge P_7$ for $\omega_3\in \Omega^3_{dR}(M)$ constructed in \cite{Moore:2004jv}. Due to the noncommutativity of these Page charges:
\begin{equation}
    [P[\omega_3],P[\omega_3']]=\frac{i}{2\pi}\int\omega_3\wedge\omega_3'\wedge G_4
\end{equation}
Then by gauge invariance $\omega_3$ is restricted to the magnetic translation group satisfying $\int_{N_{10}} \omega_3\wedge \phi_3\wedge G_4\in \mathbb Z$ for all $\phi_3\in H^3(X,\mathbb Z)$. In the presence of a single magnetic flux $\frac 1{2\pi}\int_{Z_4}G_4=m$ then this condition recovers the above constraint. \\

\subsection*{Global Features of Electric Charge}

It is useful to recall how electric flux and electric charge are typically characterized in ordinary ($p$-form) Maxwell theory, and how this can be disrupted even by simple ordinary Chern-Simons terms. In pure $p$-form Maxwell theory, there is a natural definition of electric flux in the Hamiltonian formulation given in \cite{Freed:2006yc}. They define a state $\psi(\check A_p)$ for $\check A_p\in \check H^p(M_d)$ to have definite electric flux $\check {\mathcal E}\in \check H^{d-p}(M_d)$ if it is a eigenstate under translation by a character:
\begin{equation}
    \psi(\check A_p+\check \chi_p)=e^{2\pi i \int\check{\mathcal E}\star \check \chi}\psi(\check A_p) \quad \text{for all } \quad \check \chi\in \check H^p(M_d)
\end{equation}
We typically only care about the topological class of the electric flux $e=c(\check{\mathcal E})\in H^{n-p}(M,\mathbb Z)$, which can be isolated by restricting to translation by only flat characters $\chi \in H^{p-1}(M,\mathbb{R/Z})$:
\begin{equation}
    \psi(A+\chi)=e^{2\pi i\int \chi\cup e}\psi(A)
\end{equation}
Therefore the Hilbert space admits a natural grading into electric flux sectors $\mathcal H(M_d)=\bigoplus_e\mathcal H_e$. Likewise there exists a grading by a magnetic flux $m=c(A)\in H^p(M,\mathbb Z)$ which is simply defined to be the topological class of the gauge field configuration. However as shown in \cite{Freed:2006yc} this decomposition is disrupted if we add even an ordinary Chern-Simons term $\int A\star A$. The magnetic flux is then constrained by the tadpole condition to be at most torsion, while the decomposition of the Hilbert space into electric flux is replaced by the condition that it represents a Heisenberg group of the Wilson lines. \\

Nevertheless one can still approach the characterization of electric charge in the case of 11d supergravity by studying the Gauss law in Hamiltonian quantization \cite{Diaconescu:2003bm}. This proceeds by considering the physical wavefunction $\Psi(C)$ associated to a hypersurface $N_{10}$ as a gauge-invariant section of a line bundle over the moduli space of gauge field configurations. The requirement of gauge invariance is summarized by the Gauss law:\footnote{See \cite{Diaconescu:2003bm} for a thorough analysis and further details of this construction.}
\begin{equation}
    \gamma\cdot \Psi(C)=\varphi(C,\gamma)^*\cdot e^{i\int \mathcal A}\Psi(C)
\end{equation}
where $\gamma$ denotes the action of a gauge transformation and $\varphi(C,\gamma)$ is a cocyle constructed in \cite{Diaconescu:2003bm}. The definition of this cocycle followed that proposed by Witten \cite{Witten:1996hc} as:
\begin{equation}
    \varphi(\check C,\check \chi)=\Phi(\check C+\check t\cdot \check \chi)|_{N_{10}\times S^1}
\end{equation}
where $t\in H^1(S^1,\mathbb Z)$ is the generator and we take Neveu-Schwarz spin structure on $S^1$. \\

The gauge transformations can be split into special cases. Topologically trivial transformations reproduce the Page charges \cite{Moore:2004jv} discussed above. To isolate the topological class of the electric charge \cite{Diaconescu:2003bm} considers topologically nontrivial but flat gauge transformations $\gamma_2 \in H^2(M, \mathbb{R/Z})$. Then the prefactor reduces to a homomorphism:
\begin{equation}
    \varphi:H^2(M,U(1))\to U(1)
\end{equation}
and so naturally defines a class $\Theta\in H^8(N,\mathbb Z)$ by Poincare duality such that $\varphi(C,\gamma_2)=e^{i \braket{\Theta_8,\gamma_2}}$. This class $\Theta$ is naturally interpreted as the topological class of the electric charge for a pure gauge configuration. As an integral class it gives a quadratic refinement of the cup product on integral cohomology. It’s de Rham refinement is: 
\begin{equation}\label{eq - global electric charge}
    [\Theta]_{dR}=\frac 12G_4G_4-I_8
\end{equation}
It would be interesting to understand if $\Theta$ plays the role of the topological class defining the quadratic refinement $Q(C)$ which we have used to define the non-invertible defect modeling the M5-brane charge. In this way, it would be the most accessible version of a ``Postnikov class" for characterizing $\frak G$.  We also note that this expression of the Gauss law includes only pure gauge field backgrounds. Crucially, if we study states which include M5-branes then the Gauss law will be modified due to the gauge variation of the M5-brane effective action. This will result in a generalization of the tadpole condition considered in \cite{Diaconescu:2003bm}, which necessarily relates the electric charge of an M5-brane to $\Theta$ on the grounds of requiring a gauge-invariant physical state. Our key point is that such a relation is reminiscent of our proposal that the charge of the M5-brane is defined by the half-space gauging defect whose 7d Hall response is $Q(C)$. Moreover, one could consider more general configurations including both M2 and M5 branes, where an M2 is allowed to end on an M5-brane. We hope that a careful treatment of such situations will shed more light on our proposal for the non-invertible charges of the M5-branes.

\subsection{Aspects of the Higher Structure of M-Theory Charges}\label{sec55}

We have emphasized that the gauge charges of the M2 and M5 branes do not resemble the charges of a genuine 6-group $\mathbb G^{(6)}$, but rather some unspecified structure $\frak G$. There are two obvious distinctions which separate them:
\begin{enumerate}
    \item The 3-form gauge field $C_3$ for the 6-group $\mathbb G^{(6)}$ is an ordinary 3-form abelian gauge field $\check C_3\in \check H^4(M_{11})$, while the 3-form gauge field of M-theory has a shifted flux quantization condition $G_4 = \frac 12 w_4 \mod \mathbb Z$. 
    \item The Postnikov class for the 6-group $\mathbb G^{(6)}$ is a genuine cohomology class represented by the characteristic class $G_4^2\in H^8(B^3U(1),\mathbb Z)$. On the other hand, the analogous candidate for the Postnikov class of $\frak G$ is the 8-form $\Theta$ defining the quadratic form $Q(C)$, whose de Rham refinement takes the form: 
    \begin{equation}
        [\Theta]_{dR}=\frac 12G_4^2-I_8
    \end{equation}
    This does not obviously correspond to an integral cohomology class of any classifying space. As shown in the appendix, the quadratic refinement and this corresponding 8-form are most straightforwardly defined on Spin manifolds using the index of the Dirac operator coupled to an $E_8$ gauge bundle. 
\end{enumerate}

This leaves the natural question: what is the appropriate mathematical object which captures these features? We do not offer a specific proposal here, but instead discuss various features of these objects, as well as survey some proposals which exist in the literature. 

\subsection*{Wu-Chern-Simons and Generalized Cohomology}

For simplicity let us first return to the case of the ordinary 2-group $\mathbb G^{(2)}$ defined by the Postnikov class $c_1^2\in H^3(BU(1),U(1))\cong H^4(BU(1),\mathbb Z)$. Under the Chern-Weil isomorphism this Postnikov class is represented by the characteristic class $\frac{1}{4\pi^2}\int F\wedge F$, which defines the bosonic Chern-Simons theory at level 2 appearing in the 2-group field strength: 
\begin{equation}
    H_3=dB_2+\frac{1}{2\pi}A_1dA_1
\end{equation}
However our case instead deals with spin-Chern-Simons theory at level 1. This is not defined by any characteristic class of $BU(1)$ defined by ordinary cohomology, but it does admit a classification by generalized cohomology. In fact there are multiple options for the appropriate generalized cohomology. The “minimal choice” is the $E_n$ cohomology considered in \cite{Freed:2006mx, jenquin2005classicalchernsimonsmanifoldsspin, Belov:2005ze, Freed:2017rlk}, whose spectrum is defined as a simple fibration of the Eilenberg-MacLane spectrum:
\begin{equation}
    H\mathbb Z_n\to E_n \to H\mathbb Z/2\mathbb Z_{n-2}
\end{equation}
classified by the stable cohomology operation $\beta\circ Sq^2$, and thus yields the long exact sequence in cohomology:
\begin{equation}
    \dots\to H^n(X)\to E^n(X)\to H^{n-2}(X,\mathbb Z_2)\xrightarrow{\beta \circ Sq^2}H^{n+1}(X)\to \dots
\end{equation}
There exists a homomorphism $H^4(BU(1))\to E^4(BU(1))$ which maps the generator to twice the generator, so in this sense the generator of $E^4(BU(1))$ can be considered $\frac 12 c_1^2$. This cohomology theory has only two nontrivial homotopy groups compared to the single homotopy group of the Eilenberg MacLane spectrum. In this sense the theory is minimal. The group $E^2(X)$ may be interpreted as the group of isomorphism classes of $\mathbb Z_2$-graded line bundles on $X$, while $E^3(X)$ gives the $\mathbb Z_2$-graded gerbes on $X$, etc. However this is not the unique choice. There also exists such a generator in the generalized cohomology theory given by the Anderson dual of spin cobordism $I\Omega^{spin}_{\bullet}(BU(1))$. Indeed the spectrum $E_\bullet$ may be viewed as a truncation of this spin cobordism cohomology \cite{Lee:2020ojw}. In fact an attractive feature of the spin cobordism generalized cohomology theory is that it also naturally includes generators corresponding to the gravitational class  $\frac 1{48}p_1(TM)\in I\Omega^{4}_{spin}(BU(1))$. Of course these generalized cohomology theories also admit differential refinements. However to our knowledge there is no known generalization of higher groups classified by such generalized (differential) cohomology. It would be interesting to investigate the existence of such a structure for the cohomology theory $E_\bullet$. \\

One may also consider the higher-dimensional analogues of spin-Chern-Simons theories for $(2p+1)$-form gauge fields in $(4p+3)$-dimensions. These are given by Wu-Chern-Simons theories, where the partition function both requires and depends on a choice of Wu structure. In \cite{Hopkins:2002rd}  quadratic refinements are constructed for the pairing on generalized differential cocycles $\check \chi \in \check H^{2p+2}_\nu(M)$ shifted by the Wu class $\nu_{2p+2}$. A different approach is taken in \cite{Monnier:2016jlo} which constructs a generalization of the above $E_\bullet$ cohomology theory, which is used to define a more general class of Wu-Chern-Simons theories. It would be useful to also perform an analysis of the Wu cobordism groups $\Omega^{Wu}_{\bullet}$ and the generators of their Anderson duals. While we do not include an analysis here, we hope to report on this in future work. \\

\subsection*{Hypothesis H}

There has been great effort towards various proposals for the formal structure defining the charge algebra of M-theory. This includes investigations to find the correct model of flux quantization for the electric/magnetic supergravity fluxes $G_4,G_7$, as well as the charges of M2 and M5 branes themselves. One major proposal is \textit{Hypothesis H} \cite{Sati:2013rxa, Fiorenza:2015gla, Fiorenza:2016ypo, Fiorenza:2019ckz, Fiorenza:2019usl, Fiorenza:2019ain, Fiorenza:2020hiq, Sati:2020cml, Sati:2021uhj, Grady:2025kld}. This puts forward a conjectured generalized cohomology theory for the M2 and M5 brane charges based on differential $J$-twisted degree-4 rational unstable cohomotopy. Our goal here is not to give a comprehensive overview of this proposal. Rather, we will sketch the major definitions and highlight some connections, and for more refer to the several nice expositions \cite{Sati:2024klw, Fiorenza:2019ckz, Sati:2026scd, Banerjee:2026wop}.  \\

In brief, the idea is that the generalized cohomology of the M-theory fluxes is modeled off of the homotopy type of the 4-sphere $S^4$. Let us unpack what is meant by $J$-twisted degree-4 cohomotopy. This is a form of nonabelian cohomology. In this context, nonabelian cohomology $H^\bullet(M,\mathcal A)$ with coefficients in a space $\mathcal A$ refers to the set of homotopy classes $\pi_0(\text{Map}(M,\mathcal A))\cong[M, \mathcal A]$. In the case of Hypothesis H, this space is taken to be the 4-sphere $\mathcal A=S^4$. This defines a generalized cohomology theory which is conjectured to capture the topological classes of the electric/magnetic charges defined by the 11d supergravity fluxes $(G_7,G_4)$. A specific cocycle is represented by a map $c : M\to S^4$, and its homotopy class defines the \textit{charge sector}. \\

In order to promote this topological class into local flux represented by differential forms, the proposal is to use the \textit{Sullivan model} $\mathcal M_{\mathbb R}(S^4)$ of the 4-sphere. This is a differential graded algebra assigned to $S^4$ which is freely generated by one generator for each rational homotopy group of $S^4$. It is defined as the Chevalley-Eilenberg algebra $\text{CE}(\frak l S^4)$ of the $L_{\infty}$-algebra $\frak lS^4$ which captures the rational homotopy type of $S^4$. In particular the rational Sullivan model works over the rationalization of the homotopy algebra $\pi_k(S^4_{\mathbb Q})\cong \pi_k(S^4)\otimes_{\mathbb Z}\mathbb Q$. The cocycle $c: M\to S^4$ induces a map from the Sullivan model to the de Rham complex $\mathcal M_{\mathbb R}(S^4)\to \Omega^{\bullet}_{dR}(M)$ which yields differential forms representing the homotopy generators, while translating the relations among the generators into differential relations among the fluxes. This is the character map which defines cohomotopy. \\

Finally, the $J$\textit{-twist} allows for the incorporation of the gravitational tangential contributions, such as the shifted flux quantization for $G_4$ or the contribution of $I_8$ to $dG_7$. This is defined as a twist of the cohomology theory by allowing for an action of the spacetime $Spin$ group on $S^4$ given by $\rho : SO(5)\to \text{Aut}(S^4)$. This requires a choice of $Spin(5)$ structure on spacetime which is defined by a reduction of structure group on $TM_{11}$, which is assumed to be part of the flux data. \\ 

We take this opportunity to note a possible relationship between this approach to the M-theory charge algebra and other relevant developments in categorical symmetries capturing the charges of solitons in quantum field theory \cite{Chen:2022cyw, Chen:2023czk, Pace:2023mdo, Sheckler:2025rlk, Sheckler:2025fql}. It was noted in \cite{Chen:2022cyw} that nonlinear sigma models with target space $X$ which possess solitons characterized by multiple different homotopy classes $\pi_k(X)$ may have corresponding non-invertible symmetries. This was clarified in \cite{Pace:2023mdo} by the explanation that the corresponding higher-groups appearing in the Postnikov tower of the target space $X$ were non-split with an unstable Postnikov invariant. This bears a direct resemblance to some features of the non-invertible charge algebra considered here. For example, in the case of target space $X=S^2$ then the Hopf fibration implies the existence of non-invertible symmetries generated by the minimal abelian TQFTs $\mathcal A^{N,p}$ in 4d $\sigma$-models or 5d gauge theories \cite{Sheckler:2025fql}. It would be extremely interesting to further understand the relevance of these developments for the charge algebra of M-theory.\footnote{A recent formalization of some of these developments was given in \cite{Perez-Lona:2026mlj}.}  \\

\subsection*{String and Fivebrane Structures}

At this point we can not help but mention the presence of another 6-group which has previously been discussed in the context of M-theory. This is the aptly-named 6-group $\textbf{Fivebrane}(n)$ which appears in the Whitehead tower of $SO(n)$. Just as the simply connected double cover $Spin(n)$ can be obtained as a group extension:
\begin{equation}
    1 \to \mathbb Z_2\to Spin(n)\to SO(n)\to 1
\end{equation}
with extension class $w_2\in H^2(BSO(n),\mathbb Z_2)$, the Whitehead tower organizes higher fibrations which kill higher homotopy classes. The 2-group $\textbf{String}(n)$ with trivial $\pi_3(\textbf{String}(n))=0$ is defined by Postknikov class $\frac 12 p_1\in H^4(BSpin(n),\mathbb Z)$, and the
6-group $\textbf{Fivebrane}(n)$ with trivial $\pi_7(\textbf{Fivebrane}(n))=0$ has Postnikov class $\frac 16p_2\in H^8(B\textbf{String}(n),\mathbb Z)$:

\[\begin{tikzcd}[cramped]
	{B^6U(1)} && {B\textbf{Fivebrane}(n)} && \\
	\\
	{B^2U(1)} && {B\textbf{String}(n)} && {K(\mathbb Z,8)} \\
	\\
	{K(\mathbb Z_2,1)} && {BSpin(n)} && {K(\mathbb Z,4)} \\
	\\
	&& {BSO(n)} && {K(\mathbb Z_2,2)}
	\arrow[from=1-1, to=1-3]
	\arrow[from=1-3, to=3-3]
	\arrow[from=3-1, to=3-3]
	\arrow["{\frac 16p_2}", from=3-3, to=3-5]
	\arrow[from=3-3, to=5-3]
	\arrow[from=5-1, to=5-3]
	\arrow["{\frac 12p_1}", from=5-3, to=5-5]
	\arrow[from=5-3, to=7-3]
	\arrow["{w_2}", from=7-3, to=7-5]
\end{tikzcd}\] \\

Just as spacetime may be endowed with a choice of Spin structure through a choice of priveleged trivialization of $w_2(TM_{11})$, so too can ``String structures" and ``Fivebrane structures" be chosen by trivializations of the Postnikov classes $\frac 12p_1$ and $\frac 16p_2$. Such structures were investigated in the context of M-theory in \cite{Sati:2008kz, Sati:2010ss, Sati:2009ic, Fiorenza:2012mr, Fiorenza:2012tb}, where they were related for example to anomaly cancellation on brane worldvolume theories and proposals for the flux quantization of the gauge fields. We do not offer a concrete proposal here for the relationship between the 6-group \textbf{Fivebrane} and our pseudo-6-group structure $\frak G$, however future progress on such a connection would be extremely interesting.

\section{Discussion} \label{sec6}

We have successfully constructed non-invertible topological defects which model key features of the gauge charges of M2 and M5 branes. This provides encouraging evidence that the continued study of these defects and their overarching algebraic structure may be used to further investigate the nature of these objects. This opens the door to many interesting future problems. We can partition them into two classes:
\begin{enumerate}
    \item Developing a better understanding of the overall algebraic structure governing these defects and their junctions, which we have simply referred to as $\frak G$. 
    \item Extending these constructions to related and more general regimes.
\end{enumerate}

Let us begin with the first class. We have remarked on the similarity of the structure $\frak G$ to the 6-group $\mathbb G^{(6)}$. Indeed, this has been our key point of leverage in constructing our non-invertible defects, which mirrored the construction for higher-categorical charges of finite higher groups. A clear crucial open problem is a precise characterization of this structure which elucidates the features of our construction. In \hyperref[sec55]{Section 5.5} we referred to the sophisticated advancements which have been made towards understanding the charges of M-theory. It would be fascinating to bridge these efforts with the developments being made in the field of categorical symmetries, to better uncover the compatibility with our results here. A necessary prerequisite for this is a further development of the categorical technology for continuous symmetries. Indeed, our higher-groups of interest were continuous abelian higher groups, which obstructs us from giving clean categorical interpretations of our results. \\ 

It would also be interesting to generalize and extend our results to broader cases of interest in M-theory and string theory. One specific next step which could be taken would be to  incorporate the fermionic sectors and supersymmetric structure of 11d supergravity and the brane worldvolume theories. Of course this full structure is obviously crucial for the characterization of the fluxes and branes in M-theory, but it would be useful to bridge this with developments in fermionic generalized symmetries. \\

Another clear future direction of interest would be an investigation into what degree our results may be extended to the Dp-branes and NS5 branes of type II string theory. There is a rich literature which tracks the charge quantization of the 11d M-theory $C_3$ field and branes under compactification on a circle $S^1$, under which one expects to obtain type IIA. In particular, the D-branes and RR fluxes of type II are proposed to be charge quantized by K-theory \cite{Witten:1998cd, Moore:1999gb}, and there exist many nontrivial checks that this proposal is compatible with the 11d charge quantization under $S^1$ dimensional reduction \cite{Diaconescu:2000wz, Diaconescu:2003bm, Diaconescu:2000wy, Mathai:2003mu, Golub:2026mfm}. For example, it was shown in \cite{Diaconescu:2003bm, Moore:2025tmt} that, when the $S^1$ fibration over $U_{10}$ is trivial, and $\partial Y_{11} = U_{10} \times S^1$, integrating the class $\Theta$ of  \ref{eq - global electric charge} characterizing the electric charge in M-theory recovers the K-theoretic constraint on RR-fluxes. On the other hand, dimensional reduction of generalized symmetry defects has been investigated \cite{Nardoni:2024sos, Caro-Perez:2026wxd}. It would be interesting to examine if our construction can be consistently reduced to lower dimension. \\ 

Another accessible generalization would be to consider a more general set of backgrounds. For example, one may consider M5-branes transversely probing singular backgrounds or orientifold planes. In the decoupling limit, the former correspond to a wide class of 6D $(1,0)$ SCFTs associated with $ADE$-Lie algebra known as the conformal matter \cite{DelZotto:2014hpa} , while the latter leads to the 6D $\mathfrak{d}_N$ $(2,0)$ SCFT \cite{Dasgupta:1995zm, Witten:1995em}. A necessary step in order to carry out this generalization is an understanding of the associated 7d bulk theory. In the example of the type $\frak d_N$ theories this might be derived from the fibration of the tubular neighborhood:
\begin{equation}
    \mathbb{RP}^4\to X_{10}\to W_6
\end{equation}
An analysis of the generalized symmetries of the 6d (2,0) type $\frak{d}_N$ theories was performed via holography in \cite{Bonetti:2024etn}, including novel Hanany-Witten moves. In the example of conformal matter, there is similar fibration of the tubular neighborhood:
\begin{equation}
    \widetilde{S^4/\Gamma_{\mathrm{ADE}}} \rightarrow X_{10} \rightarrow W_6
\end{equation}
where $\widetilde{S^4/\Gamma_{\mathrm{ADE}}}$ denotes the transverse $S^4/\Gamma_{\mathrm{ADE}}$ with the singularities at north pole and south pole resolved \cite{Ohmori:2014kda,Bah:2019rgq}. It would be extremely interesting to investigate if the construction given here could be generalized to M-theories on these geometries. \\

\bigskip

\textbf{Acknowledgments}
We thank John McGreevy and Shi Chen for useful discussions. We acknowledge Claude Opus 5.5 (Anthropic) for helpful discussions and feedback on the drafts.  
This work was supported by the Simons Collaboration on Global Categorical Symmetries, Simons Foundation award 568420, and by DOE award DE-SC0009919.

\appendix

\section{Quadratic and Cubic Refinements}\label{appa}

We now review relevant facts and constructions for the quadratic and cubic refinements of pairings on differential cohomology, which are used to define the Chern-Simons terms of interest in this work. \\

\subsection{Quadratic Refinements}

Differential cohomology has a natural product operation:
\begin{equation}
    \check A_p\star \check B_q : \check H^p(M)\times \check H^q(M)\to \check H^{p+q}(M) 
\end{equation}
which has the characteristic property that it takes the wedge product of curvatures $F_{A\star B}=F_A\wedge F_B $ and cup product of characteristic classes $c_{A\star B} = c_A\cup c_B$. In particular, when the dimension of the spacetime $M_d$ is $d=4p+3$ this gives rise to a natural pairing on differential cocycles in $\check H^{2p+2}(M)$:
\begin{equation}
    (\check A_p, \check B_p) \equiv \int_M \check A_{2p+1}\star \check B_{2p+1} \in U(1)
\end{equation}
This pairing gives a global definition for the Chern-Simons action. Indeed, for topologically trivial $A$ we have
\begin{equation}
    (\check A, \check A)= \frac{1}{2\pi}\int_M A_{2p+1}dA_{2p+1}
\end{equation}
Importantly, on 3-manifolds supporting a Spin structure (or more generally on $4p+3$-dim manifolds  a Wu structure for the Wu class $\nu_{2p+2}$) this pairing has a quadratic refinement $Q(\check A)$ satisfying:
\begin{equation}\label{quadref}
    Q(\check A+\check B)-Q(\check A)-Q(\check B)+Q(0)=(\check A, \check B)
\end{equation}
In 3d this quadratic refinement gives a global definition for Spin-Chern-Simons theory:
\begin{equation}
    Q(\check A)\sim \frac{1}{4\pi}\int \check A_1d\check A_1
\end{equation}
which depends on the choice of spin structure. More generally on $4p+3$-dim manifolds it defines a Wu-Chern-Simons theory, and depends on the choice of Wu structure. \\ 

There are several different constructions of Wu-Chern-Simons theories and their associated quadratic forms. A general construction was given in \cite{Hopkins:2002rd} using shifted differential cohomology. Another approach was given in \cite{Monnier:2016jlo} using an extension of the generalized cohomology $E^\bullet$ discussed in \hyperref[sec55]{Section 5.5}. In each model the Wu-Chern-Simons action is constructed using a differential refinement of the respective generalized cohomology, but each still defines a quadratic refinement for the natural pairing on differential cohomology. In the main body of this work, the key property we will use is the quadratic refinement property \hyperref[quadref]{(A.4)}. In this sense, our results should be viewed as agnostic to the underlying construction, and therefore model-independent. The only exception is when we actually discuss the derivation from M-theory. We will point out below how the quadratic refinement property can be derived directly from the definition of the M-theory Chern-Simons term. \\

To remain concrete, we will review one explicit construction of the quadratic refinement following the discussion in \cite{Hsieh:2020jpj}, see there for more details. We will focus on the dimensions of interest to our examples in \hyperref[sec4]{Section 4}: namely $d=3$ and $d=7$. As with the traditional definition of the Chern-Simons action, we could try to define $Q(A)=\int_{M_d}\text{CS}(A)$ on a manifold $M_d$ by extending $A_p$ to a bounding manifold $Z_{d+1}$ with $M_d=\partial Z_{d+1}$:
\begin{equation}
    \exp(iQ(A))\ = \exp\Big(\frac i{4\pi} \int_{Z_{d+1}}F\wedge F\Big)
\end{equation}
However a crucial feature of such a definition requires ensuring an independence of the choice of extension $Z_{d+1}$, which will require us to include additional gravitational contributions. Moreover, when we consider manifolds $M_d$ which themselves have a boundary, which we must when dealing with solitonic objects such as M5-branes, then this definition no longer becomes viable. It would be better to have an intrinsically $d$-dim definition which still maintains the above property. \\

The key will be to use the APS index theorem. Recall that the APS $\eta$ invariant on a manifold $M_d$ with Dirac-type operator $\mathcal D_M$ is defined as the regulated sum over eigenvalues $\lambda$:
\begin{equation}
    \eta(\mathcal D_M)\equiv \frac 12\Big[\sum_\lambda\text{sgn}(\lambda)\Big]_{reg}
\end{equation}
Consider a manifold $Z_{d+1}$ with Dirac operator $\mathcal D_Z$ and $\Z_2$ grading matrix $\Gamma$ satisfying $\{\mathcal D_Z,\Gamma\}=0$. Then the index of $\mathcal D_Z$ is defined as $n_+-n_-$ where $n_{\pm}$ are the number of Dirac zeromodes with $\Gamma=\pm 1$ respectively. The APS index theorem says that if $M_d=\partial Z_{d+1}$ then:
\begin{equation}
    \text{index}\ \mathcal D_Z=\int_Z\hat A(R)\text{ch}(F)+\eta(\mathcal D_Y)
\end{equation}

We can use this to define the quadratic refinement in each dimension: \\ 

$\mathbf{d=3}:$ \\

Given a $U(1)$ gauge field $\check A_1\in \check H^2(M_3)$ on a spin 3-manfiold $M_3$ then we define the quadratic refinement:
\begin{equation}
    Q(\check A_1)=-\eta(\mathcal D_M)
\end{equation}
where $\mathcal D_M$ is a Dirac operator coupled to the $U(1)$ bundle defining $\check A_1$. This is an intrinsically 3d definition as desired. However in the case that $M_3=\partial Z_4$, we can suitably extend the $U(1)$ bundle and spin structure since $\Omega_3^{spin}(BU(1))=0$, then by the APS index theorem this can be written as:
\begin{equation}
    Q(A)=\int_Z \frac 12 F\wedge F+\hat A_1(R) \mod \mathbb Z
\end{equation}
where $\hat A_k(R)$ is the $4k$-form part of $\hat A(R)$. It is then straightforward to see that $Q(\check A)$ is a quadratic refinement:
\begin{equation}
    Q(A+B)-Q(A)-Q(B)+Q(0)=(A,B)=\int_Z F_A\wedge F_B
\end{equation}

$\mathbf{d=7:}$ \\

In the case of a 3-form $U(1)$ gauge field $\check C_3\in \check H^4(M_7)$ then whenever $M_7=\partial Z_8$ we may once again define the quadratic refinement using the bulk:
\begin{equation}
    Q(A)=\int_Z \frac 12 G_4\wedge G_4-\frac 14p_1(R)\wedge G_4+28\hat A_2(R)
\end{equation}
which can be easily seen to satisfy the quadratic refinement property. We can see that $\int G\wedge G-\frac 12 p_1\wedge G$ is even on spin manifolds by combining the facts that $p_1=\frak P(w_2)+2w_4 \mod 4$ and since $w_2$ vanishes on spin manifolds then the Wu class reduces to $\nu_4=w_4$ so that $\int w_4\cup G=\int G\cup G$. However one must verify that this definition is independent of the choice of extension $Z_8$. This was shown in \cite{Hsieh:2020jpj} building on \cite{Witten:1996md} using an $E_7$ model for a 3-form gauge field in $\check H^4(M_7)$ in 7d, motivated by the observation that $\pi_k(BE_7)=\pi_k(K(\mathbb Z,4))$ for $k\leq 10$. In this model, we consider a Dirac operator coupled to an $E_7$ bundle and model $C_3\in \check H^4(M_7)$ as the Chern-Simons form for the 56-dim representation of an $E_7$ gauge field with field strength $F_{E_7}$. In particular, we identify:
\begin{equation}
    G_4=\frac{1}{24}\text{Tr}_{56}\Big(\frac{i}{2\pi}F_{E_7}\Big)^2
\end{equation}
We then define:
\begin{equation}
    Q(\check C)\equiv -\frac 12\eta(\mathcal D_Y)
\end{equation}
The APS index theorem then shows that:
\begin{equation}
    \begin{gathered}
        \text{index}\  \mathcal D_Z- \eta(\mathcal D_Y)=\int_Z\hat A(R)\text{Tr}_{56}(e^{\frac{i}{2\pi}F_{E_7}}) \\ 
        =\int G\wedge G-\frac 12 p_1\wedge G+56\hat A_2(R)
    \end{gathered}
\end{equation}
The index is even so we find: 
\begin{equation}
    Q(\check C)=-\frac 12\eta(\mathcal D_Y)=\int_Z \frac 12 G_4\wedge G_4-\frac 14p_1(R)\wedge G_4+28\hat A_2(R)\mod \mathbb Z
\end{equation}
which matches the above bulk definition. \\

\subsection{The M-Theory Chern-Simons Form and Cubic Refinements}
We now review the issue of defining the cubic Chern-Simons form appearing in the 11d supergravity action:
\begin{equation}
    \Phi(C_3)\sim \exp\Big(i\int_{M_{11}}\frac 16C_3G_4G_4-C_3I_8\Big)
\end{equation}
We follow the discussion of \cite{Diaconescu:2003bm}, see there for more details. A valid definition on a closed 11d spin manifold $M_{11}$ which can be extended to a 12d bulk $Z_{12}$ was given by Witten \cite{Witten:1996md}:
\begin{equation}
    \Phi(C_3)\equiv \exp\Big(i\int_Z\frac 16G_4^3-G_4I_8\Big)
\end{equation}
To show that this definition is independent of the choice of extension (up to a sign cancelled by the fermionic contribution), Witten used an $E_8$ model for the field $C_3$, which models the 3-form gauge field as the Chern-Simons form for an $E_8$ gauge field $C_3\sim CS(A_{E_8})-CS(g)$. Specifically, one identifies:
\begin{equation}
    G_4\cong\frac{1}{60}\text{Tr}_{248}\Big(\frac{1}{8\pi^2}F_{E_8}^2\Big)+\frac 1{32\pi^2}\text{Tr}_{11}R^2
\end{equation}
One motivation for this definition is that $\pi_k(BE_8)=\pi_k(K(\mathbb Z,4))$ for $k\leq 15$. Another key fact is that $\Omega_{11}^{spin}(K(\mathbb Z,4))=0$ \cite{Stong:1985vj}. The equivalence of this model with the model of the $C_3$ field as a shifted differential cocycle was shown in \cite{Diaconescu:2003bm}. As with the Wu-Chern-Simons theory, this Chern-Simons form also admits an intrinsically 11d definition by means of the APS index theorem, as argued in \cite{Diaconescu:2003bm}. Their definition considers the index density $i(\mathcal D_{E_8})$ of a Dirac operators $\mathcal D_{E_8}$ coupled to the $E_8$ bundle:
\begin{equation}
    i(\mathcal D_{E_8})=\text{Tr}_{248}e^{\frac 1{2\pi}F_{E_8}}\hat A(g)
\end{equation}
along with the gravitino index density:
\begin{equation}
    i(\mathcal D_{RS})=\hat A\text{Tr}(e^{\frac{1}{2\pi}R}-1)+8\hat A
\end{equation}
from which the 12-form part on a 12d manifold can be identified as:
\begin{equation}
    \frac 12\Big[i(\mathcal D_{E_8})+\frac 12i(\mathcal D_{RS})\Big]^{(12)}=\frac 16 G^3-G I_8-I_{local}
\end{equation}
Once again one may then use the index theorem to relate these to the $\xi$-invariant  $\xi(\mathcal D)=\eta(\mathcal D)+h(\mathcal D)$. Their intrinsically 11d definition of the Chern-Simons form is then:
\begin{equation}
    \Phi(C)\equiv \exp\Big[i\pi\Big(\xi(\mathcal D_{E_8})+\frac 12\xi(\mathcal D_{RS}) + 2I_{local}\Big)\Big]
\end{equation}
An intriguing property of the M-theory Chern-Simons form is that, similar to the quadratic Wu-Chern-Simons form, this term defines a cubic refinement of the trilinear form:
\begin{equation}
    (\check A_k, \check B_k,\check C_k)\equiv \int_{M_{3k-1}}\check A_k\star \check B_k \star \check C_k:\check H^k(M)\times \check H^k(M)\times \check H^k(M)\to U(1)
\end{equation}
The defining property of a cubic refinement is the following:
\begin{equation}
    \frac{\Phi(A+B+C)\Phi(A)\Phi(B)\Phi(C)}{\Phi(A+B)\Phi(B+C)\Phi(A+C)\Phi(0)}=\exp\Big[i(A,B,C)\Big]
\end{equation}
The fact that $\Phi$ satisfies this property follows easily from the 12d definition. A more thorough discussion of cubic forms is given in \cite{Freed:2019sco}, where it was shown that the definition of the M-theory Chern-Simons form can be extended to nonorientable manifolds using a differential $\frak m_c $ structure for $\check C$. 

\subsection{Quadratic from Cubic} \label{Appendix A.3 Quadratic from Cubic}

Finally, we note a trivial but useful fact that cubic forms can be used to define quadratic forms. If we fix a class $\check \lambda\in \check H^k(M)$ and define the bilinear pairing:
\begin{equation}
    \braket{\check A_k,\check B_k}_{\lambda} \equiv (\check A_k,\check B_k,\check \lambda) : \check H^k(M)\times \check H^k(M)\to U(1)
\end{equation}
Then define the combination:
\begin{equation}
    \tilde Q_\lambda(\check C)\equiv \frac{\Phi(\check C+\lambda)}{\Phi(\check C)}
\end{equation}
then it follows from the cubic refinement property that $\tilde Q_\lambda(C)$ defines a quadratic refinement of the bilinear pairing $\exp(i\braket{A,B}_\lambda)$:
\begin{equation}
    \frac{\tilde Q_\lambda(A+B)\tilde Q_\lambda(0)}{\tilde Q_\lambda(A)\tilde Q_\lambda(B)}=\exp\Big(i\braket{A,B}_\lambda\Big)
\end{equation}
In fact this gives a natural explanation for why the effective action for the M5-brane partition function has a gauge variation equivalent to that of a quadratic refinement. Upon insertion of an M5-brane, we take a decomposition of the gauge field $G_4=\tilde G_4+h_4$ where $h_4$ has unit flux around the cycle linking the worldvolume, while $\tilde G_4$ does not. We can then define the ratio $\Phi(\tilde G_4+h_4)/ \Phi(\tilde G_4)$ on a spacetime which is the total space of a 4-sphere fibration and perform the fiber integration. Since $\tilde G_4$ has no flux through the 4-sphere then this is equivalent to the fiber integration of $\Phi(\tilde G_4+h_4)$ alone. Performing the fiber reduction on both sides of the above equation shows that the resulting 7d action is guaranteed to be a quadratic refinement of the usual pairing $(A,B)$. \\

\bibliographystyle{JHEP}
\bibliography{biblio.bib}

\end{document}